\documentclass[12pt]{article}

\usepackage{mathtools}
\usepackage{amssymb}
\usepackage{caption}
\usepackage{enumitem}
\usepackage[colorlinks=true,allcolors=blue]{hyperref}
\usepackage{xcolor}
\definecolor{cmmt}{rgb}{0.0, 0.5, 0.0}
\usepackage{bm}
\usepackage{times}
\usepackage{setspace}
\usepackage{multirow}
\usepackage{threeparttable}
\usepackage{footnote}
\usepackage[sorting=none, backend=biber, url=false, style=nature, isbn=false, eprint=false]{biblatex}
\AtEveryBibitem{\clearfield{month}}
\AtEveryCitekey{\clearfield{month}}

\makesavenoteenv{tabular}
\makesavenoteenv{table}

\usepackage{algorithm,algorithmicx,algcompatible}
\usepackage[noend]{algpseudocode}

\usepackage{siunitx}

\DeclarePairedDelimiter{\norm}{\lVert}{\rVert}

\title{\huge A machine learning route between band mapping and band structure}

\author
{R. Patrick Xian,$^{1,\dagger,\ast,a}$ Vincent Stimper,$^{2,\dagger,\ast}$ Marios Zacharias,$^{1,b}$ Maciej Dendzik,$^{1,c}$\\
Shuo Dong,$^{1}$ Samuel Beaulieu,$^{1,d}$ Bernhard Sch{\"o}lkopf,$^{2}$ Martin Wolf,$^{1}$\\
Laurenz Rettig,$^{1}$ Christian Carbogno,$^{1}$ Stefan Bauer,$^{2,\ast,e}$ and Ralph Ernstorfer$^{1,\ast}$\\
\\
\normalsize{$^{1}$Fritz Haber Institute of the Max Planck Society, 14195 Berlin, Germany.}\\
\normalsize{$^{2}$Department of Empirical Inference, Max Planck Institute for Intelligent Systems,}\\
\normalsize{72076 T{\"u}bingen, Germany.}\\
\normalsize{$^{\dagger}$These authors contributed equally to this work.}\\
\normalsize{$^\ast$Correspondence authors:  xrpatrick AT gmail.com, vstimper AT tue.mpg.de,}\\
\normalsize{baue AT kth.se, ernstorfer AT fhi-berlin.mpg.de.}\\
\\
\normalsize{$^{a}$Current Address: Department of Mechanical Engineering, }\\
\normalsize{University College London, WC1E 7JE London, UK.}\\
\normalsize{$^{b}$Current Address: Université de Rennes, INSA Rennes, CNRS,}\\
\normalsize{Institut FOTON, F-35000 Rennes, France.}\\
\normalsize{$^{c}$Current Address: Department of Applied Physics, }\\
\normalsize{KTH Royal Institute of Technology, 114 19 Stockholm, Sweden.}\\
\normalsize{$^{d}$Current Address: Université de Bordeaux—CNRS—CEA, CELIA,}\\
\normalsize{UMR5107, F33405, Talence, France.}\\
\normalsize{$^{e}$Current Address: Division of Decision and Control Systems, }\\
\normalsize{KTH Royal Institute of Technology, 114 28 Stockholm, Sweden.}
}

\date{}

\begin{document} 

\baselineskip20pt

\maketitle 

\begingroup
\addtolength\leftmargini{-0.05in}
\begin{quote}
\textbf{Electronic band structure (BS) and crystal structure are the two complementary identifiers of solid state materials. While convenient instruments and reconstruction algorithms have made large, empirical, crystal structure databases possible, extracting quasiparticle dispersion (closely related to BS) from photoemission band mapping data is currently limited by the available computational methods. To cope with the growing size and scale of photoemission data, we develop a pipeline including probabilistic machine learning and the associated data processing, optimization and evaluation methods for band structure reconstruction, leveraging theoretical calculations. The pipeline reconstructs all 14 valence bands of a semiconductor and shows excellent performance on benchmarks and other materials datasets. The reconstruction uncovers previously inaccessible momentum-space structural information on both global and local scales, while realizing a path towards integration with materials science databases. Our approach illustrates the potential of combining machine learning and domain knowledge for scalable feature extraction in multidimensional data.}
\end{quote}
\endgroup

\section*{Introduction}
\begin{figure}[htb!]
	\begin{center}
		\includegraphics[width=\textwidth]{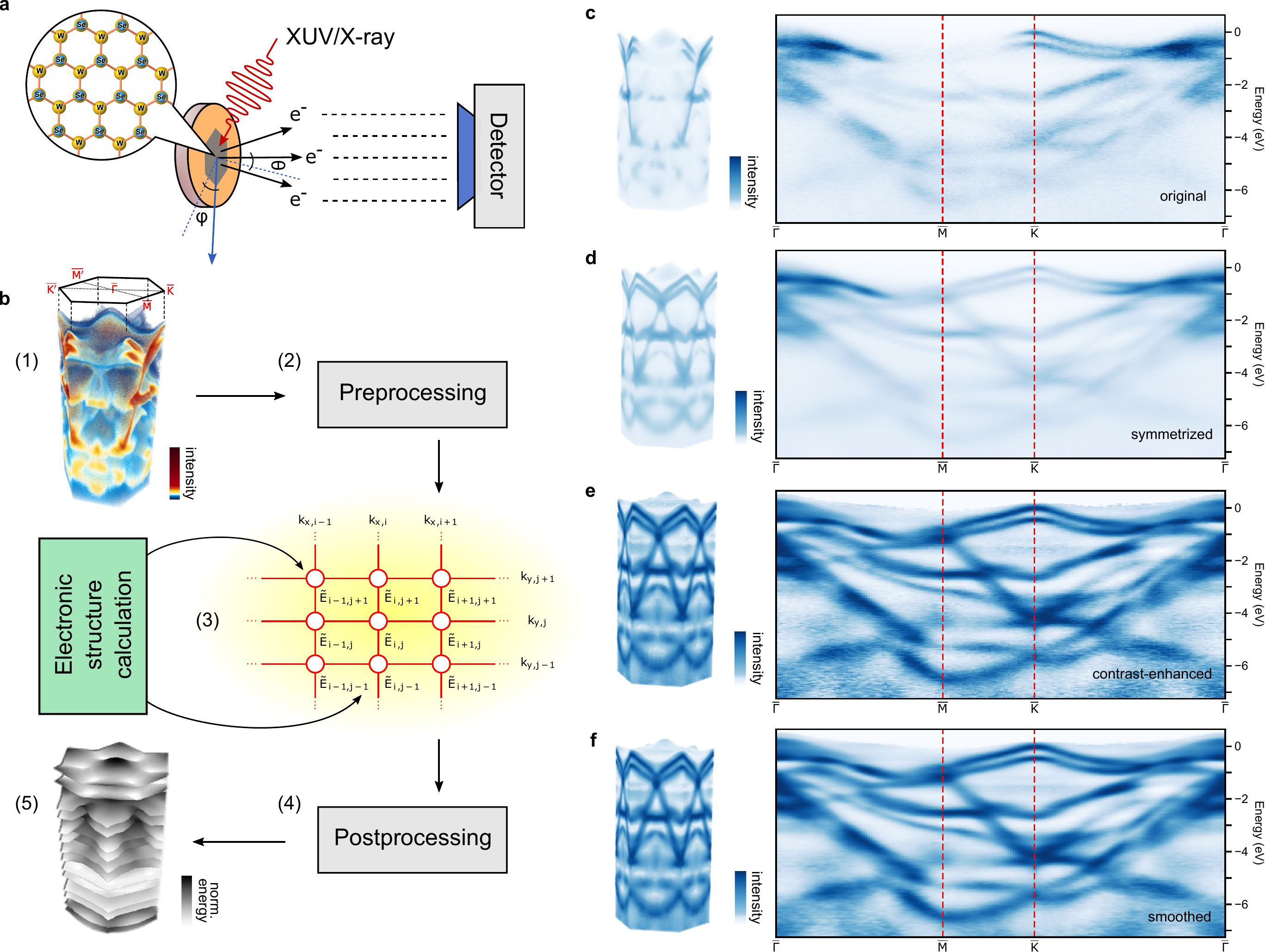}
	\end{center}
    \caption{\textbf{From band mapping to band structure.} \textbf{a}, Schematic of a photoemission band mapping experiment. The electrons from a crystalline sample's surface are liberated by extreme UV (XUV) or X-ray pulses and collected by a detector through either angular scanning or time-of-flight detection schemes. \textbf{b}, Overview of the computational framework for reconstruction of the photoemission (or quasiparticle) band structure: (1) The volumetric data obtained from a band mapping experiment first (2) go through preprocessing steps, then are (3) fed into the probabilistic machine learning algorithm along with electronic structure calculations as initialization of the optimization. The reconstruction algorithm for volumetric band mapping data is represented as a 2D probabilistic graphical model with the band energies represented as nodes, leading to tens of thousands of nodes in practice. (4) The outcome of the reconstruction is postprocessed (e.g. symmetrization) to (5) yield the dispersion surfaces (i.e. energy bands) of the photoemission band structure ordered by band indices. \textbf{c}-\textbf{f}, Effects of the intensity transforms in preprocessing viewed in both 3D and along high-symmetry lines of the projected Brillouin zone (hexagonal as in \textbf{b}(1)), from the original data (\textbf{c}) through intensity symmetrization (\textbf{d}), contrast enhancement \cite{Stimper2019} (\textbf{e}) and Gaussian smoothing of intensities (\textbf{f}). The intensity data in \textbf{c}-\textbf{f} are normalized individually for visual comparison.}
    \label{fig:overview}
\end{figure}

The modelling and characterization of the electronic BS of materials play an essential role in materials design \cite{Isaacs2018} and device simulation \cite{Marin2018}. The BS lives in the momentum space, $\Omega(k_x, k_y, k_z, E)$ and imprints the multidimensional and multi-valued functional relations between energy ($E$) and momenta ($k_x, k_y, k_z$) of periodically confined electrons \cite{Bouckaert1936}. Photoemission band mapping \cite{Chiang2001} (see Fig. \ref{fig:overview}a) using momentum- and energy-resolved photoemission spectroscopy (PES), including angle-resolved PES (ARPES) \cite{Damascelli2003,Zhang2022} and multidimensional PES \cite{Schonhense2015,Medjanik2017} measures the BS as an intensity-valued multivariate probability distribution directly in $\Omega$. The proliferation of band mapping datasets and their public availability brought about by recent hardware upgrades \cite{Schonhense2015,Medjanik2017,Puppin2019,Gauthier2020} have ushered in the possibilities of comprehensive benchmarking between theories and experiments, which become especially challenging for multiband materials with complex band dispersions \cite{Riley2014,Bahramy2018,Schroter2019}. The available methods for interpreting the photoemission spectra fall into two categories: Physics-based methods require least-squares fitting of 1D lineshapes, named energy or momentum distribution curves (EDCs or MDCs), to analytical models \cite{Valla1999,Damascelli2003,Levy2014}. Although physics-informed data models guarantee high accuracy and interpretability, upscaling the pointwise fitting (or estimation) to large, densely sampled regions in the momentum space (e.g. including $10^4$ or more momentum locations) presents challenges due to limited numerical stability and efficiency. Therefore, their use is limited to selected momentum locations determined heuristically from physical knowledge of the materials and the experimental settings. Image processing-based methods apply data transformations to improve the visibility of dispersive features \cite{Zhang2011,He2017,Peng2020,Kim2021}. They are more computationally efficient and can operate on entire datasets, yet offer only visual enhancement of the underlying band dispersion. They don't allow reconstruction and therefore are insufficient for truly quantitative benchmarking or archiving.

A method balancing the two sides will extract the band dispersion with sufficiently high accuracy and be scalable to multidimensional datasets, therefore providing the basis for distilling structural information from complex band mapping data and for building efficient tools for annotating and understanding spectra. In this regard, we propose a computational framework (see Fig. \ref{fig:overview}b) for global reconstruction of the photoemission (or quasiparticle) BS as a set of energy (or electronic) bands, formed by energy values (i.e. band loci) connected along momentum coordinates. This local connectedness assumption is more valid than using local maxima of photoemission intensities because local maxima are not always good indicators of band loci \cite{Moser2017}. We exploit the connection between theory and experiment in our framework, based on a probabilistic machine learning \cite{Murphy2012,Ghahramani2015} model to approximate the intensity data from band mapping experiments. The gist of the model is rooted in Bayes rule,
\begin{equation}
    p(\bm{X}|\mathcal{D}) \propto p(\mathcal{D}|\bm{X}) p(\bm{X}),
    \label{equ:bayes}
\end{equation}
where $\bm{X}$ are the random variables to be inferred and the data $\mathcal{D}$ are mapped directly onto unknowns and experimental observables. We assign the energy values of the photoemission BS as the model's variables to extract from data, and a nearest-neighbor (NN) Gaussian distribution as the prior, $p(\bm{X})$, to describe the proximity of energy values at nearby momenta. The EDC at every momentum grid point relates to the likelihood, $p(\mathcal{D}|\bm{X})$, when we interpret the photoemission intensity probabilistically. The optimum is obtained via \textit{maximum a posteriori} (MAP) estimation in probabilistic inference \cite{Murphy2012} (see Methods and Supplementary Fig. 2). Given the form of the NN prior, the posterior, $p(\bm{X}|\mathcal{D})$, in the current setting forms a Markov random field (MRF) \cite{Murphy2012,Wang2013,Comer2022}, which encapsulates the energy band continuity assumption and the measured intensity distribution of photoemission in a probabilistic graphical model. For one benefit, the probabilistic formulation can incorporate imperfect physical knowledge algebraically in the model or numerically as initialization (i.e. warm start, see Methods) of the MAP estimation, without requiring \textit{de facto} ground truth and training as in supervised machine learning \cite{Kaufmann2020}. For another, the graphical model representation allows convenient optimization and extension to other dimensions (see Supplementary Fig. 1 and Section 1).

To demonstrate the effectiveness of the method, we have first reconstructed the entire 3D dispersion surfaces, $E(k_x, k_y)$, of all 14 valence bands within the projected first Brillouin zone (in ($k_x$, $k_y$, $E$) coordinates) of the semiconductor tungsten diselenide (WSe$_2$), spanning $\sim$ 7~eV in energy and $\sim$ 3 $\text{\AA}^{-1}$ along each momentum direction. Furthermore, we adapt informatics tools to BS data to sample and compare the reconstructed and theoretical BSs globally. The accuracy of the reconstruction is validated using synthetic data and the extracted local structural parameters in comparison with pointwise fitting. The available data and BS informatics enable detailed comparison of band dispersion at a resolution of $<$ 0.02 \textup{\AA}$^{-1}$. Besides, we performed various tests and benchmarking on datasets of other materials and simulated data, where ground truth is available to evaluate the accuracy and computational efficiency.

\section*{Results}
\noindent\textbf{Band structure reconstruction and digitization.} Our primary example is the 2D layered semiconductor WSe$_2$ in the hexagonal lattice with a bilayer stacking periodicity (denoted as 2$H$-WSe$_2$) is a model system for band mapping experiments \cite{Traving1997,Finteis1997,Riley2014}. Earlier valence band mapping and reconstruction in ARPES experiments on WSe$_2$ have demonstrated a high degree of similarity between theory and experiments \cite{Traving1997,Finteis1997,Riley2014}, but a quantitative assessment within the entire (projected) Brillouin zone is still lacking. The valence BS of 2$H$-WSe$_2$ contains 14 strongly dispersive energy bands, formed by a mixture of the $5d^4$ and $6s^2$ orbitals of the W atoms and the $4p^4$ orbitals of the Se atoms in its hexagonal unit cell. The strong spin-orbit coupling due to heavy elements produces large momentum- and spin-dependent energy splitting and modifications to the BS \cite{Riley2014,Kormanyos2015}.

We use a 2D MRF to model the loci of an energy band within the intensity-valued 3D band mapping data, regarded as a collection of momentum-ordered EDCs. It is graphically represented by a rectangular grid overlaid on the momentum axes with the indices $(i, j)$ ($i$, $j$ are nonnegative integers), as shown in step (3) of Fig. \ref{fig:overview}b. The undetermined band energy of the EDC at $(i,j)$, with the associated momentum coordinates $(k_{x,i}, k_{y,j})$, is considered a random variable, $\Tilde{E}_{i,j}$, of the MRF. Together, the probabilistic model is characterized by a joint distribution, expressed as the product of the likelihood and the Gaussian prior, in Eq. \eqref{equ:bayes}. To maintain its simplicity, we don't explicitly account for the intensity modulations of various origins (such as imbalanced transition matrix elements \cite{Moser2017}) in the original band mapping data, which cannot be remediated by upgrading the photon source or detector. Instead, we preprocess the data to minimize their effects on the reconstruction (see Fig. \ref{fig:overview}c-f). The preprocessing steps include (1) intensity symmetrization, (2) contrast enhancement \cite{Stimper2019}, followed by (3) Gaussian smoothing (see Methods), whereafter the continuity of band-like features is restored. The EDCs from the preprocessed data, $\Tilde{I}$, are used effectively as the likelihood to calculate the MRF joint distribution,
\begin{equation}
	p(\{\Tilde{E}_{i,j}\}) = \frac{1}{Z}\prod_{ij} \Tilde{I}(k_{x,i}, k_{y,j}, \Tilde{E}_{i,j}) \cdot \prod_{(i,j)(l,m)|\text{NN}} \exp \left[ - \frac{(\Tilde{E}_{i,j}-\Tilde{E}_{l,m})^2}{2\eta^2}\right].
	\label{equ:joint_p_mrf}
\end{equation}
Here, $Z$ is a normalization constant, $\eta$ is a hyperparameter defining the width of the Gaussian prior, $\prod_{ij}$ denotes the product over all discrete momentum values sampled in the experiment and $\prod_{(i,j)(l,m)|\text{NN}}$ the product over all the NN terms. A detailed derivation of Eq. \eqref{equ:joint_p_mrf} is given in Supplementary Section 1. Reconstruction of the bands in the photoemission BS is carried out sequentially and relies on local optimization of the MRF's variables, $\{\Tilde{E}_{i,j}\}$.

To optimize over large graphical models, we adopt multiple parallelization schemes to achieve efficient operations on scalable computing hardware. A single band reconstruction involving optimization over $10^4$ random variables is achieved within seconds and hyperparameter tuning within tens of minutes (see Methods, Supplementary Figs. 3-4). In comparison, pointwise fitting often requires hand-tuning individually and therefore difficult to scale up to whole bands accordingly within a meaningful timeframe. To correctly resolve band crossings and nearly degenerate energies, we further inject relevant physical knowledge in the optimization by using density functional theory (DFT) band structure calculation with semi-local approximation \cite{Perdew2001} as a starting point for the reconstruction. The calculation qualitatively entails such physical symmetry information for WSe$_2$, albeit not quantitatively reproducing the experimental quasiparticle BSs at all momentum coordinates. As shown with four DFT calculations with different exchange-correlation functionals~\cite{Perdew2001} to initiate the reconstruction for WSe$_2$ and in various cases using synthetic data with known ground truth (see Methods, Supplementary Table \ref{Table_1} and Supplementary Figs. 4-8), the reconstruction algorithm is not particularly sensitive to the initialization as long as the information about band crossings is present. The current framework can also support the initialization from more advanced electronic structure methods, such as $GW$ \cite{Golze2019} or that including electronic self-energies renormalized by electron-phonon coupling \cite{Zacharias2020}, when semi-local approximation yields not only quantitatively, but also qualitatively wrong quasiparticle BSs compared with the experiment. However, a systematic benchmark of theory and experiment goes beyond the scope of this work.
\begin{figure}[htbp!]
    \begin{center}
		\includegraphics[width=\textwidth]{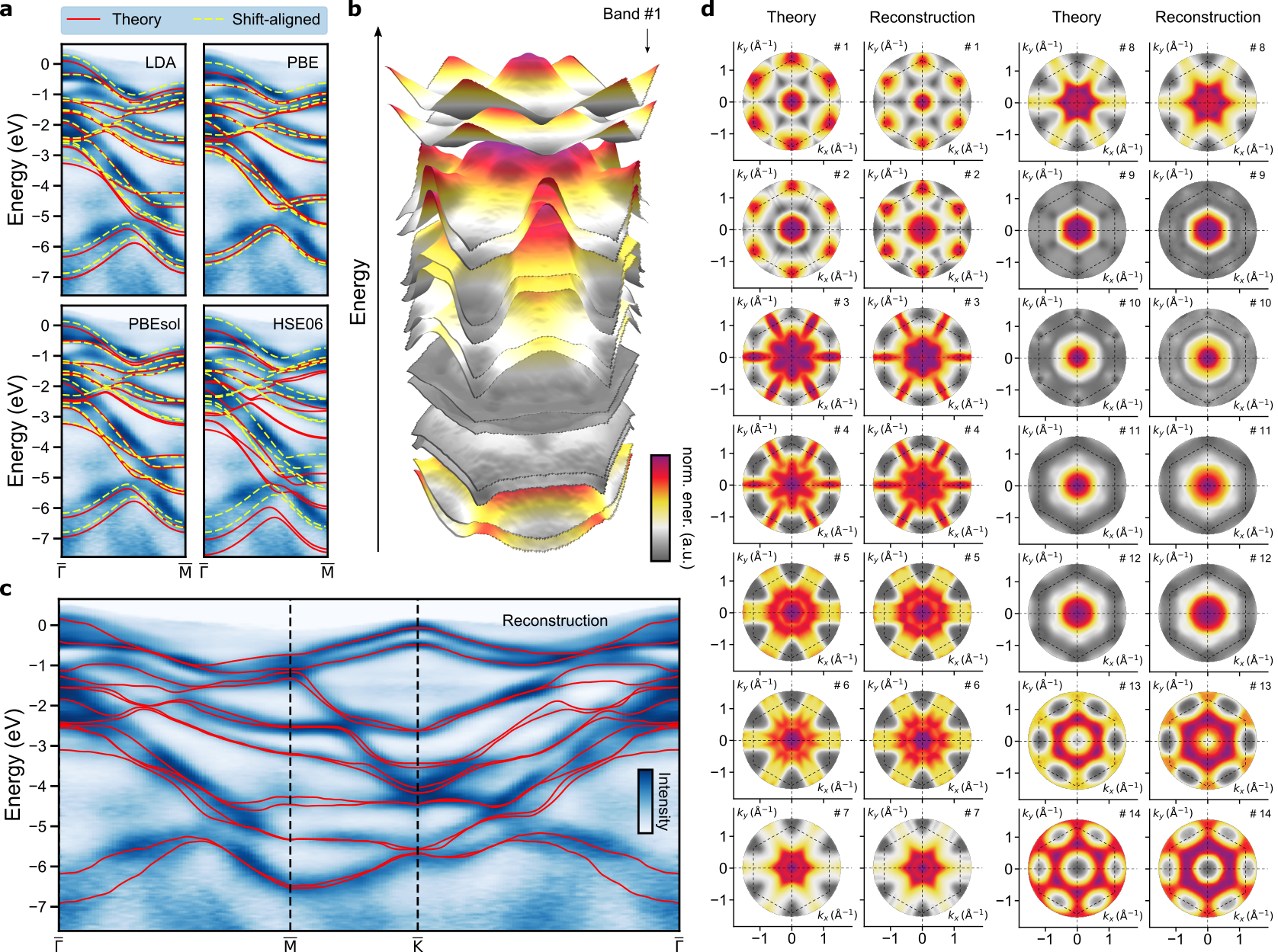}
	\end{center}
    \caption{\textbf{Band reconstruction from WSe$_2$ photoemission data.} \textbf{a}, Comparison between the preprocessed WSe$_2$ valence band photoemission data along $\overline{\Gamma}$-$\overline{\text{M}}$ direction, DFT band structure calculated with different exchange-correlation functionals (solid red lines), and their final positions after band-wise rigid-shift alignment (dashed yellow lines) as part of hyperparameter tuning. The energy zero of each DFT calculation is set at the $\overline{\mathrm{K}}$ point (not shown). \textbf{b}, Exploded view (with enlarged spacing between bands for better visibility) of reconstructed energy bands of WSe$_2$. \textbf{c}, Overlay of reconstructed band dispersion (red lines) on preprocessed photoemission band mapping data cut along the high-symmetry lines in the hexagonal Brillouin zone of WSe$_2$. \textbf{d}, Band-wise comparison between LDA-level DFT (LDA-DFT) calculation used to initialize the optimization and the reconstructed 14 valence bands of WSe$_2$ (symmetrized in postprocessing). The dashed hexagons trace out the boundaries of the first Brillouin zone. The band indices on the upper right corners in \textbf{d} follow the ordering of the electronic orbitals in this material obtained from LDA-DFT. \textbf{b} and \textbf{d} are paired plots (see Methods) that share the same colorbar, which shows the per-band normalized energy (norm. ener.) in arbitrary units (a. u.).}
    \label{fig:recon}
\end{figure}

\begin{figure}[htbp!]
    \begin{center}
		\includegraphics[width=\textwidth]{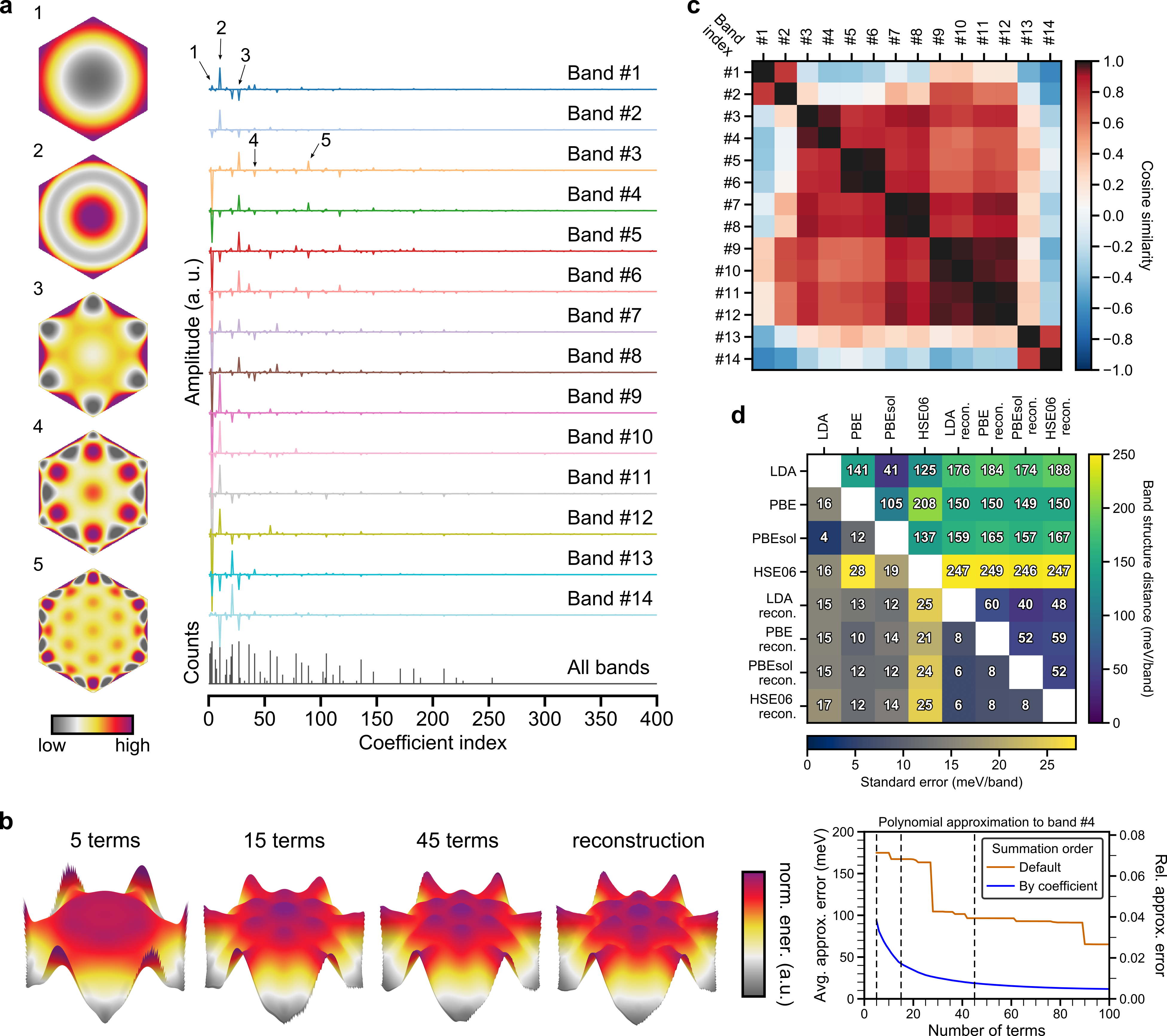}
	\end{center}
    \caption{\textbf{Digitization and comparison of WSe$_2$ band structures.} \textbf{a}, Decomposition of the 14 energy bands of WSe$_2$ into hexagonal Zernike polynomials (ZPs) with selected major terms displayed on the left. The zero spatial frequency term in the decomposition is subtracted for each band. The counts of large ($> 10^{-2}$ by absolute value) coefficients of all 14 bands are accumulated at the bottom row of the decomposition to illustrate their distribution, which decrease in value towards higher-order terms. \textbf{b}, Approximation of the shape (or dispersion) of the fourth energy band using different numbers of hexagonal ZPs. \textbf{c}, Cosine similarity matrix for pairwise comparison of the reconstructed band dispersion in Fig. \ref{fig:recon}. The band indices follow those in Fig. \ref{fig:recon}d. \textbf{d}, Two-part similarity matrix showing band structure distances (in the upper triangle) and their corresponding standard errors (in the lower triangle) between the computed and reconstructed band structures of WSe$_2$. The abbreviation ``LDA recon." denotes reconstruction with LDA-level DFT band structure as the initialization.}
    \label{fig:repr}
\end{figure}

The reconstructed 14 valence bands of WSe$_2$ initialized by LDA-level DFT are shown in Fig. \ref{fig:recon}b-d and Supplementary Videos. To globally compare the computed and reconstructed bands at a consistent resolution, we expand the BS in orthonormal polynomial bases \cite{Zhang2004}, which are global shape descriptors and unbiased by the underlying electronic detail. The geometric featurization of band dispersion allows multiscale sampling and comparison using the coefficient (or feature) vectors \cite{Khotanzad1990}. We choose Zernike polynomials (ZPs) to decompose the 3D dispersion surfaces (see Fig. \ref{fig:repr} and Methods) because of their existing adaptations to various boundary conditions \cite{Mahajan2007}.
In Fig. \ref{fig:repr}a-b, the band dispersions show generally decreasing dependence (seen from the magnitude of coefficients) on basis terms with increasing complexities (see Fig. \ref{fig:repr}a), and the majority of dispersion is encoded into a subset of the terms (see Fig. \ref{fig:repr}b). This observation implies that moderate smoothing may be applied to remove high-frequency features to improve the reconstruction in case of limited-quality data (acquired without sufficient accumulation time), which is often unavoidable when materials exhibit vacuum degradation, or during experimental parameter tuning. The example in Fig.~\ref{fig:repr}b and additional numerical evidence in Supplementary Fig. \ref{fig:approx} illustrate the approximation capability of the hexagonal ZPs. Concisely, these coefficients act as geometric fingerprints of the energy band dispersion, which enable the use of similarity or distance metrics (see Methods) for their comparison \cite{Khotanzad1990}. In Fig.~\ref{fig:repr}c, the positive cosine similarity confirms the strong shape (or dispersion) resemblance of the 7 pairs of spin-split energy bands in the reconstructed BS of WSe$_2$, while the low negative values, such as those between bands 1-2 and 13-14, reflect the opposite directions of their respective dispersion (see Fig. \ref{fig:recon}d). These observations are consistent with the outcome obtained from DFT calculations (see Supplementary Fig. \ref{fig:bandhzps}).\\

\noindent\textbf{Computational metrics and performance.} To quantify the computational advantages of the machine learning-based reconstruction approach, we examine the outcome from diverse perspectives in consistency, accuracy and cost. To assess the consistency of reconstruction in its entirety, we introduce a BS distance metric (see Methods), invariant to the global energy shift frequently used to adjust the energy zero, to quantify the differences in band dispersion and the relative spacing between bands, which are the two major sources of variation between theories and experiments. The distance is calculated using the geometric fingerprints to bypass interpolation errors while reconciling the coordinate spacing difference between reconstructed and theoretical BSs, essential for differentiating BS data from heterogeneous sources in materials science databases \cite{Himanen2019,Horton2021}. The results in Fig.~\ref{fig:repr}d refer to the valence BS of WSe$_2$ discussed in this work, where the distances and their spread (i.e. standard errors) are displayed in the upper and lower triangles, respectively. A high degree of consistency exists among the reconstructions (pairwise distance no larger than 60$\pm$8 meV/band) regardless of the level of DFT calculation used for initialization, indicating the robustness of the probabilistic reconstruction algorithm, whereas the distances between the DFT calculations are much larger, both in energy shifts and their spread. As shown in Fig.~\ref{fig:repr}d and Supplementary Fig. 5, the learning algorithm can effectively reduce the epistemic uncertainty \cite{Kiureghian2009} between theories to obtain a consistent reconstruction.
\begin{figure}[htbp!]
    \begin{center}
		\includegraphics[width=\textwidth]{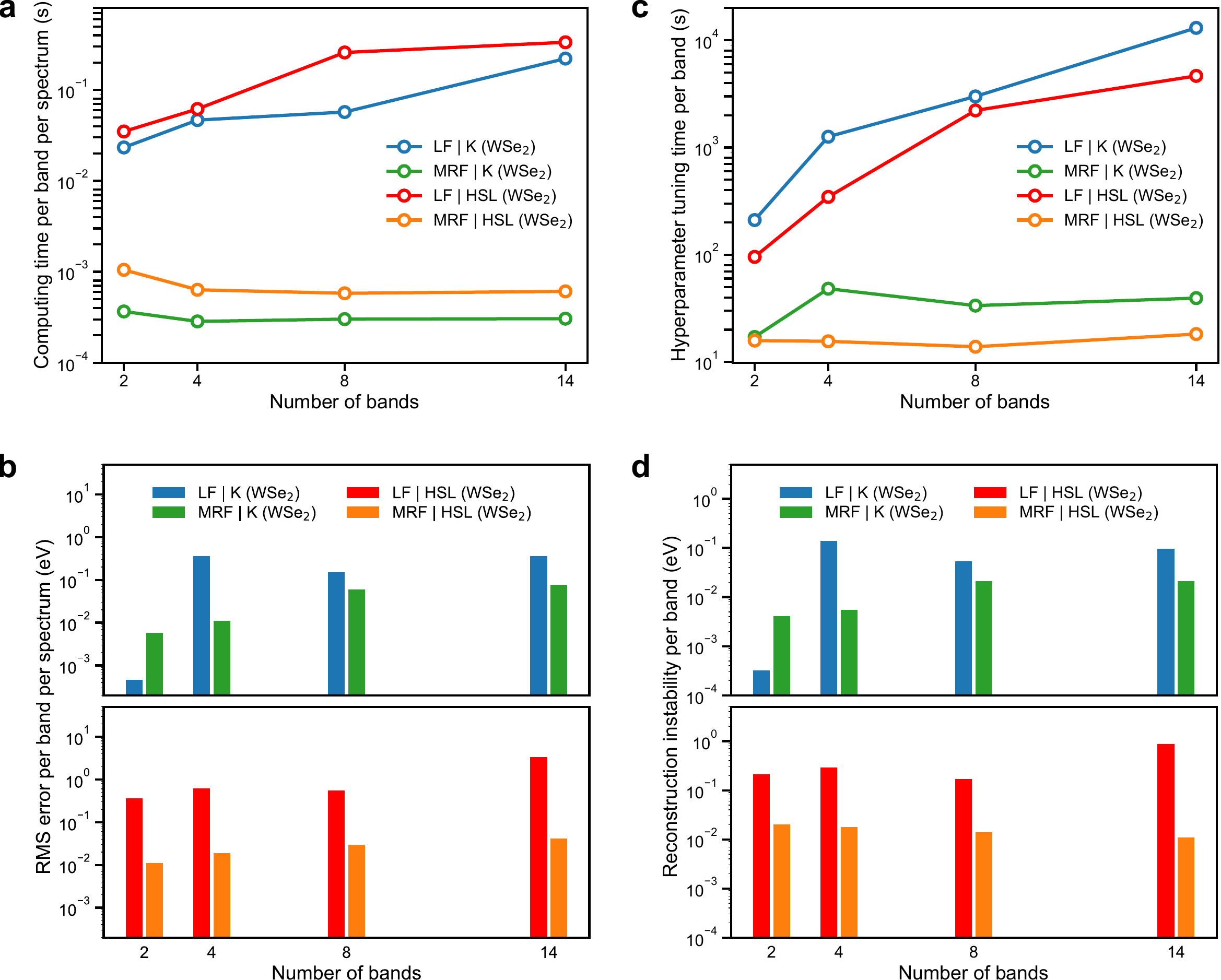} 
	\end{center}
    \caption{\textbf{Performance evaluation on benchmarks.} Visual summary of the benchmarking outcomes for band structure reconstruction using normalized metrics that are able to compare across datasets. These include \textbf{a}, the computing time and \textbf{b}, root-mean-square error (reconstruction error), both normalized to the per-band, per-spectrum level \cite{Xian2021}. The other metrics, including \textbf{c} the hyperparameter tuning time and \textbf{d}, the reconstruction instability are normalized to the per-band level. The methods used in reconstruction include pointwise line fitting (LF) and the Markov random field (MRF) approach presented in this work, while the synthetic data are around the K point and along the high-symmetry line (HSL) of the WSe$_2$ band structure. The benchmarks were run with synthetic datasets terminated at fixed energy ranges that contain the specified number of bands (2, 4, 8, and 14, the maximum band index in the dataset) shown in \textbf{a}-\textbf{d}.}
    \label{fig:comp_metrics}
\end{figure}

To demonstrate the computational advantage of the MRF reconstruction over traditional line fitting methods, we benchmarked the outcome over selected regions in synthetic photoemission data. The regions are chosen based on their importance and we limit the size to have a manageable computing time (about an hour on our computing cluster at maximum for a single run), determined by the slower method, and allow for hyperparameter tuning, which requires tens of runs. The line fitting approach uses the Levenberg-Marquardt least squares optimization \cite{Nocedal2006} with bound constraints for multicomponent photoemission spectra composed of a series of lineshape functions. We used the benchmark established in \cite{Xian2021} for pointwise line fitting employing high-performance computing and two synthetic datasets with known ground truth dispersion, representing the local and global settings of the band structure reconstruction problem (see Supplementary Section 2.5). The synthetic data were based on band structure at the LDA-DFT level around the K point and along the high-symmetry line of the Brillouin zone. To level the hardware requirements, we used only distributed multicore-CPU computing for performance benchmarking. The estimated computing times are normalized to the per-band per-spectrum level \cite{Xian2021}. The accuracy of the reconstruction is calculated using root-mean-squared (RMS) error, while the stability is quantified by the standard deviation of the residuals, which measures surface roughness \cite{Smith2014}. The benchmarking results are compiled in Fig. \ref{fig:comp_metrics} and Supplementary Table \ref{tab:benchmark}. They show that, compared with pointwise line fitting, the MRF reconstruction offers a considerable reduction in both normalized computing time and hyperparameter tuning time, while achieving consistently higher accuracy and stability in all but the two-band case. The combination of accuracy and stability in MRF reconstruction is due to the connectivity built into the prior, whereas in the pointwise fitting approach, information is not explicitly shared among neighbors. Since the number of bands reflects the complexity of multicomponent spectra, a near-constant normalized computing time and hyperparameter tuning time (see Fig. \ref{fig:comp_metrics}a-b) in MRF reconstruction regardless of the number of bands (or spectral components) allow us to scale up the computation to datasets comprising 10$^4$-10$^5$ or more spectra. The substantial gain in computational efficiency is a result of the inherent divide-and-conquer strategy in our BS reconstruction problem formulation and the associated distributed optimization method in the algorithm design. Comparatively, the distributed pointwise fitting exhibits a quasi-linear computational scaling with respect to the number of bands. When hyperparameter tuning is taken into account, in practice, it is only feasible for fitting small datasets with up to 10$^3$ multicomponent spectra \cite{Xian2021}.\\

\begin{figure}[htb!]
    \centering
    \includegraphics[width=\textwidth]{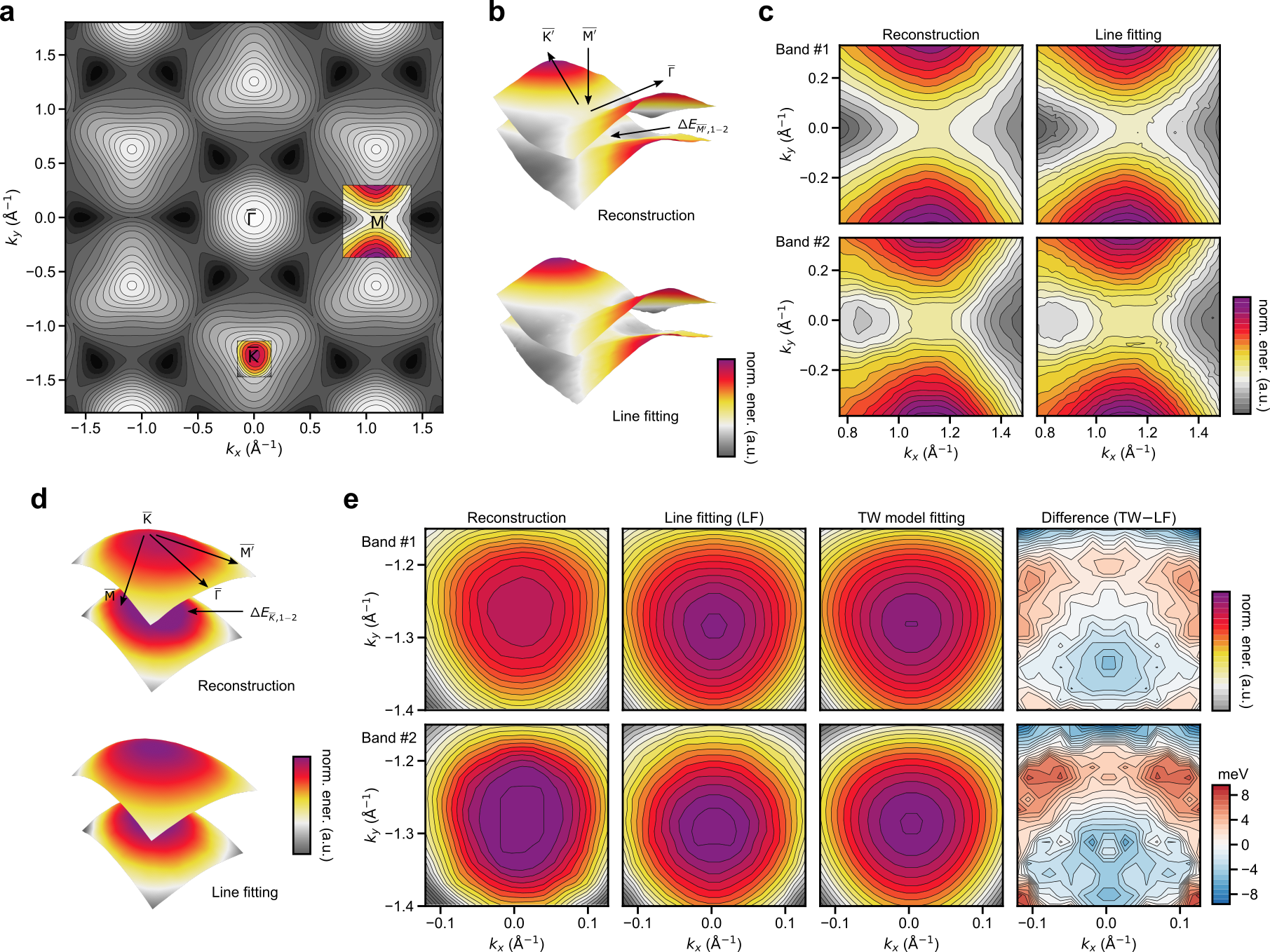}
    \caption{\textbf{Local band structure parameters of WSe$_2$}. \textbf{a}, The first valence band of 2$H$-WSe$_2$ with constant-energy contours. The patches around high-symmetry points $\overline{\text{K}}$ and $\overline{\text{M}^{\prime}}$ from reconstruction (with LDA-DFT as the initialization) are overlaid in color. \textbf{b},\textbf{c}, Patch around the $\overline{\text{M}^{\prime}}$-point, a saddle point in the dispersion surface, visualized in 3D (\textbf{b}) and 2D (\textbf{c}), respectively. The energy gap at $\overline{\text{M}^{\prime}}$ due to spin-orbit coupling (SOC) results in the energy difference $\Delta E_{\overline{\text{M}^{\prime}},1-2}$. \textbf{d},\textbf{e}, Patch around the $\overline{\text{K}}$-point, the energy maximum of the valence band, visualized in 3D (\textbf{d}) and 2D (\textbf{e}), respectively. The SOC results in the energy gap $\Delta E_{\overline{\text{K}},1-2}$. The outcome of fitting to a trigonal warping (TW) model around $\overline{\text{K}}$ from \textbf{k}$\cdot$\textbf{p} theory \cite{Kormanyos2015} is shown in \textbf{e}.}
    \label{fig:localparams}
\end{figure}

\noindent\textbf{Extended use cases and applications.} The band dispersions recovered from photoemission data are often examined locally near dispersion extrema. We show in Fig. \ref{fig:localparams} that, besides providing global structural information, the reconstruction improves the robustness of traditional pointwise lineshape fitting in extended regions of the momentum space, when used as initial guess, because BS calculations may exhibit appreciable momentum-dependent deviations from experimental data that prevent them from being a sufficiently good starting point. Pointwise fitting in turn acts as the \textit{refinement} of local details not explicitly included in the probabilistic reconstruction model, which prioritizes efficiency. This sequential approach recovers large regions in the Brillouin zone at high energy resolution without laborious hand-tuning of the fitting parameters per photoemission spectrum. Adopting this approach to WSe$_2$,  we recovered (i) a compendium of local band structure parameters (see Supplementary Table \ref{tab:bsparams}). The trigonal warping parameters of the first two valence bands around the $\overline{\text{K}}$-point are 5.8 eV$\cdot\text{\AA}^3$ and 3.9 eV$\cdot\text{\AA}^3$, respectively, confirming the magnitude difference between these spin-split bands predicted by theory \cite{Kormanyos2015}. The warping signature extends further to high-energy bands. (ii) Dispersion fitting around the saddle point $\overline{\text{M}^{\prime}}$ (and $\overline{\text{M}}$) of the band structure reveals that the gap opened by spin-orbit interaction extends beyond it anisotropically on the dispersion surfaces with the minimum gap at 338 meV, markedly larger than DFT results, which predict degeneracy \cite{Kormanyos2015}. We expect this observation to contribute to the spin-dependent optical absorption due to the association of the saddle point in energy dispersion with a van Hove singularity \cite{Kormanyos2015,Guo2015}.

\begin{figure}[hbtp!]
    \centering
    \includegraphics[width=0.95\textwidth]{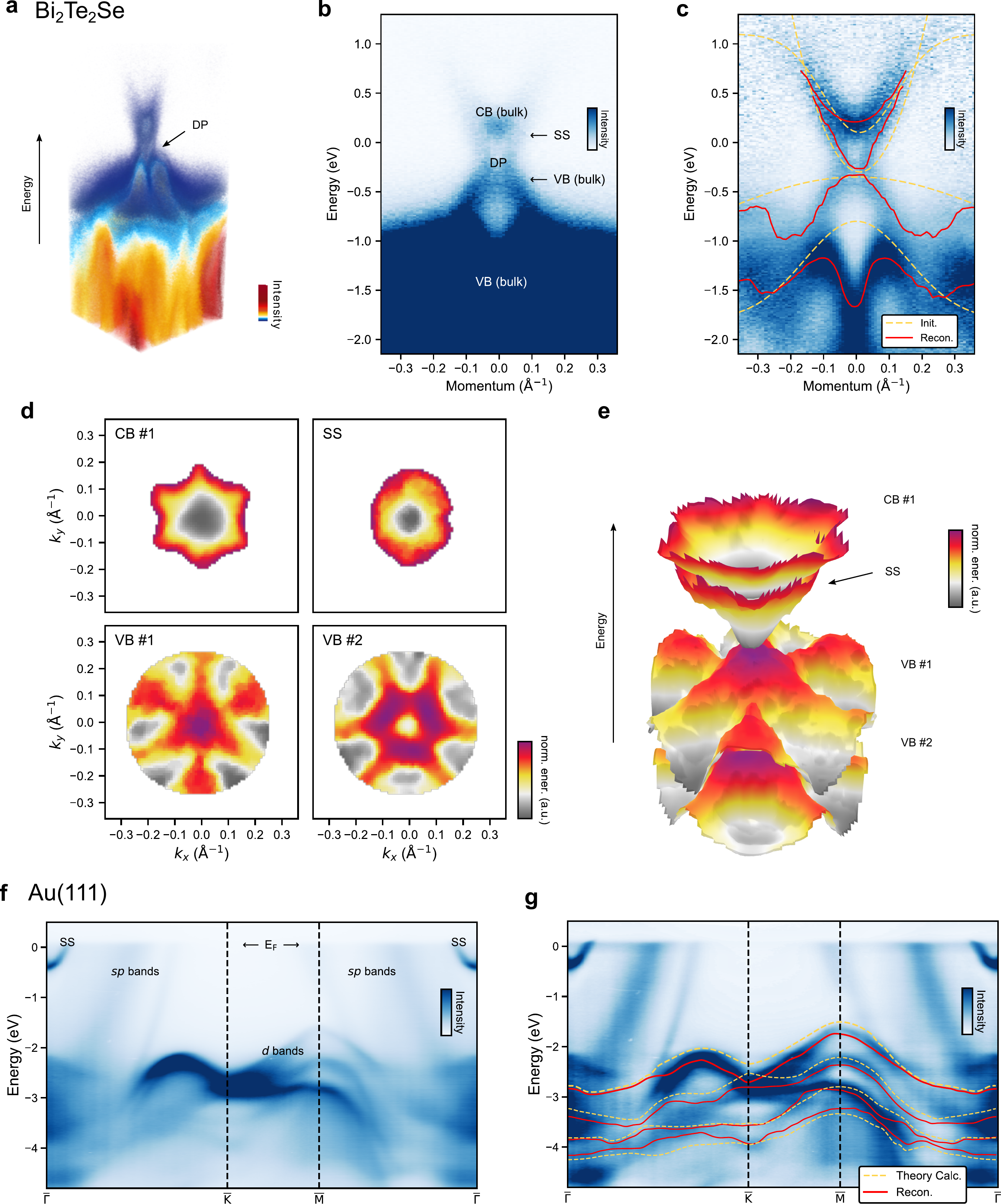} 
    \caption{\textbf{Band reconstruction for Bi$_2$Te$_2$Se and Au(111)}. \textbf{a}, 3D view of the photoemission band mapping data of the topological insulator Bi$_2$Te$_2$Se around the Dirac point (DP). The energy bands near the DP are labeled in \textbf{b} in a 2D slice through the DP. The outcome of reconstruction (after smoothing) is superimposed on the preprocessed data in \textbf{c}. Momentum-resolved reconstruction is shown in 2D (\textbf{d}) and 3D (\textbf{e}), where the color map represents the energy values within each band. The experimental photoemission data for Au(111) is shown in \textbf{f} with designations of the band structures labeled. Reconstruction of some of the $d$ bands are shown in \textbf{g} along with the theoretical calculations used for initialization.}
    \label{fig:othermat}
\end{figure}

In addition to WSe$_2$, we have performed BS reconstruction on two other photoemission datasets from other classes of materials: (1) Bismuth tellurium selenide (Bi$_2$Te$_2$Se), a topological insulator, measured using the same laboratory photoemission setup (see Fig. \ref{fig:othermat}a-e) as for the WSe$_2$ dataset. Although we used only simple numerical functions (Gaussian and paraboloid) to initialize the MRF reconstruction, the outcome demonstrates correct discrete momentum-space symmetry and details of energy dispersion down to the concave-shaped hexagonal warping in the band energy contours around the Dirac point \cite{Heremans2017}. Four energy bands, including the two low-energy valence bands, a surface-state energy band, and a partially occupied conduction band, were recovered using our approach for Bi$_2$Te$_2$Se. (2) Bulk gold (Au) photoemission dataset measured at a synchrotron X-ray source (see Fig. \ref{fig:othermat}f-g). We used DFT calculations as the initialization to reconstruct four of the bulk energy bands, which are usually very challenging to extract by hand tracing or parametric function fitting, due in part to blurring ($k_z$ dispersion) from the 3D characteristics of the electrons in the metallic bulk. Further discussions on these two materials and their band reconstructions are provided in Supplementary Section 3.

\section*{Discussion}
The reconstruction approach described here provides a quantitative connection between empirical band dispersion ($E_b^{\mathrm{emp}}$) obtained from photoemission band mapping and their theoretical counterparts ($E_b^{\mathrm{theory}}$) through various orders of momentum-dependent ``perturbations'' ($\Delta E_b^{(n)}$). The connection may be expressed as,
\begin{align}
    E_b^{\mathrm{emp}}(\textbf{k}, \Sigma) &\approx E_b^{\mathrm{theory}}(\textbf{k}, \Sigma) + \Delta E_b^{(0)} + \Delta E_b^{(1)}(\textbf{k}, \Sigma) + \Delta E_b^{(2)}(\textbf{k}, \Sigma) + ... \nonumber\\
    &= E_b^{\mathrm{theory}}(\textbf{k}, \Sigma) + \sum_{n} \Delta E_b^{(n)}(\textbf{k}, \Sigma) = E_b^{\mathrm{theory}}(\textbf{k}, \Sigma) + \Delta E_b(\textbf{k}, \Sigma).
    \label{eq:viewpoint}
\end{align}
In Eq. \eqref{eq:viewpoint}, $b$ is the band index, $\Sigma$ represents electron self-energy, the zeroth-order term ($\Delta E_b^{(0)}$) means a rigid shift, while higher-order terms have increasing momentum-dependent nonlinearities. Our results here demonstrate that this formulation leads to practical band reconstruction, which recovers the accumulated ``perturbations'' ($\Delta E_b$) in Eq. \eqref{eq:viewpoint} for every experimentally resolvable energy band. The outcome with current reconstruction accuracy and stability should assist interpretation of deep-lying bands, parametrizing multiband Hamiltonian models \cite{Ehrhardt2014}. The data size reduction by over 5000 times from 3D band mapping data to geometric features vectors (see Methods) facilitates database integration \cite{Horton2021,Scheffler2022}.

Apart from the benefits, we want to outline three limitations of our reconstruction approach. Firstly, the reconstruction approach doesn't work \textit{ab initio} and requires knowing the number of energy bands, $N_b$, as implicated by the correspondence in Eq. \eqref{eq:viewpoint}. Although in simple datasets with up to several bands, $N_b$ can be estimated using prior knowledge of the material or from visual inspection, correctly estimating $N_b$ in complex datasets still requires calculated band structures. Secondly, when the electron self-energy modulation is significant, separating the so-called bare-band dispersion (i.e. single-particle dispersion) from the quasiparticle dispersion is needed for understanding the materials physics \cite{Kordyuk2005}. This requires re-evaluating the band structure reconstruction concept and consider the full spectral function (see Supplementary Section 1.1) explicitly to account for nonstandard lineshapes. Nevertheless, the outcome of our current approach may act as a trial solution for disentangling the bare-band dispersion relation from the electron self-energy \cite{Kordyuk2005}. Because the local connectedness assumption in Eq. \eqref{equ:joint_p_mrf} remains largely valid, our reconstruction may still recover the quasiparticle dispersion. We demonstrate this in Supplementary Fig. \ref{fig:kinked} using simulated photoemission data with a kink anomaly, a strong modification of dispersion from electron self-energy \cite{Damascelli2003,Zhang2022}. Thirdly, an appropriate initialization may be expensive or impossible to obtain, either due to the computational cost, if higher-level theories (such as DFT with hybrid functionals and $GW$) are required, or due to the complexity of the materials system, including undetermined microscopic interactions, sample defects or structural disorder, creating strong intensity blurring from $k_z$ dispersion, etc. These scenarios will remain challenging for band reconstruction.

Besides our demonstrations, we anticipate additional use cases that include (i) online monitoring \cite{Noack2021} of band mapping experiments in the study of materials phase transitions \cite{Beaulieu2021} or functioning devices \cite{Curcio2020}, where changes in atomic structure or carrier mobility are often accompanied by detectable changes in the electronic structure (including band dispersion), resulting in $I(\mathbf{k}, E, t)$ with time ($t$) dependence in addition to momentum ($\mathbf{k}$) and energy. (ii) Spatial mapping of electronic structure variations for electronic devices via scanning photoemission measurements \cite{Wilson2017,Ulstrup2019}, resulting in $I(\mathbf{k}, E, \mathbf{x})$ with spatial ($\mathbf{x}$) dependence. In cases (i)-(ii), a fast reconstruction and evaluation framework may be used in a feedback loop to steer or optimize experimental conditions. (iii) Implementation of the reconstruction across various materials and to band-mapping data \cite{Schonhense2015} conditioned on external parameters, including temperature, photon energy, dynamical time delay, and spin as resolved quantities, will generate comprehensive knowledge about the (non)equilibrium electronic structure of materials to benchmark theories. Moreover, the reconstruction method is (iv) transferable to extracting the band dispersion of other quasiparticles (e.g. phonons \cite{Ewings2016}, polaritons \cite{Whittaker2018}, etc \cite{Frolich2014}) in periodic systems, given the availability of corresponding multidimensional datasets. (v) The analogy between band mapping and spatially-resolved spectral imaging, which produces location-dependent spectra, or $I(x, y, E)$ suggests that the reconstruction algorithm may find use in teasing out the spatial ($x$, $y$) variation of the spectral shifts, complementary to the outcome of clustering algorithms \cite{Amenabar2017}.

The increasing amount of publicly accessible and reusable datasets from materials science communities \cite{Scheffler2022} motivate future extensions to the model with other types of informative priors that account for the full complexity of the physical signal while maintaining computational efficiency. Overall, the multidisciplinary methodology provides an example for building next-generation high-throughput materials characterization toolkits combining learning algorithms with physical knowledge \cite{VonRueden2021} to arrive at a comprehensive understanding of materials properties unattainable before.

\printbibliography[keyword=maintext, heading=subbibliography, title=References]

\subsection*{Methods}
\begin{enumerate}[wide, labelwidth=!, labelindent=-8pt]
\item[] \textbf{Band mapping measurements of WSe$_2$.} Multidimensional photoemission spectroscopy experiments were conducted with a laser-driven, high harmonic generation-based extreme UV light source \cite{Puppin2019} operating at 21.7 eV and 500 kHz and a METIS 1000 (SPECS GmbH) momentum microscope featuring a delay-line detector coupled to a time-of-flight drift tube \cite{Medjanik2017,Oelsner2001}. The experiment captures photoelectrons directly in their 3D coordinates, $(k_x, k_y, E)$ \cite{Schonhense2015,Medjanik2017}. Single crystal samples of WSe$_2$ ($>$ 99.995\% pure) were purchased from HQ graphene and were used directly for measurements without further purification. Before measurements, the WSe$_2$ samples were attached to the Cu substrate by conductive epoxy resin (EPO-TEK H20E). The samples were cleaved by cleaving pins attached to the sample surface upon transfer into the measurement chamber, which operates at an ambient pressure of $10^{-11}$ mbar during photoemission experiments. No effect of surface termination has been observed in the measured WSe$_2$ photoemission spectra, similar to previous experimental observations \cite{Traving1997,Riley2014}. For the valence band mapping experiments, the energy focal plane of the photoelectrons within the time-of-flight drift tube was set close to the top valence band. Although effects of sample degradation has also been reported \cite{Finteis1997} during the course of long-duration angular scanning in ARPES measurements, with our high-repetition-rate photon source \cite{Puppin2019} and the fast electronics of the momentum microscope, band mapping of WSe$_2$ achieves sufficient signal-to-noise ratio for valence band reconstruction within only tens of minutes of data acquisition, without the need for angular scanning and subsequent reconstruction from momentum-space slices.

\item[] \textbf{Data processing and reconstruction.} The raw data, in the form of single-electron events recorded by the delay-line detector, were preprocessed using home-developed software packages \cite{Xian2020a}. The events were first binned to the $(k_x, k_y, E)$ grid with a size of 256$\times$256$\times$470 to cover the full valence band range in WSe$_2$ within the projected Brillouin zone, which amounts to a pixel size of $\sim$ 0.015 $\mathrm{\AA}^{-1}$ along the momentum axes and $\sim$ 18 meV along the energy axis. The bin sizes are within the limits of the momentum resolution ($<$ 0.01 $\mathrm{\AA}^{-1}$) and energy resolution ($<$ 15 meV) of the photoelectron spectrometer \cite{SPECSGmbH2019}.

Data binning is carried out in conjunction with the necessary lens distortion correction \cite{Xian2019a} and calibrations as described in \cite{Xian2020a}. The outcome provides a sufficient level of granularity in the momentum space to resolve the fine features in band dispersion while achieving higher signal-to-noise ratio than using single-event data directly. Afterwards, we applied intensity symmetrization to the data along the sixfold rotation symmetry and mirror symmetry axes \cite{Riley2014} of the photoemission intensity pattern in the $(k_x, k_y)$ coordinates, followed by contrast enhancement using the multidimensional extension of the contrast limited adaptive histogram equalization (MCLAHE) algorithm, where the intensities in the image are transformed by a look-up table built from the normalized cumulative distribution function of local image patches \cite{Stimper2019}. Finally, we applied Gaussian smoothing to the data along the $k_x$, $k_y$ and $E$ axes with a standard deviation of 0.8, 0.8 and 1 pixels (or about 0.012 $\mathrm{\AA}^{-1}$, 0.012 $\mathrm{\AA}^{-1}$, and 18 meV), respectively.

After data preprocessing, we sequentially reconstructed every energy band of WSe$_2$ from the photoemission data using the \textit{maximum a posteriori} (MAP) approach described in the main text. The reconstruction requires tuning of three hyperparameters: (1) the momentum scaling and (2) the rigid energy shift to coarse-align the computed energy band, e.g. from density functional theory (DFT), to the photoemission data, and (3) the width of the nearest-neighbor Gaussian prior ($\eta$ in Eq. \eqref{equ:joint_p_mrf}). The hyperparameter tuning is also carried out individually for each band to adapt to their specific environment. An example of hyperparameter tuning is given in Supplementary Fig. \ref{fig:hypertuning}. The MAP reconstruction method involves optimization of the band energy random variables, $\{\Tilde{E}_{i,j}\}$ to maximize the posterior probability $p = p(\{\Tilde{E}_{i,j}\})$ or to minimize the negative log-probability loss function, $\mathcal{L}:=-\log p$, obtained from Eq. \eqref{equ:joint_p_mrf} as is used in our actual implementation.
\begin{equation}
    \mathcal{L}(\{\Tilde{E}_{i,j}\})
    = -\sum_{i,j} \log I(k_{x,i}, k_{y,j}, \Tilde{E}_{i,j}) + \sum_{(i,j),(l,m)|\mathrm{NN}} \frac{(\Tilde{E}_{i,j}-\Tilde{E}_{l,m})^2}{2\eta^2} + \mathrm{const.}
    \label{equ:loss}
\end{equation}
We implemented the optimization using a parallelized version of the iterated conditional mode (ICM) \cite{Kittler1984} method in Tensorflow \cite{Abadi2016} in order to run on multicore computing clusters and GPUs. The parallelization involves a checkerboard coloring scheme (or coding method) of the graph nodes \cite{Li2009} and subsequent hierarchical grouping of colored nodes, which allows alternating updates on different subgraphs (i.e. subsets of the nodes) of the Markov random field during optimization. Typically, the optimization process in the reconstruction of one band converges within and therefore is terminated after 100 epochs, which takes $\sim$ 7 seconds on a single NVIDIA GTX980 GPU for the above-mentioned data size. Details on the parallelized implementation are provided in Supplementary Section 1. In addition, because symmetry information is not explicitly included in the MRF model, the reconstructed bands generally requires further symmetrization as refinement or post-processing to be ready for database integration.

We described our approach of using band structure calculations to initialize the MAP optimization as a warm start. The term "warm start" in the context of numerical optimization generally refers to the initialization of an optimization using the outcome of an associated and yet more solvable problem (e.g. surrogate model) obtained beforehand that yields an approximate answer, instead of starting from scratch (i.e. cold start). Warm-starting an optimization improves the effective use of prior knowledge and its convergence rate \cite{Nocedal2006}. In the current context, we regard the band structure reconstruction from photoemission band mapping data as the optimization problem to warm start, and the outcome from an electronic structure calculation can produce a sufficiently good approximate to the solution of the optimization problem. For WSe$_2$, straightforward DFT calculations with semi-local approximation (which in itself involves explicit optimizations such as geometric optimization of the crystal structures) are sufficient, but our approach is not limited to DFT. Therefore, the use of "warm start" in our application is conceptually well-aligned with the origin of the term.

To validate the MAP reconstruction algorithm in a variety of scenarios, we used synthetic photoemission data where the nominal ground-truth band structures are available. The band structures are constructed using analytic functions, model Hamiltonians or DFT calculations. The initializations are generated by tuning the numerical parameters used to generate the ground-truth band structures. The procedures and results are presented in Supplementary Section 2. In simple cases, such as single or well-isolated bands, the reconstruction yields a close solution to the ground truth even with a flat band initialization. In the more general multiband scenario with congested bands and band crossings (or anti-crossings), an approximate dispersion (or shape) of the band and the crossing information is required in the initialization (i.e. warm start) in order to converge to a realistic solution. We further tested the robustness of the initializations by (1) scaling the energies of the ground truth and by (2) using DFT calculations with different exchange-correlation (XC) functionals, in order to capture sufficient variability of available band structure calculations in the real world. We quantify the variations in the initializations and the performance of the reconstruction using the average error (Eq. \eqref{eq:avgapprox}, or Fig. \ref{fig:repr}b), calculated with respect to the ground truth. Among the different numerical experiments, we find that the optimization converges consistently to a set of bands that better matches the experimental data than the initialization. This is manifest in that the average errors of the initializations are reduced to a similar level in the corresponding reconstruction outcomes, a trend seen over all bands regardless of their dispersion. In the synthetic data with an energy spacing of $\sim$ 18 meV, the average error in the reconstruction is on the order of 40-50 meV for each band, which amounts to an average inaccuracy of $<$ 3 bins along the energy dimension at a momentum location. The inaccuracy is, however, dependent on the bin sizes used in the preprocessing and the fundamental resolution in the experiment. We have made the code for the MAP reconstruction algorithm and the synthetic data generation publicly accessible from the online repository \textsf{fuller} \cite{Stimper2019a} for broader applications.

\item[] \textbf{Visualization strategies.} Band mapping and band structure data contain unique multidimensional data structures in materials science that are often presented with specific visualizations motivated by the underlying solid state physics and symmetry properties. In this work, we select a fixed set of 2D and 3D visualization techniques to illustrate their link and allow comparison with other photoemission studies of the same materials. Typically, ARPES data \cite{Zhang2022} of the form $I(E, k)$ are sampled and visualized along a particular path (i.e. k-path \cite{Hinuma2017}) in the momentum space \cite{Finteis1997,Traving1997} where only specific high-symmetry positions are labeled with capital letters \cite{Bouckaert1936}. A canonical k-path exists for each symmetry setting \cite{Hinuma2017}. Photoemission band mapping generates datasets with a dimensionality of three or higher and often contains a lower symmetry (in intensity $I$) as a result of the photoemission matrix elements \cite{Moser2017} and the experimental condition. These factors lead to more flexibility in data representation \cite{Xian2020a} and motivate the use of alternative k-paths that capture the complexity of the photoemission spectra. In Fig. \ref{fig:overview}c-f for WSe$_2$ and Fig. \ref{fig:othermat}a-c for Bi$_2$Te$_2$Se, we combine 3D volumetric rendering and 2D k-path views to illustrate both the data symmetry and the intensity modulations present in the data.

To visualize band dispersion surfaces, $E_b(k_x, k_y) \, (b = 1, 2, ...)$, we combine 3D stacked surfaces and 2D image sequences, as exemplified in Fig. \ref{fig:recon}b, d for WSe$_2$ and Fig. \ref{fig:othermat}d, e for Bi$_2$Te$_2$Se. This paired approach balances the strengths and shortcomings of different viewpoints to achieve a comprehensive representation of the data type. The 3D stacked surface representation highlights the entirety and complexity of the data, but often contains occluded regions imperceptible from a fixed viewing direction. The 2D image sequence representation includes all energy dispersion information, yet loses the interrelationship on the energy scale between energy bands, which matter in the event of (anti)crossings. In combining these two approaches, we typically choose the same color map and scale to maintain referenceability between the two representations. For each energy band, the full color scale is used to cover its energy range, becoming the normalized energy scale, which illustrates the local detail of the dispersion that otherwise may be hard to discern.

\item[] \textbf{Band structure calculations.} Electronic band structures were calculated within (generalized) DFT using the local density approximation (LDA) \cite{Ceperley1980, Perdew1992}, the generalized-gradient approximation (GGA-PBE) \cite{Perdew1996} and GGA-PBEsol \cite{Perdew2008}), and the hybrid XC functional HSE06 \cite{Heyd2003}, which incorporates a fraction of the exact exchange. All calculations were performed with the all-electron, full-potential numeric-atomic orbital code, FHI-aims \cite{Blum2009}. They were conducted for the geometries obtained by fully relaxing the atomic structure with the respective XC-functional to keep the electronic and atomic structures consistent. Spin-orbit coupling was included in a perturbational fashion \cite{Huhn2017}. The momentum grid used for the calculation was equally sampled with a spacing of 0.012 $\mathrm{\AA}^{-1}$ in both $k_x$ and $k_y$ directions that covers the irreducible part of the first Brillouin zone at $k_z$ = 0.35 $\mathrm{\AA}^{-1}$, estimated using the inner potential of WSe$_2$ from a previous measurement \cite{Riley2014}. The calculated band structure is symmetrized to fill the entire hexagonal Brillouin zone to be used to initialize the band structure reconstruction and synthetic data generation. We note here that for the MAP reconstruction, the momentum grid size used in theoretical calculations (such as DFT at various levels used here) need not be identical to that of the data (or instrument resolution) and in those cases an appropriate upsampling (or downsampling) should be applied to the calculation to match their momentum resolution. Further details are presented in Supplementary Section 4.

\item[] \textbf{Band structure informatics.} The shape feature space representation of each electronic band is derived from the decomposition,
\begin{equation}
    E_b(\textbf{k}) = \sum_l a_l \phi_l(\textbf{k}) = \textbf{a} \cdot \textbf{$\Phi$}
    \label{eq:decomposition}
\end{equation}
Here, $\textbf{k} = (k_x, k_y)$ represents the momentum coordinate, $E_b(\textbf{k})$ is the single-band dispersion relation (e.g. dispersion surface in 3D), $a_l$ and $\phi_l(\textbf{k})$ are the coefficient and its associated basis term, respectively. They are grouped separately into the feature vector, $\textbf{a} = (a_1, a_2, ...)$, and the basis vector, $\textbf{$\Phi$} = (\phi_1, \phi_2, ...)$. The orthonormality of the basis is guaranteed within the projected Brillouin zone (PBZ) of the material.
\begin{equation}
    \int_{\textbf{k} \in \Omega_{\text{PBZ}}} \phi_m(\textbf{k}) \phi_n(\textbf{k}) \, d\textbf{k} = \delta_{mn}
\end{equation}
For the hexagonal PBZ of WSe$_2$, the basis terms are hexagonal Zernike polynomials (ZPs) constructed using a linear combination of the circular ZPs via Gram-Schmidt orthonormalization within a regular (i.e. equilateral and equiangular) hexagon \cite{Mahajan2007}. A similar method can be used to generate ZP-derived orthonormal basis adapted to other boundary conditions \cite{Mahajan2007}. The representation in feature space \cite{Khotanzad1990} provides a way to quantify the difference (or distance) $d$ between energy bands or band structures at different resolutions or scales without additional interpolation. To quantify the shape similarity between energy bands $E_{b}$ and $E_{b^{\prime}}$, we calculate the cosine similarity using the feature vectors,
\begin{equation}
    d_{\mathrm{cos}}(E_{b}, E_{b^{\prime}}) = \frac{\textbf{a} \cdot \textbf{a}^{\prime}}{|\textbf{a}|\cdot|\textbf{a}^{\prime}|},
    \label{equ:dcos}
\end{equation}
The cosine similarity is bounded within $\left[-1, 1\right]$, with a value of 0 describing orthogonality of the feature vectors and a value of 1 and -1 describing parallel and anti-parallel relations between them, respectively, both indicating high similarity. The use of cosine similarity in feature space allows comparison of dispersion while being unaffected by their magnitudes. In comparing the dispersion between single energy bands using Eq. \eqref{equ:dcos}, the first term in the polynomial expansion, or the hexagonal equivalent of the Zernike piston \cite{Wyant1992}, is discarded as it only represents a constant energy offset (with zero spatial frequency) instead of dispersion, which is characterized by a combination of finite and nonzero spatial frequencies.

The electronic band structure is a collection of energy bands $E_{B} = \{ E_{b_i} \}$ $(i=1, 2, ...)$. To quantify the distance between two band structures, $E_{B_1} = \{ E_{b_{1,i}} \}$ and $E_{B_2} = \{ E_{b_{2,i}} \}$, containing the same number of energy bands while ignoring their global energy difference, we first subtract the energy grand mean (i.e. mean of the energy means of all bands within the region of the band structure for comparison). Then, we compute the Euclidean distance, or the $\ell^2$-norm, for the $i$th pair of bands, $d_{b,i}$.
\begin{equation}
    d_{b,i}(E_{b_{1,i}}, E_{b_{2,i}}) = \norm{\Tilde{\textbf{a}}_{1,i} - \Tilde{\textbf{a}}_{2,i}}_2 = \sqrt{\sum_{l} (\Tilde{a}_{1,il} - \Tilde{a}_{2,il})^2}.
    \label{eq:euclidean}
\end{equation}
Here, $\Tilde{\textbf{a}}$ denotes the feature vector after subtracting the energy grand mean so that any global energy shift is removed. We define the band structure distance as the average distance over all $N_b$ pairs of bands, or $d_B(E_{B_1}, E_{B_2})$ = $\sum_i^{N_b}d_{b,i}(E_{b_1,i}, E_{b_2,i})/N_b$. The values of $d_B(E_{B_1}, E_{B_2})$ are shown in the upper triangle of Fig. \ref{fig:repr}d and their corresponding standard errors (over the 14 valence bands of WSe$_2$) in the lower triangle. The distance in Eq. \eqref{eq:euclidean} is independent of basis and allows energy bands calculated on different resolutions or from different materials with the same symmetry (e.g. differing only by Brillouin zone size) to be compared.

We use same-resolution error metrics to evaluate the approximation quality of the expansion basis and to quantify the reconstruction outcome with a known ground-truth band structure. Specifically, we define the average approximation error (with energy unit), $\eta_{\mathrm{avg}}$, for each energy band using the energy difference at every momentum location,
\begin{equation}
    \eta_{\mathrm{avg}}(E_{\mathrm{approx}}, E_{\mathrm{recon}}) = \sqrt{\frac{1}{N_k}\sum_{\textbf{k} \in \Omega_{\mathrm{PBZ}}}(E_{\mathrm{approx}, \textbf{k}} - E_{\mathrm{recon}, \textbf{k}})^2},
    \label{eq:avgapprox}
\end{equation}
where $N_k$ is the number of momentum grid points and the summation runs over the projected Brillouin zone. In addition, we construct the relative approximation error, $\eta_{\mathrm{rel}}$, following the definition of the normwise error \cite{Watkins2010} in matrix computation,
\begin{equation}
    \eta_{\mathrm{rel}}(E_{\mathrm{approx}}, E_{\mathrm{recon}}) = \frac{{\norm{E_{\mathrm{approx}} - E_{\mathrm{recon}}}}_2}{\norm{E_{\mathrm{recon}}}_2}.
    \label{eq:relapprox}
\end{equation}
Eq. \eqref{eq:avgapprox}-\eqref{eq:relapprox} are used to compute the curves in Fig. \ref{fig:repr}b as a function of the number of basis terms included in the approximation. The relevant code for the representation using hexagonal ZPs and the computation of the metrics is also accessible in the public repository \textsf{fuller} \cite{Stimper2019a}.

\item[] \textbf{Data reduction.} The raw data and intermediate results are stored in the HDF5 format \cite{Xian2020a}. The file sizes quoted here for reference are calculated from storage as double-precision floats or integers (for indices). The photoemission band mapping data of WSe$_2$ (256$\times$256$\times$470 bins) have a size of about 235 MB (240646 kB) after binning from single-event data (7.8 GB or 8176788 kB). The reconstructed valence bands at the same resolution occupy about 3 MB (3352 kB) in storage, and the size further decreases to 46 kB when we store the shape feature vector associated with each band. If only the top-100 coefficient (ranked by the absolute values of their amplitudes) and their indices in the feature vectors are stored, the data amounts to 24 kB. For the case of WSe$_2$, the top-100 coefficients can approximate the band dispersion with a relative error (see Eq. \eqref{eq:relapprox}) of $< 0.8\%$ for every energy band, as shown in Supplementary Fig. \ref{fig:approx}.
\end{enumerate}

\printbibliography[keyword=methods, heading=subbibliography, title=References]

\subsection*{Acknowledgments}
We thank M.~Scheffler for fruitful discussions and S.~Sch{\"u}lke, G.~Schnapka at Gemeinsames Netzwerkzentrum (GNZ) in Berlin and M.~Rampp at Max Planck Computing and Data Facility (MPCDF) in Garching for support on the computing infrastructure. The work was partially supported by BiGmax, the Max Planck Society's Research Network on Big-Data-Driven Materials-Science, the European Research Council (ERC) under the European Union's Horizon 2020 research and innovation program (Grant No.~740233 and Grant No. ERC-2015-CoG-682843), the German Research Foundation (DFG) through the Emmy Noether program under grant number RE 3977/1, the SFB/TRR 227 ``Ultrafast Spin Dynamics" (project-ID 328545488, projects A09 and B07), and the NOMAD pillar of the FAIR-DI e.V. association. We thank M. Bremholm for providing the Bi$_2$Te$_2$Se samples, Ph. Hofmann and M. Bianchi for their support in obtaining Au(111) photoemission data. M. Dendzik acknowledges support from the Göran Gustafssons Foundation. S. Beaulieu acknowledges the financial support of the Banting Fellowship from the Natural Sciences and Engineering Research Council (NSERC) in Canada.

\subsection*{Authors contributions}
R.P.X. and R.E. conceived the project. The photoemission band mapping experiments were supervised by L.R., R.E., and M.W.. S.D. and Sa.B. acquired the data on WSe$_2$, M.D. acquired the data on Bi$_2$Te$_2$Se and Au(111). M.Z., M.D., and C.C. performed the DFT band structure calculations. R.P.X. and M.D. processed the raw data. R.P.X. devised the band structure digitization, algorithm validation schemes, metrics, and performed computational benchmarking. V.S. designed and implemented the machine learning algorithm under the supervision of St.B. and B.S. along with inputs from R.P.X.. R.P.X., V.S. co-wrote the first draft of the manuscript with contributions from M.Z. and M.D.. All authors contributed to discussion and revision of the manuscript to its final version.

\subsection*{Data availability}
Source data for Figs. \ref{fig:overview}-\ref{fig:othermat} are available with this manuscript. The electronic structure calculation of WSe$_2$ are available from the NOMAD repository (\href{http://dx.doi.org/10.17172/NOMAD/2020.03.28-1}{10.17172/NOMAD/2020.03.28-1}). The raw and processed photoemission datasets used in this work for WSe$_2$ (\href{http://dx.doi.org/10.5281/zenodo.7314278}{10.5281/zenodo.7314278}), Bi$_2$Te$_2$Se (\href{http://dx.doi.org/10.5281/zenodo.7317667}{10.5281/zenodo.7317667}), and Au(111) (\href{http://dx.doi.org/10.5281/zenodo.7305241}{10.5281/zenodo.7305241} including DFT calculation) are available on Zenodo.

\subsection*{Code availability}
The code developed for band structure reconstruction including examples is available at \\ \href{https://github.com/mpes-kit/fuller}{https://github.com/mpes-kit/fuller}.

\subsection*{Competing interests}
The authors declare no competing interests in the content of the article.

\clearpage


\renewcommand{\figurename}{Supplementary Figure}
\renewcommand{\tablename}{Supplementary Table}
\setcounter{figure}{0}

\begin{quote}
    \centering
    \huge Supplementary Information \\
    \vspace{0.5em}
    A machine learning route between band mapping and band structure
\end{quote}
\vspace{1em}

\tableofcontents

\clearpage
\doublespacing
\section{Band structure reconstruction}
\subsection{Physical foundations}
The three quantities of common interest for the interpretation of photoemission spectra are (1) the bare band energy, $\epsilon_\textbf{k}$, (2) the complex-valued electron self-energy, $\Sigma(\textbf{k}, E) = \mathrm{Re}\Sigma(\textbf{k}, E) + i\mathrm{Im}\Sigma(\textbf{k}, E)$, and (3) the transition matrix elements connecting the final ($f$) and initial ($i$) electronic states, $M_{f,i}(\textbf{k}, E)$. An established interface between theory and experiment for quantitating and interpreting the photoemission signal is the formalism of an experimental observable: the single-particle spectral function \cite{Hufner2003,Damascelli2003}, $A(\textbf{k}, E)$. For a single energy band of a many-body electronic system,
\begin{equation}
    A(\textbf{k}, E) = \frac{1}{\pi} \frac{\text{Im}\Sigma(\textbf{k}, E)}{\left[E - \epsilon_{\textbf{k}} - \text{Re}\Sigma(\textbf{k}, E)\right]^2 + \left[\text{Im}\Sigma(\textbf{k}, E)\right]^2}.
    \label{eq:spsf}
\end{equation}
Within this framework, the band loci of the photoemission (or quasiparticle) band structure (BS), $b(\textbf{k}, E) = \epsilon_{\textbf{k}} + \mathrm{Re}\Sigma(\textbf{k}, E)$, correspond to the bare band dispersion modulated by the real part of the electron self-energy, and they occupy the local maxima of the spectral function evaluated at different momenta. However, in the photoemission process, the intensity counts registered by the detector are modulated by the transition matrix elements \cite{Moser2017}, the Fermi-Dirac occupation function, $f_{\mathrm{FD}}(E)$, and the resolution of the measuring instrument, $G(E, \sigma_E, \sigma_{\textbf{k}})$, typically a multidimensional Gaussian function. This leads to the expression of the photoemission intensity, $I(\textbf{k}, E)$, registered on an energy- and momentum-resolved detector,
\begin{equation}
    I(\textbf{k}, E) \propto |M_{f,i}(\textbf{k}, E)|^2 f_{\text{FD}}(E) A(\textbf{k}, E) \otimes G(E, \sigma_E, \sigma_{\textbf{k}}).
    \label{eq:pes_signal}
\end{equation}
For a multiband electronic structure, band mapping measurements, in principle, have access to the spectral functions of at least all valence bands. The photoemission intensities are combined in summation to form the multiband (MB) counterpart of the single-band formula.
\begin{align}
    I_{\mathrm{MB}}(\textbf{k}, E) &= \sum_j I_j(\textbf{k}, E) \propto \sum_j |M_{f_j,i_j}(\textbf{k}, E)|^2 f_{\mathrm{FD}}(E) A_j(\textbf{k}, E) \otimes G(E, \sigma_E, \sigma_{\textbf{k}})
    \label{eq:multiband_I}\\
    &\sim \sum_j A_j(\textbf{k}, E) \otimes G(E, \sigma_E, \sigma_{\textbf{k}}), \quad (\mathrm{when } |M_{f_j,i_j}(\textbf{k}, E)| \rightarrow 1, f_{\mathrm{FD}}(E) \rightarrow 1).
    \label{eq:multiband_Isim}
\end{align}
The condition $f_{\mathrm{FD}}(E) \rightarrow 1$ applies to valence bands, while $|M_{f_j,i_j}(\textbf{k}, E)| \rightarrow 1$ may be achieved through nonlinear intensity normalization or contrast enhancement in data processing. The expression of the multiband photoemission intensity in Eqs. \eqref{eq:multiband_I}-\eqref{eq:multiband_Isim} provides the physical foundation and inspiration for the approximate generation of band mapping data (see Supplementary Section 2) that we employ to validate the reconstruction algorithm introduced in this work.

\subsection{Markov random field modeling}
The Markov random field (MRF) model for the photoemission band structure in photoemission band mapping data can be constructed similarly for data in multiple dimensions. In traditional angle-resolved photoemission spectroscopy (ARPES), photoemission intensities are measured in the $(k, E)$ coordinates, the proximity of the momentum positions in the band structure can be modeled using an MRF composed of a 1D chain of random variables as shown in Supplementary Fig. \ref{fig:multidim_mrf}a. Band mapping data in $(k_x, k_y, E)$ coordinates, as described in the main text, can be modelled using a 2D MRF. In addition, the algorithm can be extended to higher dimensions involving coordinates beyond energy and momenta. For example, time-resolved photoemission data recorded in $(k_x, k_y, E, t)$ coordinates can be modelled using a 3D MRF as shown in Supplementary Fig. \ref{fig:multidim_mrf}c. In the following, we provide a brief introduction to the theory underlying MRF and provide a simplified derivation of the 2D MRF model introduced in the main text.

Deriving the MRF amounts to determining the joint distribution of the random variables associated with its graphical representation. In probabilistic graphical model theory \cite{Bishop2006}, a graph is constructed from the fundamental components called cliques. Each clique $C$ of a graph is a subset of nodes that shares an edge with another node in $C$, with the total number of nodes in $C$ defined as its size. The MRFs in Supplementary Fig. \ref{fig:multidim_mrf}a-c that model the photoemission data are built out of cliques of sizes 1--2 shown in Supplementary Fig. \ref{fig:multidim_mrf}d. Although larger cliques can be constructed similarly \cite{Bishop2006}, their parent graphical models are described by more complex joint distributions with drastically higher computational costs in optimization, therefore are not used in our MRFs. Mathematically, each clique is represented by a so-called potential function, $\psi_C$, which is used to derive the joint distribution that characterizes the MRF. The potential function only depends on the node configuration in the cliques, $\mathbf{X}_C$, and satisfies $\psi_C(\mathbf{X}_C)\geqslant 0$. According to the Hammersley-Clifford theorem \cite{Hammersley1971,Besag1974,Bishop2006}, the joint distribution of a vector of random variables, $\mathbf{X}$, can be written in the factorized form,
\begin{equation}
    p(\mathbf{X}) = \frac{1}{Z}\prod_{C\in\mathcal{C}}\psi_C(\mathbf{X}_C).
    \label{equ:joint_p_gen}
\end{equation}
Here, $\mathcal{C}$ is the set of all cliques in the graph, and the partition function $Z$ is a normalization constant given by
\begin{equation*}
    Z = \sum_\mathbf{X}\prod_{C\in\mathcal{C}}\psi_C(\mathbf{X}_C).
\end{equation*}
\begin{figure}[htb!]
    \centering
	\includegraphics[width=\textwidth]{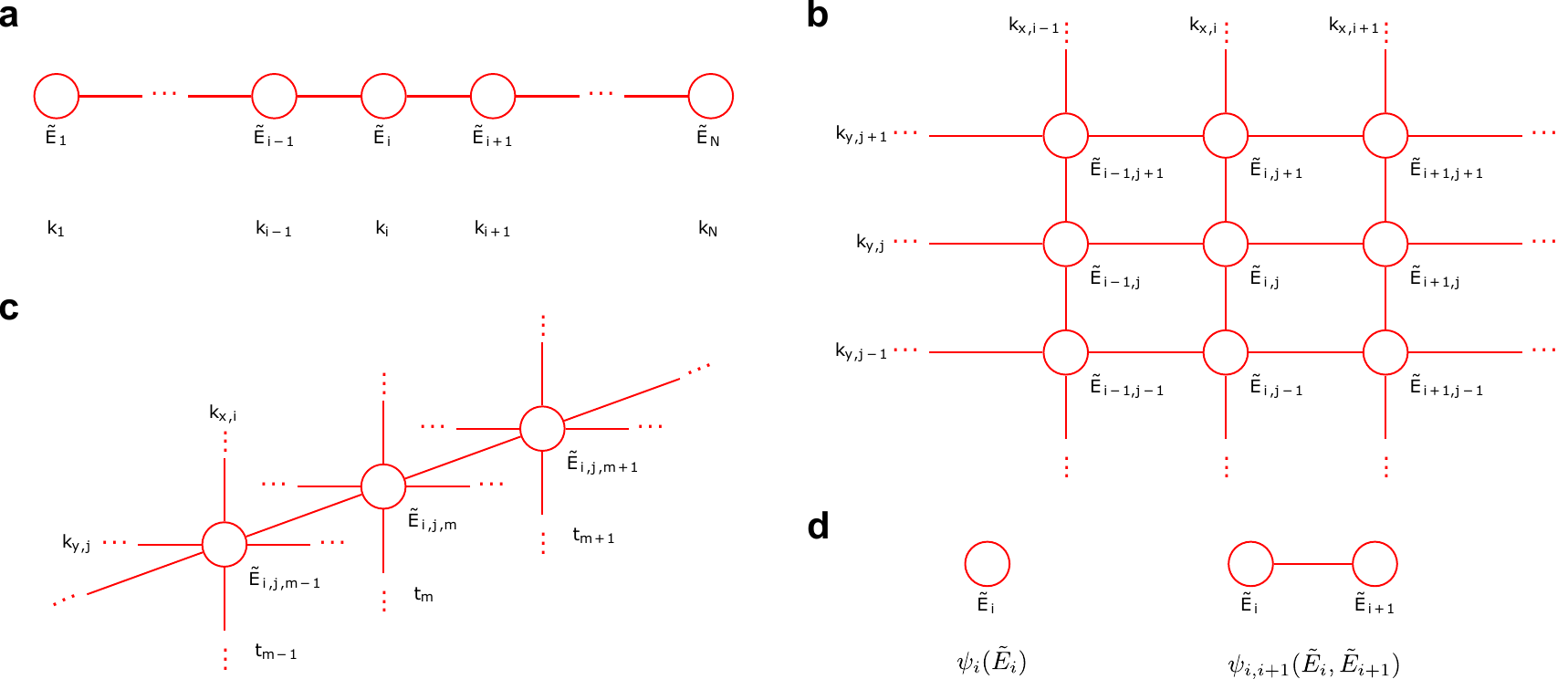}
	\caption{\textbf{Examples of the MRF models for photoemission spectroscopy data}. \textbf{a}, 1D MRF model for data in $(k, E)$ coordinates, represented as a chain of random variables $\Tilde{E}_i$. $N$ is the number of measured momentum values. \textbf{b}, 2D MRF model of photoemission data in $(k_x, k_y, E)$ coordinates as introduced and demonstrated for use in the main text, with the random variables $\Tilde{E}_{i,j}$ connected on two dimensions $k_x$ and $k_y$. \textbf{c}, 3D MRF model for time- and momentum-resolved photoemission spectroscopy data in $(k_x, k_y, E, t)$ coordinates. The random variables $\Tilde{E}_{i,j,m}$ are first connected in the graph to the neighboring momentum positions as in \textbf{b}, then subsequently along the neighboring time points. The time variable in \textbf{c} may also be replaced with other variables without changes in the structure of the graphical model. In \textbf{a}-\textbf{c}, the MRFs are constructed using components (cliques) with sizes 1 (left) and 2 (right) in \textbf{d}, with their respective potential functions written below the illustrations.}
	\label{fig:multidim_mrf}
\end{figure}
The graphical representation of the MRFs relevant to this work are rectangular grids shown in Supplementary Fig. \ref{fig:multidim_mrf}. The respective potential functions of the size-1 and size-2 cliques are interpreted as the likelihood and prior of the probabilistic graphical model, respectively. To cast the band structure reconstruction problem into this framework, we assign the band energies as the random variables in the model, and the potential function of each node (size-1 clique) as the (preprocessed) photoemission intensity at the respective grid position. For simplicity and computational efficiency, this formulation doesn't explicitly account for the intensity modulations described in Eq. \eqref{eq:pes_signal} and preprocessing steps are required to neutralize their effects. The continuity assumption (i.e. no sharp jump) of the band energies along momentum directions means that the potential function of size-2 cliques can be represented by a Gaussian on adjacent momentum grid positions. Intuitively, this means that the closer the two adjacent energies is, the more probable they are the actual band loci, and \textit{vice versa}.

In the 1D case (see Supplementary Fig. \ref{fig:multidim_mrf}a), the potential function of each node (containing one band energy random variable $\Tilde{E}_i$) is given by
\begin{equation}
    \psi_i(\Tilde{E}_i) = \Tilde{I}(k_i, \Tilde{E}_i),
    \label{equ:pot_1_clq}
\end{equation}
where $\Tilde{I}$ is the photoemission intensity after preprocessing. The potential function of two connected nodes (describing the similarity between two neighboring band energy random variables) is given by
\begin{equation}
    \psi_{j,j+1}(\Tilde{E}_j, \Tilde{E}_{j+1}) = \exp\left[{-\frac{(\Tilde{E}_j - \Tilde{E}_{j+1})^2}{2\eta^2}}\right].
    \label{equ:pot_2_clq}
\end{equation}
Plugging Eqs. \eqref{equ:pot_1_clq}-\eqref{equ:pot_2_clq} into Eq. \eqref{equ:joint_p_gen} yields
\begin{align}
    p(\Tilde{E}_1, ..., \Tilde{E}_N) =& \frac{1}{Z}\prod_{i=1}^N\psi_i(\Tilde{E}_i)\cdot \prod_{j=1}^{N-1} \psi_{j,j+1}(\Tilde{E}_j,\Tilde{E}_{j+1}) \nonumber\\
    =& \frac{1}{Z}\prod_{i=1}^N\Tilde{I}(k_i,\Tilde{E}_i)\cdot \prod_{j=1}^{N-1} \exp\left[-\frac{\left(\Tilde{E}_j - \Tilde{E}_{j+1}\right)^2}{2\eta^2}\right] \label{equ:posterior_1D}
\end{align}
as the joint distribution of the 1D MRF, with $N$ being the total number of momentum grid points. Analogously, we can derive the joint distribution of the 2D MRF as given in the main text, and that for the 3D MRF in the $(k_x, k_y, E, t)$ coordinates is
\begin{equation*}
    p(\{\Tilde{E}_{i,j,m}\}) = \frac{1}{Z}\prod_{i,j,m} \Tilde{I}(k_{x,i}, k_{y,j}, t_m, \Tilde{E}_{i,j,m}) \cdot \prod_{(i,j,m),(l,o,q)|\mathrm{NN}} \exp \left[ - \frac{(\Tilde{E}_{i,j,m}-\Tilde{E}_{l,o,q})^2}{2\eta^2}\right].
\end{equation*}
The MRF models in different dimensions discussed here follow the same Bayesian interpretation as the 2D MRF (Eq. (1) in the main text).

In practice, a 4D dataset of the kind, $I(E, k_x, k_y, k_z)$, and comparable spacing along the momentum dimensions ($\Delta k_x$ $\sim$ $\Delta k_y$ $\sim$ $\Delta k_z$) should be treated together to best use the connectivity encoded in the structured prior of the Markov random field model, which becomes a 3D grid of random variables that accounts for the connectedness along the momentum directions. When the fourth dimension (such as $k_z$) in the data is not sampled as densely as the other dimensions ($\Delta k_x$ $\sim$ $\Delta k_y$ $\ll$ $\Delta k_z$), which may be the case for synchrotron-based photoemission instruments (resulting in 3.5 D or quasi-4D datasets), the dataset can also be treated individually per scanned photon energy, since the local connectedness assumption along the third momentum direction is no longer retained.

\subsection{Optimization procedure}
Optimization of the MRF model is a local minima-finding process \cite{Bishop2006}. The following procedures are described using the 2D MRF in the main text as an example, but the approach can be extended to arbitrary dimensions. Due to the large number of random variables ($\sim$ $10^4$ for the 2D MRF in the main text) and their complex dependence structure in the MRF, we solved it numerically using iterated conditional mode (ICM) \cite{Kittler1984} procedure and implemented with efficient parallelization schemes, including the coding method and the hierarchical grouping of random variables. Next, we discuss the motivations and clarify the details of these three aspects. We provide the associated pseudocode in Algorithm \ref{alg:optimization}.

\begin{enumerate}[label={\textbf{\arabic*.}}, wide, labelwidth=!, labelindent=0pt]
\item \textbf{Iterated conditional mode}: Originally developed for similar optimization problems arising in image denoising \cite{Geman1984,Besag1986,Bishop2006}, ICM is applicable to optimizing MRF at any dimension. The ICM procedure includes (i) initialization of the random variables (e.g. $\{ \Tilde{E}_{i,j} \}$ in 2D MRF) and (ii) selection of a single random variable to optimize in the loss function $\mathcal{L}$ while keeping all the other random variables fixed. Each round in (ii) requires computing at most five terms in the loss (Eq. (3) in the main text Methods) which depend on the selected random variable $\Tilde{E}_{i,j}$. We can simply evaluate these terms at the energy axis values measured in the experiment to determine the energy associated with the lowest loss. (iii) iterate over all other random variables using the same procedure in (ii).
\begin{figure}[htb!]
    \centering
    \includegraphics[width=\textwidth]{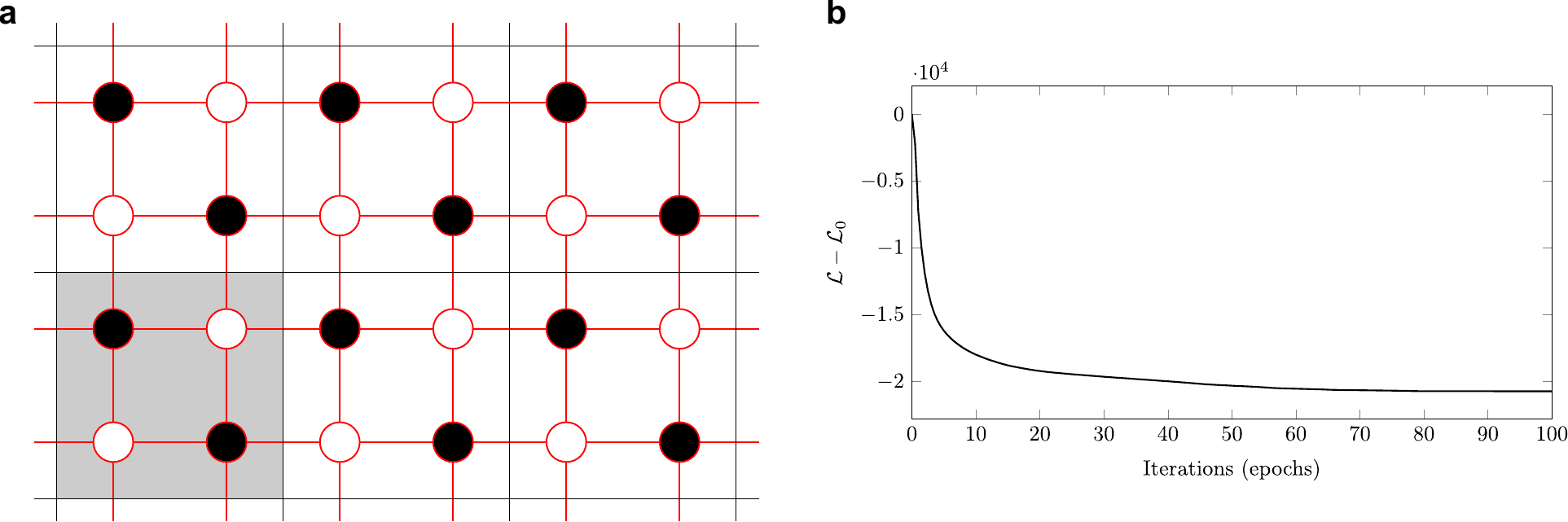}
    \caption{\textbf{Numerical optimization of the MRF model}. \textbf{a}, Schematic of the checkerboard parallelization (or coding method) and hierarchical grouping schemes for speeding up the ICM. The nodes of the MRF are alternately colored white and black (checkerboard parallelization) and each set of four neighboring nodes are group into a unit as colored in grey (hierarchical grouping). The updates in optimization are carried out first at the four-node unit level, then alternately on the white or black nodes within the units. \textbf{b}, An example loss curve for reconstructing the second valence band of WSe$_2$ using the 2D MRF model and parallelized ICM implementation. $\mathcal{L}_0$ is the initial value of the loss at the start of the optimization. Within an epoch in the parallelized scheme, the white nodes and subsequently the black nodes are separately updated, therefore each band energy random variable is effectively updated once. The loss decreases rapidly in the beginning and reaches a minimum after about 90 epochs.}
    \label{fig:optimization_2D}
\end{figure}
\item \textbf{Coding method}: The ICM procedure described above operates sequentially over every $\Tilde{E}_{i,j}$, which is inefficient for the MAP optimization involving a large number of parameters. To improve the optimization performance, we implement the ICM with a checkerboard parallelization scheme (or coding method) \cite{Li2009} that scales favorably on multicore computing clusters. The scheme assigns the nodes of the MRF alternately with white and black colors, as shown in Supplementary Fig. \ref{fig:optimization_2D}a. If the white nodes are blocked, the black nodes are no longer connected through paths (i.e. sequences of connected edges and nodes). This property is called d-separation \cite{Pearl1988,Bishop2006}. Analogously, blocking the black nodes d-separates the white nodes. Since the MRF models satisfy the Hammersley-Clifford theorem \cite{Hammersley1971}, d-separation is equivalent to conditional independence, meaning that the random variables represented by the black nodes are independent if we condition on those represented by the white nodes. Therefore, conditioning on the nodes of one color allows us to compute the terms in the log-probability loss (Eq. (3) in main text Methods) that depends on the nodes of another color in parallel, which means that the nodes associated with different colors can be updated alternately. Further details and proofs related to the coding method have been elaborated in \cite{Besag1972,Besag1974}.
\begin{spacing}{0.8}
\begin{algorithm}
\caption{Optimization procedure for reconstructing a single energy band.}
\label{alg:optimization}
\textbf{Input}: I (3D momentum-resolved photoemission data), E$_0$ (2D initialization from density functional theory calculation), E (1D energy axis)\\
\textbf{Parameter}: $\eta$ (hyperparameter of the Markov random field), N (number of epochs)\\
\textbf{Output}: E$_{\mathrm{rec}}$ (Reconstructed 2D energy band)
\begin{algorithmic}[1]
    \Statex {\color{cmmt} \textsf{\# Initialize the momentum index grid for an energy band}}
    \State size\_kx, size\_ky, size\_E = size(I)
    \State ind\_x, ind\_y = meshgrid(range(size\_kx, step=2), range(size\_ky, step=2))
    \Statex {\color{cmmt} \textsf{\# Divide data into four-node units. E$_u$(i,j,...), I$_u$(i,j,...) are the band energies and\newline \# photoemission intensities for the node (i,j) in a unit (u) in Supplementary Fig. \ref{fig:optimization_2D}, respectively}}
    \For{i \textbf{in} [0, 1]}
    \For{j \textbf{in} [0, 1]}
    \State E$_u$[i, j, :, :] = E$_0$[ind\_x + i, ind\_y + j]
    \State log\_I$_u$[i, j, :, :, :] = log(I[ind\_x + i, ind\_y + j, :])
    \EndFor
    \EndFor
    \Statex {\color{cmmt} \textsf{\# Iterative optimization of energy values}}
    \For{n \textbf{in} range(N)}
    \Statex \hspace{1em} {\color{cmmt} \textsf{\# Update white nodes}}
    \State E$_u$[0, 0, :, :] = update\_E(0, 0, log\_I$_u$, E$_u$, E)
    \State E$_u$[1, 1, :, :] = update\_E(1, 1, log\_I$_u$, E$_u$, E)
    \Statex \hspace{1em} {\color{cmmt} \textsf{\# Update black nodes}}
    \State E$_u$[0, 1, :, :] = update\_E(0, 1, log\_I$_u$, E$_u$, E)
    \State E$_u$[1, 0, :, :] = update\_E(1, 0, log\_I$_u$, E$_u$, E)
    \EndFor
    \Statex {\color{cmmt} \textsf{\# Assemble reconstruction from all nodes in the units}}
    \For{i \textbf{in} [0, 1]}
    \For{j \textbf{in} [0, 1]}
    \State E$_{\mathrm{rec}}$[ind\_x + i, ind\_y + j] = E$_u$[i, j, :, :]
    \EndFor
    \EndFor
    \vspace{1em}
    \Statex {\color{cmmt} \textsf{\# Function to update the energy of the element (i, j) within a four-node unit}}
    \Function{update\_E}{i, j, log\_I$_u$, E$_u$, E}
    \Statex \hspace{1em} {\color{cmmt} \textsf{\# Calculate the difference between current and all possible energies}}
    \State squ\_diff = (E$_u$ - E) ** 2 / (2 * $\eta$ ** 2)
    \Statex \hspace{1em} {\color{cmmt} \textsf{\# Calculate all possible $\log p$ values, start with log-likelihood}}
    \State log\_p = log\_I$_u$[i, j, :, :, :]
    \Statex \hspace{1em} {\color{cmmt} \textsf{\# Substract by energy differences from nearest neighbor nodes within unit}}
    \State log\_p -= squ\_diff[(i + 1) \% 2, j, :, :, :]
    \State log\_p -= squ\_diff[i, (j + 1) \% 2, :, :, :]
    \Statex \hspace{1em} {\color{cmmt} \textsf{\# Substract by energy differences from nearest neighbor nodes of the neighboring unit}}
    \State log\_p -= shift(squ\_diff[(i + 1) \% 2, j, :, :, :], 2 * i - 1, axis=2)
    \State log\_p -= shift(squ\_diff[i, (j + 1) \% 2, :, :, :], 2 * j - 1, axis=3)
    \Statex \hspace{1em} {\color{cmmt} \textsf{\# Return optimal energy values}}\\
    \hspace{1em} \Return E[argmax(log\_p)]
    \EndFunction
\end{algorithmic}
\end{algorithm}
\end{spacing}

\item \textbf{Hierarchical grouping}: The introduction of the checkerboard parallelization scheme reduces the translation symmetry of the original graph (originally symmetric by translation of an arbitrary number of nodes, now only symmetric by a translation of two nodes in each direction), which complicates the matrix operations needed to update the loss. However, we can restore the translation symmetry and carry out the computation on a higher level by grouping a set of four neighboring nodes into a unit, as illustrated in Supplementary Fig. \ref{fig:optimization_2D}a. In this way, updating the loss requires only standard matrix operations at the unit level followed by consecutive updates of the nodes within the units. During the optimization, the loss is updated by two sets of operations concerning (i) the nearest neighbor nodes within the unit (line 18-19 in Algorithm \ref{alg:optimization}) and (ii) the nearest neighbor nodes of the neighboring unit (line 20-21 in Algorithm \ref{alg:optimization}). The latter operations are carried out by shifting the higher-level rectangular grid formed by the units by one step vertically or horizontally, followed by an operation on nodes of the respective units of the original and the shifted grid. The procedure is implemented in the open-source \textsf{fuller} package \cite{Stimper2019a} using Tensorflow \cite{Abadi2016}. Supplementary Fig. \ref{fig:optimization_2D}b shows an example loss curve (i.e. loss as a function of iteration) in reconstruction of an energy band, where the optimization is essentially complete within $\sim$ 90 iterations.

\item \textbf{Robust initialization}: Since the current MRF model doesn't include any explicit regularization on the outcome with respect to the initialization, the optimizer is free to explore a large range of values. In other words, the initial band dispersion is able to freely deform to fit to the band loci embedded in the data. This design improves the robustness of the algorithm to initialization. As a result, in scenarios with only non-crossing energy bands, the MAP optimization can simply be initialized with constant energy values to yield consistent results. In general situations involving band crossings, the optimization procedure requires an initialization with approximate energy values that preserves the band-crossing information, such as those provided by electronic structure calculations. In this scenario, the robustness of the algorithm is manifest in the fact that it can tolerate a certain amount of deviation in the initialization and still converges to a satisfactory reconstruction, which, in realistic settings, is closer to the real band structure contained in photoemission data than the initialization (e.g. from electronic structure calculations). Quantitative examples demonstrating the robustness of initialization are provided using synthetic data in Supplementary Figs. \ref{fig:synthband1}-\ref{fig:synthband2} (see Supplementary Section 2).
\end{enumerate}

\subsection{Hyperparameter tuning}
The optimization process in the band structure reconstruction involves the tuning of three kinds of hyperparameters, which are the momentum scaling parameter, the rigid energy shift and the width of the nearest-neighbor Gaussian prior. A flowchart presented in Supplementary Fig. \ref{fig:tuningdiag} illustrates the general steps in obtaining a desirable reconstruction including where the tuning of each hyperparameter fits in.
\begin{figure}[htb!]
    \centering
    \includegraphics[width=0.7\textwidth]{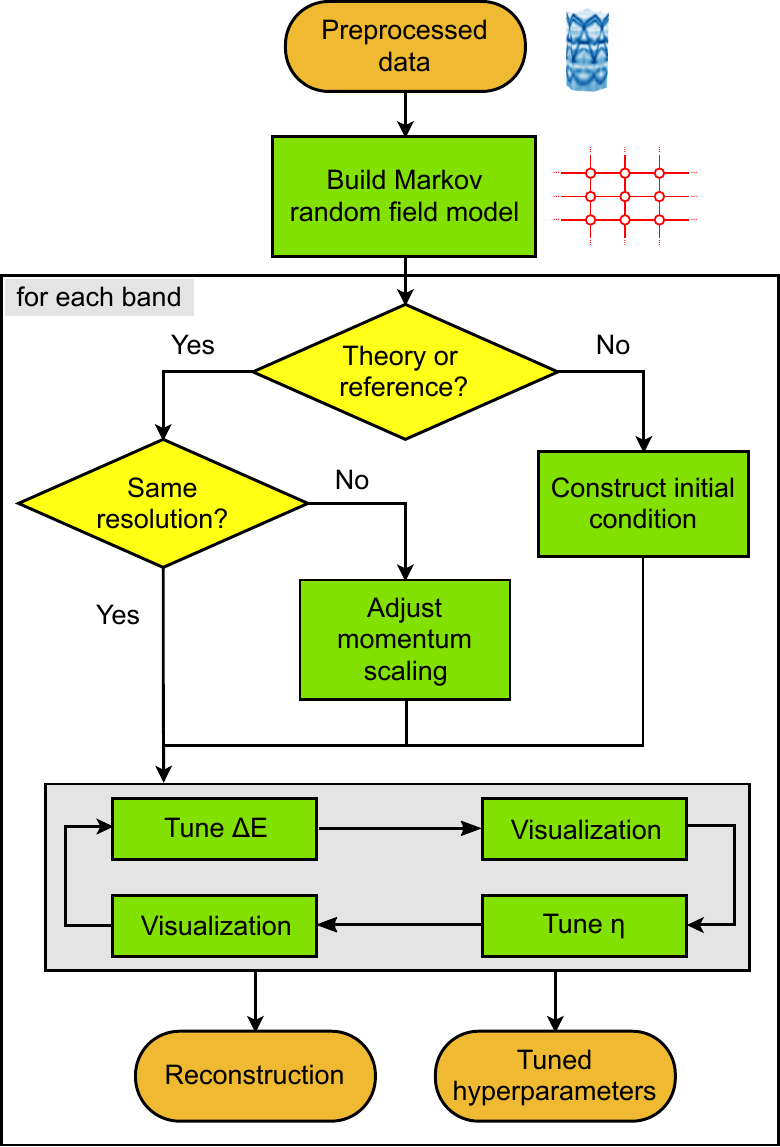}
    \vspace{1em}
    \caption{\textbf{Flowchart for reconstruction tuning}. Illustration of the steps for tuning the reconstruction starting from preprocessed data (outcome from the procedure illustrated in the main text Fig. \ref{fig:overview}c-f). Tuning of the three hyperparameters -- the momentum scaling, energy shift ($\Delta$E) and nearest-neighbor Gaussian width ($\eta$), are placed in sequence within the workflow. The workflow outputs reconstruction of a single band with tuned hyperparameters at the end. For reconstructing the dispersion of multiple energy bands, the workflow is repeated over each band.}
    \label{fig:tuningdiag}
\end{figure}
\begin{enumerate}[label={\textbf{\arabic*.}}, wide, labelwidth=!, labelindent=0pt]
    \item \textbf{Momentum scaling}: applied to equalize the momentum scale and resolution between the BS calculation (e.g. conducted on relaxed unit cells, see Supplementary Table \ref{Table_1}) and the experimental data (measured on real materials). In our reconstruction procedure, the scaling factor is fixed in the reconstruction of all energy bands using a particular level of density functional theory (DFT) calculation as initialization.
    
    \item \textbf{Rigid energy shift ($\Delta$E)}: separately applied to each energy band in the calculated BS to coarse-align to the band mapping data. In our case, the shift is chosen manually by visual inspection of the theoretical band energies overplotted on photoemission data (usually in the energy-momentum slices). In practice, the necessary energy shifts vary between bands and also depend on the level of approximation in the BS calculation used as initialization, as illustrated in Fig. 2a of the main text.
    
    \item \textbf{Width of the nearest-neighbor Gaussian prior ($\eta$)}: The value of the parameter $\eta$ is chosen manually from an initial estimate and subsequently optimized by visual inspection of the reconstruction outcome. In the case of WSe$_2$, the momentum grid of the experimental data has a spacing of $\Delta k_x = \Delta k_y \approx$ 0.015 $\mathrm{\AA}^{-1}$, we used $\eta \in$ [0.05, 0.2] eV. Generally speaking, the initial estimate of $\eta$ has the order of magnitude proportional to the momentum grid spacing times the dispersion due to the following argument: To obtain a consistent reconstruction, we expect the posterior to stay relatively constant and be independent of the momentum grid spacing, which should be sufficiently fine to ensure band continuity. Since after preprocessing the data, the intensity (i.e. the likelihood) is normalized and stays constant with respect to the momentum grid spacing, the nearest-neighbor Gaussian prior term should stay constant correspondingly. For example, for two nearest-neighbor energy variables along the $k_x$ axis, the reasoning above requires,
    \begin{equation}
        \mathrm{const} \approx \frac{(\Tilde{E}_{i + 1, j} - \Tilde{E}_{i, j})^2}{\eta^2} \approx \left(\frac{\partial E}{\partial k_x}\right) ^2 \frac{\Delta k_x^2}{\eta^2}.
    \end{equation}
    Thereby, we obtain $\eta \propto \frac{\partial E}{\partial k_x} \Delta k_x$, which provide an order-of-magnitude estimate of $\eta$. The same lines of reasoning apply to the $k_y$ axis, for detector systems with relatively constant momentum resolution. As the grid spacing is the same in both $k_x$ and $k_y$ directions, a single $\eta$ is used for reconstructing each band in the case of WSe$_2$, but the best $\eta$ differs somewhat between energy bands due to their various amounts of dispersion and how they are connected to the neighboring bands (i.e. their environment), hence the range of $\eta$ as specified earlier.
    \begin{figure}[htbp!]
        \centering
        \includegraphics[width=0.92\textwidth]{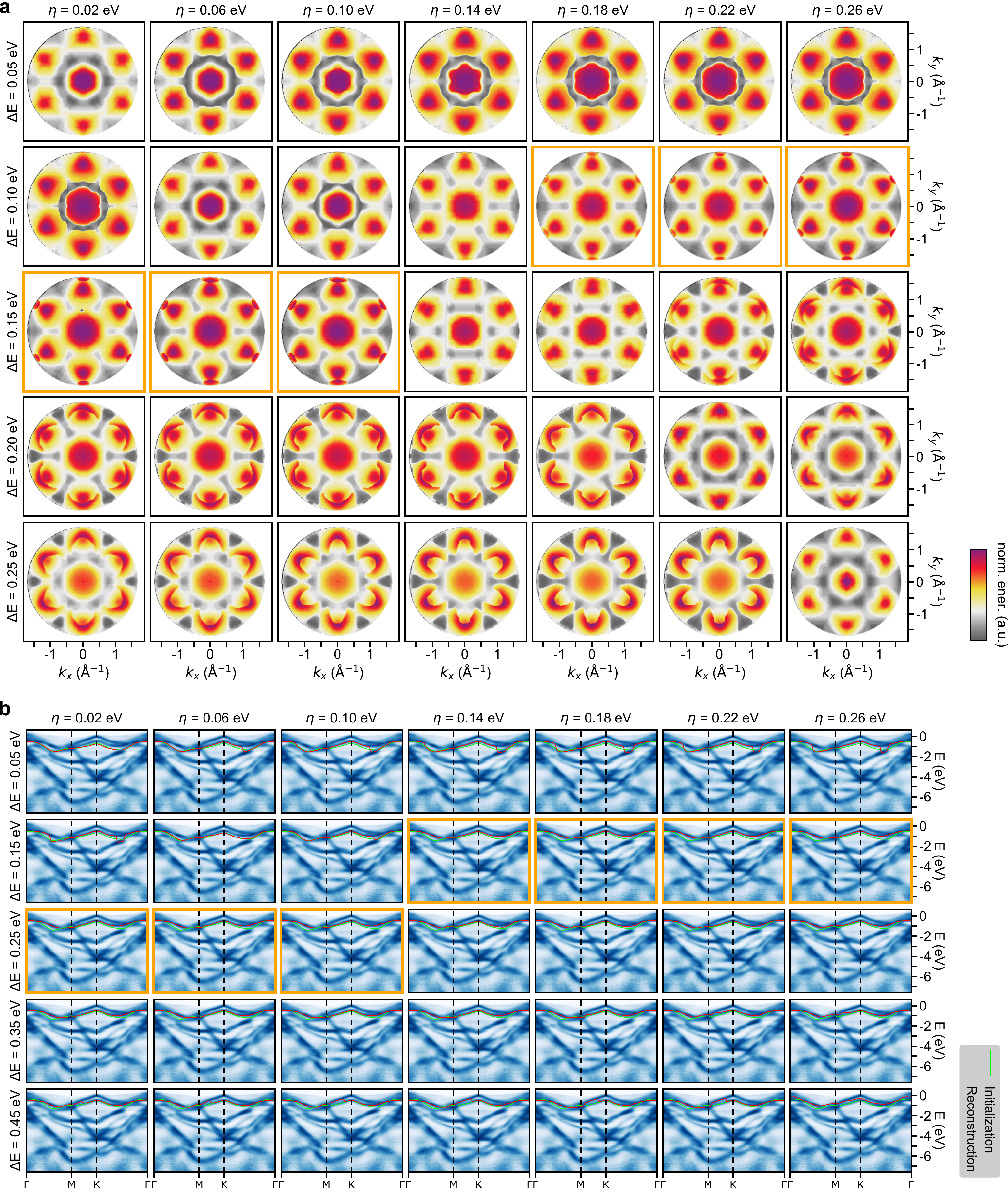} 
        \caption{\textbf{Demonstration of hyperparameter tuning}. An example of tuning the hyperparameters, the rigid energy shift ($\Delta$E) and the width of the nearest-neighbor Gaussian prior ($\eta$), for reconstructing the second valence band of WSe$_2$. \textbf{a}, Evolution of reconstructed energy band during hyperparameter tuning. \textbf{b}, Evolution of the initialization and reconstructed band along high-symmetry directions of the hexagonal lattice of WSe$_2$. The energy bands are overlaid on top of preprocessed data from photoemission band mapping of WSe$_2$ (Fig. 1f in the main text). In \textbf{a},\textbf{b}, the images showing the optimal region for the hyperparameters identified by the scientists are emphasized with orange-colored frames.}
        \label{fig:hypertuning}
    \end{figure}
    \begin{figure}[htbp!]
        \centering
        \includegraphics[width=\textwidth]{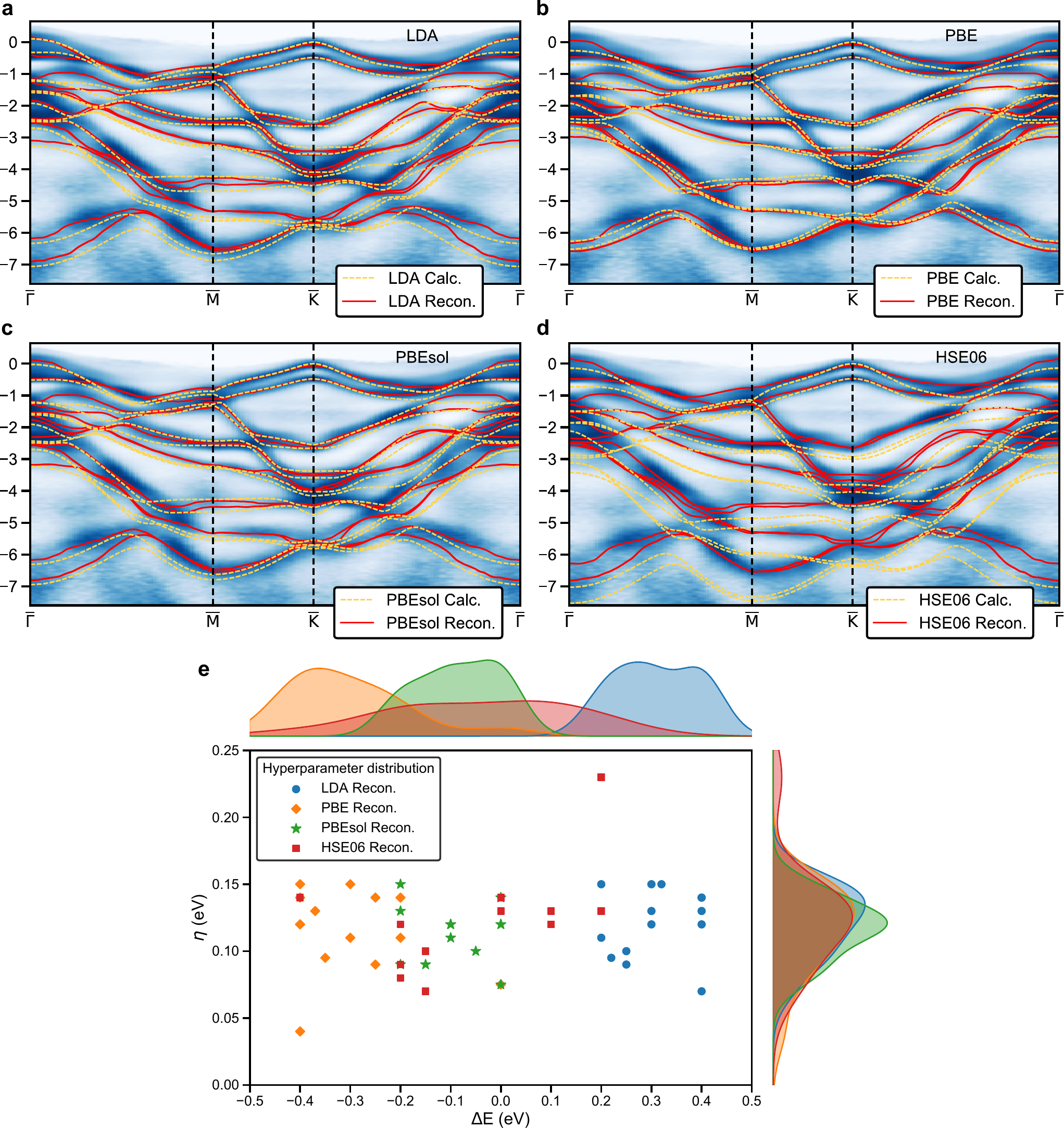}
        \caption{\textbf{Band structure reconstructions with different theory initializations}. Comparisons between reconstructed photoemission band structures (abbreviated as recon.) and calculated band structures (abbreviated as calc.) from density functional theory (DFT) with different exchange-correlation functionals, including \textbf{a,} local density approximation (LDA); \textbf{b,} PBE generalized gradient approximation (GGA); \textbf{c,} PBEsol GGA; \textbf{d,} HSE06 hybrid functional. For each set of DFT band structure, the same energy shift (as in Supplementary Fig. \ref{fig_band_strs}) is applied globally to all bands to align the energy zero at the $\overline{\mathrm{K}}$ point with the reconstruction. \textbf{e,} The distribution of hyperparameters used for the reconstruction in \textbf{a}-\textbf{d}.}
        \label{fig:theoryvsrecon}
    \end{figure}

    To demonstrate the process of hyperparameter tuning, we provide an example showing the reconstruction of the second valence band of WSe$_2$ (see Supplementary Fig. \ref{fig:hypertuning}), visualized in the top view of the reconstruction outcome and in the momentum path along high-symmetry lines of the projected Brillouin zone. The orange-framed subfigures represent the range of hyperparameter settings that yield a good reconstruction, which represents a relatively broad acceptance range to yield a good reconstruction. Although this aspect is dependent on the data, in our experience, the hyperparameter tuning may be carried out in a semi-automated fashion guided by visualization and heuristics. Typically, 10-20 trials are sufficient to yield a good reconstruction, although a grid search may also be carried out for completeness. For a given dataset, the hyperparameters typically fall within a similar range, therefore, determining the range of hyperparameters need only be carried out once. The choice of hyperparameters is more flexible for reconstructing more isolated bands or those with fewer crossings, and \textit{vice versa}. The band-wise reconstruction and the computational efficiency of the algorithm also enable further parallelization in hyperparameter tuning by distributing the optimization tasks in a high-performance computing infrastructure.
\end{enumerate}

\subsection{Reconstructions using different theories as initializations}
Comparison between reconstructed and theoretical band structures for 2$H$-WSe$_2$ are presented as a similarity matrix in the main text. To provide more intuitive visual guidance in interpreting the BS distance metric used in constructing the similarity matrix, we compare these band structures along the high-symmetry lines of the Brillouin zone in Supplementary Fig. \ref{fig:theoryvsrecon}.

Here, the comparison between reconstructed and calculated band structures show that the HSE06 and PBE have, respectively, the largest and smallest overestimation of total band width of WSe$_2$ among the four initializations, though HSE06 has a higher level of chemical accuracy than PBE \cite{Perdew2001}. The calculated band structures are closer to the reconstruction near the $\overline{\text{K}}$-point than elsewhere in the projected Brillouin zone, reflecting the difference in electronic dimensionality between $\overline{\text{K}}$ (nearly ideally 2D) and elsewhere \cite{Riley2014}.

\section{Generation of and validation on synthetic data}
The advantage of using synthetic data is that the underlying band structure (i.e. ground truth) is exactly known so it can be used for benchmarking the performance of the MAP reconstruction algorithm described in this work. Benchmarking includes numerical experiments on two interrelated aspects: (1) testing the robustness of the reconstruction algorithm using different initializations and comparing the deviations of the outcome from the ground-truth; (2) testing the accuracy of reconstruction by determining the closest-possible reconstruction outcome from a given initialization. In the following, we first describe the workflow of generating the band structure, the photoemission data and the initializations, which provide all essential components to carry out the tests. Then we present the benchmarking results on various cases.

\subsection{Generation of band structure data}
We have adopted two approaches to generate band structure data to meet the needs for testing the reconstruction algorithm. Firstly, we used analytic functions to describe the band dispersion (see Supplementary Fig. \ref{fig:synthband1}). They are computationally efficient, contain tunable parameters, can be produced at any resolution, and are easily extendable to higher dimensions. In 2D momentum space, we constructed a multi-sinusoidal band and two double-crossing parabolic bands. In 3D momentum space, we constructed a scaled version of the strongly oscillating second-order Griewank function \cite{Locatelli2003} and the tight-binding formulation of the two-band graphene band structure \cite{Bena2009} as model band dispersion surfaces. The modified Griewank function takes the form,
\begin{equation}
    E_{\mathrm{griewank}}(k_x, k_y) = \frac{1}{16000} (k_x^2 + k_y^2) - \cos(2k_x)\cos(\sqrt{2}k_y).
    \label{equ:band_griewank}
\end{equation}
The two-band tight-binding model of graphene has energy dispersion relations,
\begin{equation}
    E_{\pm}(k_x, k_y) = \pm \sqrt{3 + 2\cos\left(\sqrt{3}k_ya\right) + 4\cos\left(\frac{\sqrt{3}}{2}k_ya\right)\cos\left(\frac{3}{2}k_xa\right)}.
    \label{equ:band_graphene}
\end{equation}
Here, $E_{+}$ and $E_{-}$ refer to the conduction band and the valence band, respectively.
\begin{figure}[hbt!]
    \centering
    \includegraphics[width=\textwidth]{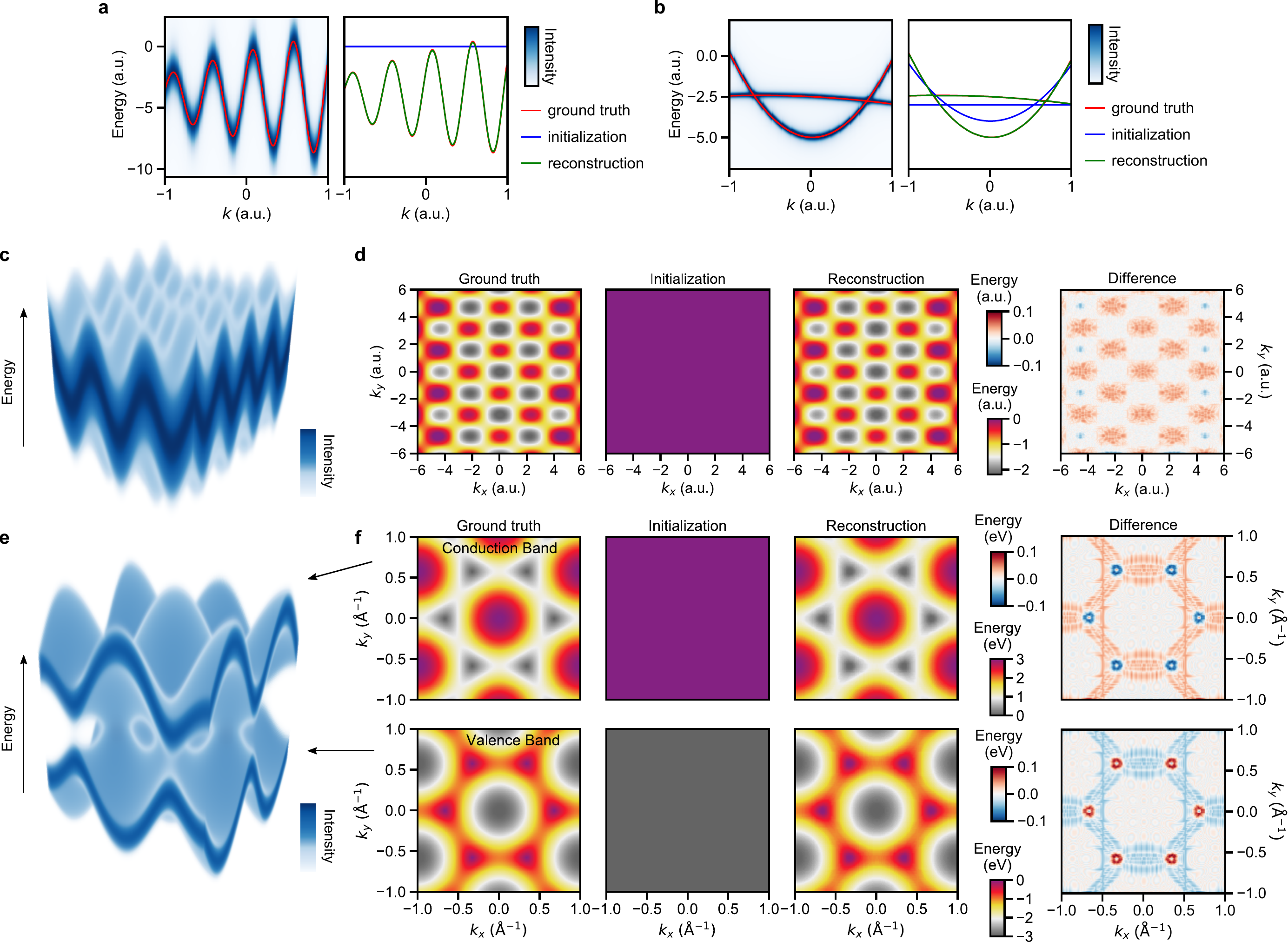}
    \caption{\textbf{Validations on 2D and 3D synthetic data}. Test results for the reconstruction algorithm on band structures generated with analytic functions. \textbf{a}, Reconstruction of a multi-sinusoidal band. \textbf{b}, Reconstruction of two double-crossing parabolic bands. \textbf{c},\textbf{d}, Reconstruction of a multi-extrema band with dispersion following the second-order Griewank function (see Eq. \eqref{equ:band_griewank}) \cite{Locatelli2003}. \textbf{e},\textbf{f}, Reconstruction of the two bands of graphene nearby its Fermi level (\textbf{e},\textbf{f}) formulated in the tight-binding model (see Eq. \eqref{equ:band_graphene}) \cite{Bena2009}. The volumetric renderings in \textbf{c},\textbf{e}, display the synthetic data. The initialization for the reconstruction in \textbf{a} is a flat line, while 2D flat bands are used to initialize the cases in \textbf{d},\textbf{f}. In \textbf{b}, two double-crossing curves are needed as initialization to preserve the crossing in the reconstruction. The values in the difference plots in \textbf{d},\textbf{f} are calculated by subtracting the ground-truth band energies from the reconstructed ones.}
    \label{fig:synthband1}
\end{figure}
Secondly, we used numerical band structures from DFT calculations with different exchange-correlation functionals (see Supplementary Section 4). They are more physically realistic, but also require more computation to obtain than generating bands from analytic functions.

\subsection{Initialization tuning}
For simple bands constructed using analytic functions, tuning can be achieved by modifying the parameters in the functions. In complex multiband situations such as that of WSe$_2$, we tuned the initialization of the reconstruction algorithm by scaling or perturbing the coefficient amplitudes of the constituent bases of the band structure. In our case, the bases are the terms of the hexagonal Zernike polynomials (ZPs) \cite{Mahajan2006,Mahajan2007}. Although unconstrained basis tuning is prone to unrealistic results, it achieves a level of ad hoc control for the efficient generation of a large amount of distinct initializations. For more physically realistic tuning, we used DFT calculations with different exchange-correlation functionals (see Supplementary Section 4).

\subsection{Approximate generation of photoemission data}
We approximately synthesized momentum-resolved photoemission data for each energy band by plugging the band energy and linewidth parameter at each momentum position into the Voigt profile \cite{VandeHulst1947} (with Gaussian and Lorentzian parameters $\sigma$ and $\gamma$, and amplitude $B$) computed using the Faddeeva function $W$ \cite{Zaghloul2011}. The Voigt profile approximates the convolution of a single-particle spectral function (see Supplementary Section 1), describing the photoemission observable, with a Gaussian energy resolution function. The synthetic photoemission intensity, $I_{\mathrm{synth}}$, for a band structure composed of a set of energy bands, $E_B = \{E_{b_i}\}$, is generated by combining multiple Voigt profiles in summation, similar to Eqs. \eqref{eq:multiband_I}-\eqref{eq:multiband_Isim}.
\begin{equation}
    I_{\mathrm{synth}}(k_x, k_y, E) = \sum_j \frac{B_j(k_x, k_y)}{\sigma_j\sqrt{2\pi}} \mathrm{Re}\left[W\left(\frac{E-E_{b_j}(k_x, k_y)+\textsf{i}\gamma_j(k_x, k_y)}{\sigma_j\sqrt{2}}\right)\right]
    \label{eq:voigt}
\end{equation}
Without loss of generality, we assume the energy resolution in detection for all bands to be the same ($\sigma_j = \sigma$). For the cases shown in Supplementary Figs. \ref{fig:synthband1}-\ref{fig:synthband2}, the linewidth parameter $\gamma$ are set to a constant throughout the band. In all synthetic data, we omitted the inhomogeneous intensity modifications in realistic photoemission data due to experimental factors such as the experimental geometry, sample condition, matrix element effect, photon energy, etc. This omittance relies on the assumption that the essential preprocessing step, such as symmetrization and contrast enhancement \cite{Stimper2019} in our workflow (see main text Methods), can sufficiently restore the intensity continuity along the energy bands. The momentum resolution effect is also not accounted for because the instrument (such as METIS 1000 \cite{Medjanik2017,SPECSGmbH2019}) has a higher momentum resolution than the momentum spacing used in data binning or generation.

\subsection{Validation of the reconstruction algorithm}
Using synthetic data generated from analytic functions of varying complexities as the band structure, we test out the accuracy of the reconstruction algorithm (see Supplementary Fig. \ref{fig:synthband1}); Using synthetic multiband data generated from the LDA-level DFT (LDA-DFT) band structures of WSe$_2$ (see Supplementary Section 4), we tested out the sensitivity of reconstruction to the initialization (see Supplementary Fig. \ref{fig:synthband2}). In this case, to capture sufficient physical realism similar to the photoemission band mapping of WSe$_2$ presented in the main text, we set the energy resolution parameter of $\sigma$ = 100 meV, the lineshape parameter $\gamma$ = 50 meV \cite{Dong2021}, and the energy spacing of data to $\sim$ 18 meV, identical to the energy bin size for the experimental data. The tests include four sets of numerical experiments summarized below:

\begin{enumerate}[wide, labelwidth=!, labelindent=0pt]
\item[\textbf{1.}] \textbf{Reconstructing non-crossing bands}: For isolated bands, we tested synthetic data constructed from a multi-sinusoidal band (Supplementary Fig. \ref{fig:synthband1}a), the band generated by the Griewank function (Supplementary Fig. \ref{fig:synthband1}c-d), and the two-band tight-binding model of graphene (Supplementary Fig. \ref{fig:synthband1}e-f). In these cases, initialization with a flat band without any initial knowledge of the band dispersion (i.e. cold start) is sufficient to recover its shape, regardless of the complexity of the dispersion.
\begin{figure}[hbtp]
    \centering
    \includegraphics[width=\textwidth]{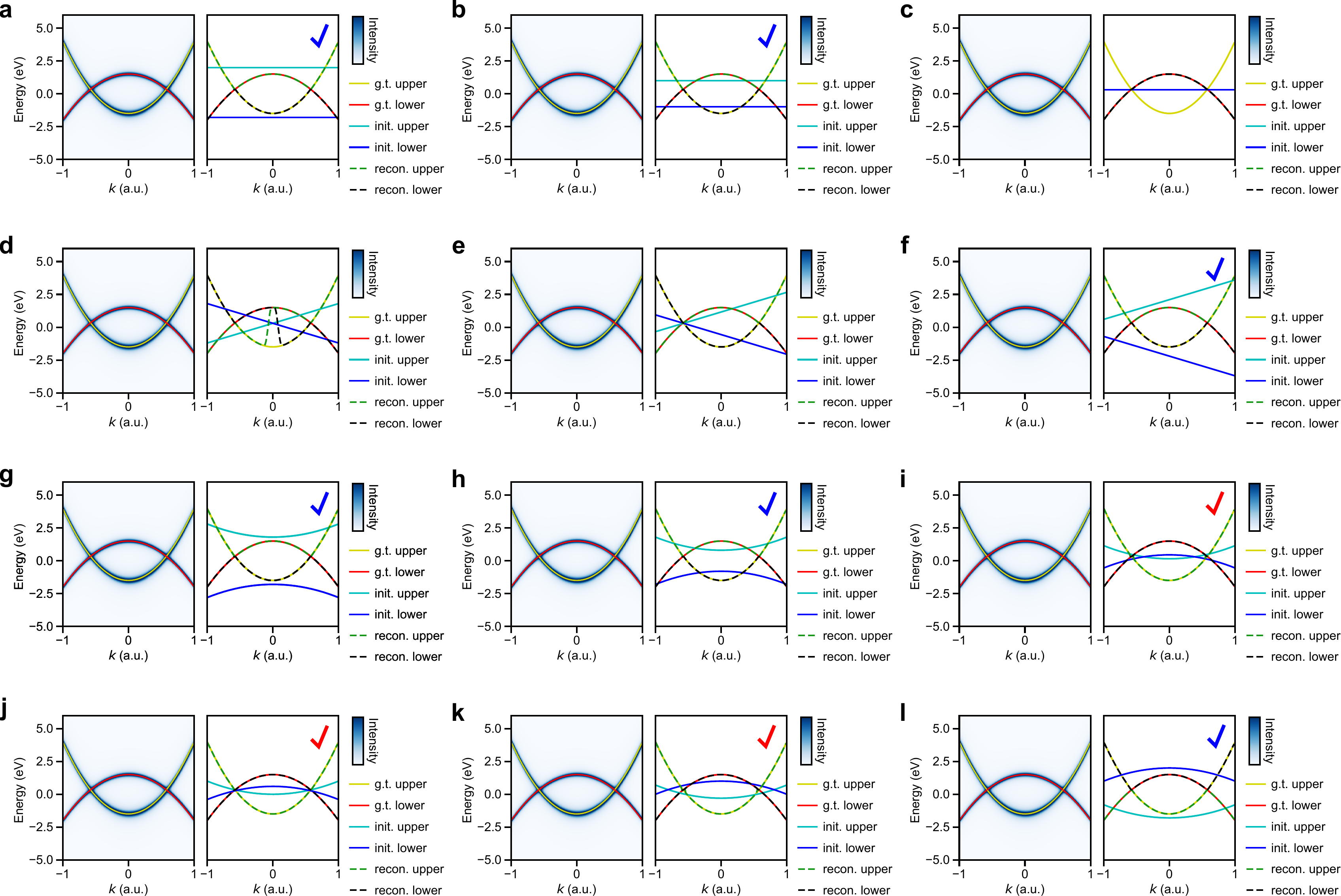} 
    \caption{\textbf{Essential information in initialization for reconstruction}. Results from a series of numerical experiments for demonstrating the effects of band-crossing information in the initialization. For clarity, the results are compared against the ground truth (g.t.) band dispersion -- double-crossing parabolas -- by overplotting in dashed lines. The tuning involves initializing the reconstruction with three sets of common curves: \textbf{a}-\textbf{c}, parallel straight lines, \textbf{d}-\textbf{f}, single-crossing straight lines, and \textbf{g}-\textbf{l}, double parabolas. The red check marks (\textcolor{red}{$\checkmark$}) label the reconstructions with correct crossings, while the blue check marks (\textcolor{blue}{$\checkmark$}) label those with anti-crossings. All numerical experiments used the same simulated data from a toy model with double-crossing parabolas containing only the second and zeroth-order terms. For reconstruction experiments, the nearest-neighbor Gaussian width hyperparameter ($\eta$) in the MRF model is tuned, while the relative position of the initial conditions is shifted to each configuration.}
    \label{fig:inittuning}
\end{figure}

\item[\textbf{2.}] \textbf{Reconstructing crossing bands}: We tested the simplest case of crossing bands with two parabolas of opposite directions of opening (Supplementary Fig. \ref{fig:synthband1}b), a recurring pattern in band structures. To recover the dispersion without band index scrambling, the knowledge of crossing needs to be included numerically in the initialization. This means, operationally, that the initialization requires crossing bands at nearby energy values, or that the reconstruction needs a warm-start optimization. For the double-crossing parabolas, the initializations that yield feasible outcomes are generated by slight tuning of the parabola parameters in the range that retains the crossing.
\begin{figure}[hbtp]
    \centering
    \includegraphics[width=\textwidth]{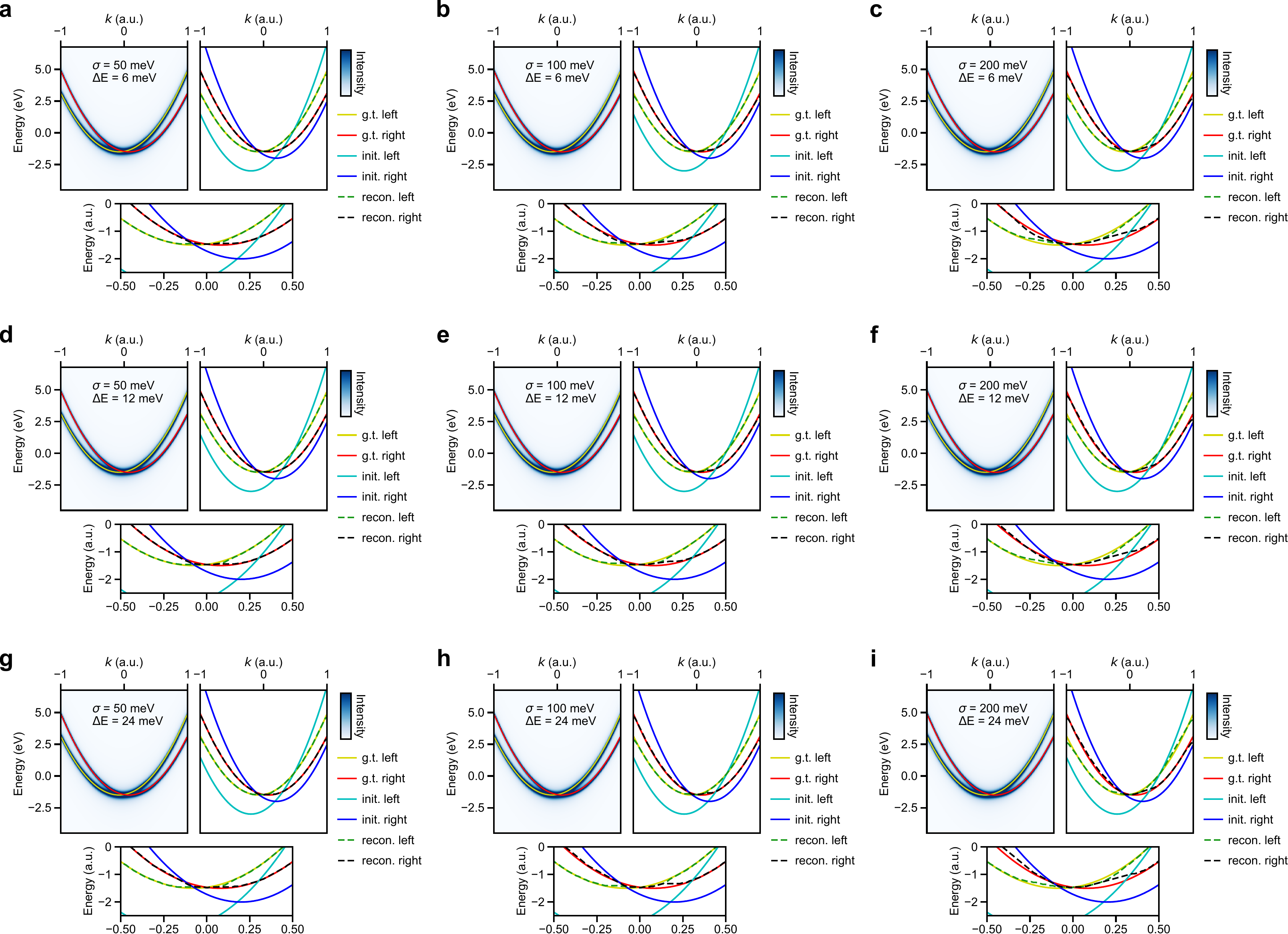} 
    \caption{\textbf{The effects of data resolution on reconstruction}. Results from a series of numerical experiments for demonstrating the effects of data resolution in either instrument resolution ($\sigma$) or energy spacing of data ($\Delta E$) on the reconstruction accuracy. The results are compared against the ground truth (g.t.) band dispersion -- displaced parabolas -- by overplotting in dashes lines. The $\sigma$ parameter is tuned to 50 meV, 100 meV and 200 meV, while the $\Delta E$ parameter to 6 meV, 12 meV and 24 meV. In \textbf{a}-\textbf{i}, the synthetic data with ground truth dispersion is shown on the left, the reconstruction outcome is displayed on the right, along with a zoomed-in view near the crossing placed at the bottom. Quantitative values of the reconstruction error are given in Supplementary Table \ref{tab:restuning}.}
    \label{fig:restuning}
\end{figure}
A careful examination over possible scenarios largely confirms this intuition: (1) Initialization with parallel straight lines (without any crossing) only results in non-crossing bands in the reconstruction (Supplementary Fig. \ref{fig:inittuning}a-c). When the initial straight line contains the crossings in the ground truth, a symmetry breaking in the reconstruction takes place (Supplementary Fig. \ref{fig:inittuning}c), depending on the data and the Gaussian width hyperparameter ($\eta$). (2) Initialization with two straight lines containing a single crossing yields a reconstruction with at most a single crossing (Supplementary Fig. \ref{fig:inittuning}d-f). (3) Initialization with double parabolas yields a reconstruction with at most the same number of crossings within the range of the data (Supplementary Fig. \ref{fig:inittuning}g-l). When the reconstruction is successful, the crossings in the initialization are close to the intersection between the two parabolas. Besides, double-crossing parabolas with other parameters from those in Supplementary Fig. \ref{fig:inittuning} are tested and similar outcomes are obtained.
\begin{table}[hbtp]
\renewcommand{\arraystretch}{1.2}
\centering
    \captionof{table}{{\bf Reconstruction error in resolution tuning experiments.} For each band, the reconstruction error is the root-mean-square error per momentum spacing (unit in meV) between reconstruction and the ground truth, according to Eq. \eqref{eq:avgapprox}. In each numerical experiment, the tabulated reconstruction error is averaged over the corresponding two parabolic bands shown in Supplementary Fig. \ref{fig:restuning}. The columns are the instrument resolution ($\sigma$) and the rows are the energy spacing ($\Delta E$) used to generate the intensity data.}
    \label{tab:restuning}
    \begin{tabular}{|c|c|c|c|}
        \hline
         & $\sigma$ = 50 meV & $\sigma$ = 100 meV & $\sigma$ = 200 meV  \\ \hline
        $\Delta E$ = 6 meV & 2.1 & 3.7 & 7.5 \\ \hline
        $\Delta E$ = 12 meV & 2.3 & 3.8 & 7.9 \\ \hline
        $\Delta E$ = 24 meV & 2.4 & 4.6 & 10.4 \\ \hline
    \end{tabular}
\end{table}

The crossing-band model is also an effective test case for resolution effects of the reconstruction algorithm. In this case, a momentum shift is introduced to two parabolic bands to produce the crossing, similar to the Rashba-split surface states of Au \cite{LaShell1996}, which if often used to calibrate experimental resolution in photoemission studies. We conduct a series of numerical experiments using different widths of the instrument resolution and energy spacing to simulate the resolution effect in the synthetic data, using reasonable parameter values. All numerical experiments use a nearest-neighbor Gaussian width hyperparameter $\eta$ within $\left[0.08, 0.11\right]$ for the reconstruction and no rigid energy shift is introduced. We tabulate the outcomes visually in Supplementary Fig. \ref{fig:restuning} as gridded figures and quantitatively in Supplementary Table \ref{tab:restuning} using the reconstruction error (root-mean-square error between ground truth and reconstruction) with the unit in meV. These results show that the reconstruction accuracy, as quantified by the error, has the same trend as the data resolution, which is determined by both the instrument resolution and energy sampling. The instrument resolution appears to have a larger effect on the reconstruction than the energy spacing. In other words, the worse the data resolution ($\sigma$ = 200 meV and $\Delta E$ = 24 meV being the worst case), the higher the reconstruction error. From visual inspection of the reconstruction in Supplementary Fig. \ref{fig:restuning}, including the zoomed-in region where the crossing is present, it appears that these changes in reconstruction accuracy create essentially no difference in the band dispersion away from the band crossing and only a marginal difference in the vicinity of the band crossing.
\begin{figure}[hbtp]
    \centering
    \includegraphics[width=\textwidth]{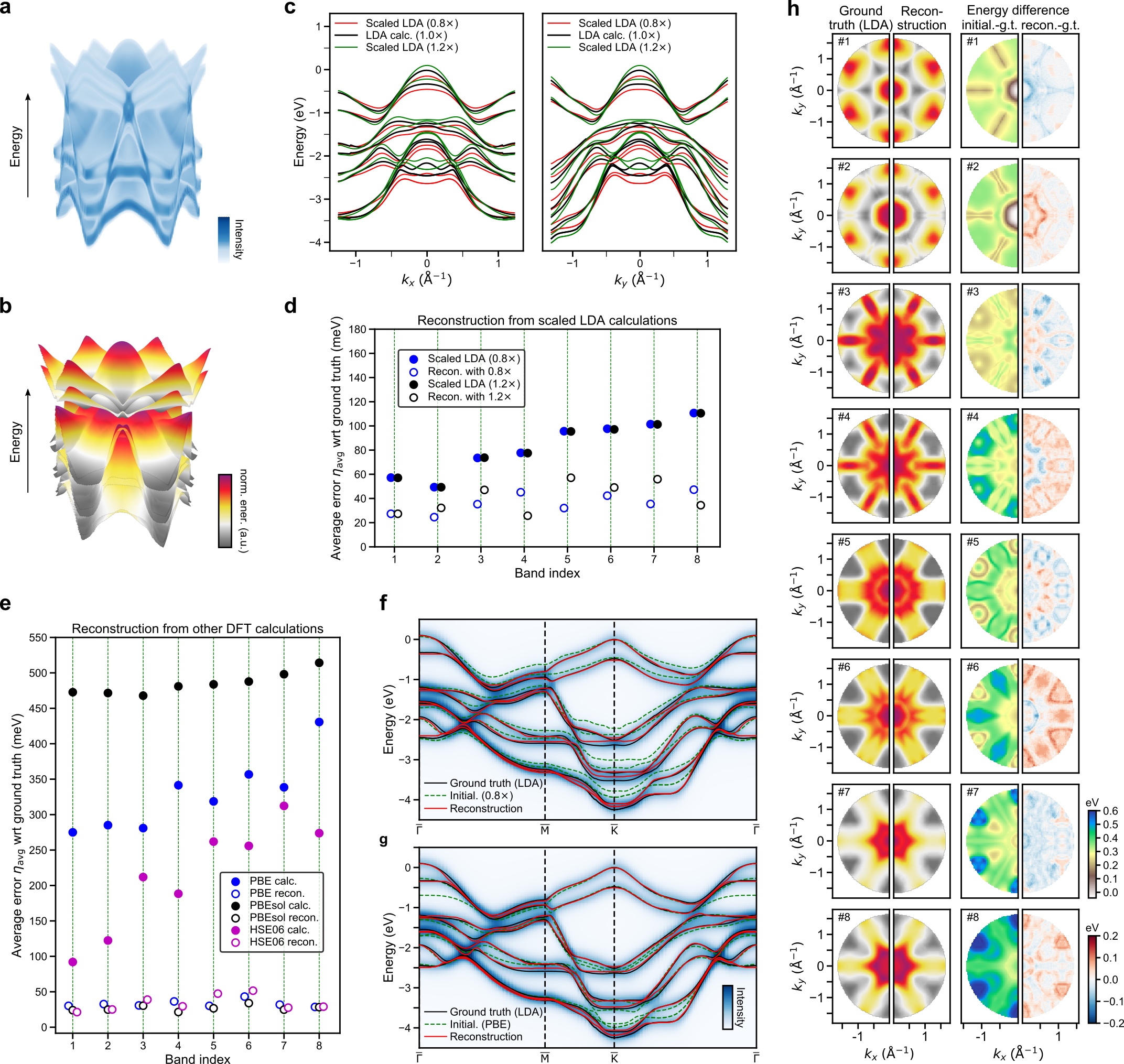} 
    \caption{\textbf{Validation on 3D synthetic multiband photoemission data}. \textbf{a}, Synthetic photoemission data with \textbf{b}, the underlying band structure obtained from LDA-level DFT calculation of WSe$_2$ (only the first 8 valence bands are used here). \textbf{c}, Comparison of two sets of differently scaled (by 0.8 and 1.2 times, respectively) initial conditions with respect to the ground-truth band structure calculation (LDA calc.), shown for a $k_x$-$E$ (left) and a $k_y$-$E$ (right) slice. \textbf{d},\textbf{e}, Comparison of the average error $\eta_{\mathrm{avg}}$ for energy bands used as initializations (solid dots) and reconstructions (hollow dots). The initializations are constructed by scaling the ground-truth band energies (\textbf{d}) or by using other DFT calculations (\textbf{e}). The reconstructions all have reduced $\eta_{\mathrm{avg}}$ compared with the initialization and $\eta_{\mathrm{avg}}$ is consistent across all energy bands. \textbf{f},\textbf{g}, Reconstruction, ground truth (LDA), and initialization overlaid on the synthetic data along high-symmetry lines of the hexagonal Brillouin zone, corresponding to two of the cases in \textbf{d} and \textbf{e}, respectively. The energy zeros of the initialization in \textbf{d}-\textbf{e} are aligned with the ground truth via a global shift. \textbf{h}, Comparisons of ground truth (LDA), reconstructed bands, and the differences between initialization (PBE), reconstruction and ground truth (g.t.) for each band.}
    \label{fig:synthband2}
\end{figure}

\item[\textbf{3.}] \textbf{Sensitivity of reconstruction to scaled energies as initialization}: We scaled the energies of the LDA-DFT band structure of WSe$_2$ (using the first 8 valence bands) around the mean energy of each band (see Supplementary Fig. \ref{fig:synthband2}c) for use as the initialization. The accuracy of the reconstruction outcome is evaluated by its average error $\eta_{\mathrm{avg}}$ (Eq. (8) in the main text Methods), calculated with respect to the ground-truth band energies. The results displayed in Supplementary Fig. \ref{fig:synthband2}d,f show that the average error and its spread in the reconstruction are reduced from the corresponding values in the initialization. Quantitatively, in the reconstruction, $\eta_{\mathrm{avg}}$ is within the range 20-65 meV, while in the initialization, $\eta_{\mathrm{avg}}$ varies within 45-100 meV for all 8 valence bands.

\item[\textbf{4.}] \textbf{Sensitivity of reconstruction to differently calculated band structures as initialization}: We used DFT band structure calculations of WSe$_2$ with PBE, PBEsol and HSE06 exchange-correlation functionals (see Supplementary Section 4) to initialize the reconstruction. The accuracy of the reconstruction is quantified similarly as in the previous numerical experiment using $\eta_{\mathrm{avg}}$. The results displayed in Supplementary Fig. \ref{fig:synthband2}e,g,h show that, despite the huge spread in the average error for the different levels of DFT calculations (used as initialization without global shift alignment of energy zero), the corresponding reconstructions all have average errors at around or below 40 meV for every band. The value of $\eta_{\mathrm{avg}}$ varies by up to $\sim$ 30 meV (i.e. between band \#1 and \#6) in each set of reconstructed bands, much lower than those in the initialization. The former can be improved by casting the experimental data into finer bins in the preprocessing stage (single-electron events can be binned into various sizes) or interpolating between existing bins, while the latter can be improved by using a continuous probabilistic model \cite{Bishop2006} to formulate the reconstruction problem, albeit at the cost of much increased computational demand.

The results of the above numerical experiments demonstrate that the reconstruction by MAP optimization converges to a consistent range in the tested scenarios and initializations. It should be noted here that the fundamental accuracy in reconstruction reported here is still limited by the coordinate spacings of the data along all dimensions and the discrete nature of the MRF model (the output is centered only at the bin locations).

\subsection{Computational benchmarks}
We used the available synthetic photoemission datasets based on the computed band structure of WSe$_2$ to construct benchmarks. The synthesis made use of the approach described in Supplementary Section 2.1. The two datasets used here, taken from \cite{pesarxiv}, exhibit different characteristics, which may be qualitatively described using the energy range of an energy band. The more overlap in the energy range between two energy bands, the more likely they have crossings (or anti-crossings). The dataset-specific information is as follows:
\begin{itemize}[wide, labelwidth=!, labelindent=0pt]
\item The synthetic dataset of the WSe$_2$ K-point shows close proximity in energies between neighboring momentum locations. The energy ranges of all energy bands have no or up to a moderate degree of overlap. The dataset size is 30 $\times$ 30 $\times$ 500 and contains 900 photoemission spectra.
\item The synthetic high-symmetry line dataset of WSe$_2$ exhibits large dispersion. Since the high-symmetry line often represents the direction with the most dispersion in the band structure, the energy ranges of all energy bands are strongly overlapping. The dataset size is 186 $\times$ 500 and contains 186 photoemission spectra.
\end{itemize}
In both cases, the ground-truth band dispersions are taken from the LDA-DFT calculation, including all 14 valence bands, while the initializations for benchmarking both band reconstruction approaches are the PBE-DFT calculation (partial example see Supplementary Fig. \ref{fig:synthband2}g). Using these two datasets, we compare the reconstruction algorithms based on pointwise fitting (using the code in \cite{Xian2021}) and MRF as introduced in this work. The hyperparameters for the pointwise fitting involve only the band-wise relative shifts applied in each band initialization (14 hyperparameters in total for 14 bands), which were tuned for each band sequentially from band \#1 to band \#14 using an expanding window approach (introduced in \cite{Xian2021}). For the MRF reconstruction, the hyperparameters (including the band-wise shift and the width of the nearest-neighbor Gaussian prior, 28 hyperparameters in total for 14 bands) were tuned individually while reconstructing each band. The hyperparameter tuning made use of grid search through a range of preset values, using the root-mean-squared (RMS) error for determining the final choice. All benchmarks were run on an on-premises computing server (Dell PowerEdge R840), equipped with four Intel Xeon Gold 6150 multicore CPUs.
\begin{table}[hbt!]
\renewcommand{\arraystretch}{1.2}
\setlength\tabcolsep{3.2pt}
\centering
\begin{threeparttable}
    \captionof{table}{{\bf Algorithm comparison using benchmark datasets.} Two synthetic datasets with different number of spectra ($N_{\mathrm{spec}}$) and the range of band indices ($R_{\mathrm{band}}$) are used for benchmarking the algorithm performance. The per-band, per-spectrum reconstruction error ($\eta_{\mathrm{bk}}$) is calculated using Eq. \eqref{eq:avgbanderr}. The instability ($r_{\mathrm{band}}$) quantifies the variation of the fitting residuals among all spectra within a dataset using the standard deviation of residuals, as in Eq. \eqref{eq:recon_instability}. The single-run time ($t_{\mathrm{mono}}$) is the averaged elapsed time in a single execution of fitting, while the tuning time ($t_{\mathrm{tune}}$) is the total time used for tuning the parameters to reach the final outcome. Both methods use DFT calculation as the initialization for the band positions.}
    \label{tab:benchmark}
    \begin{tabular}{|c|c|c|c|c|c|c|c|c|c|c|}
        \hline
        \multirow{3}{*}{Dataset \tnote{1}} & \multirow{3}{*}{$N_{\mathrm{spec}}$} & \multirow{3}{*}{$R_{\mathrm{band}}$} & \multicolumn{4}{c|}{\begin{tabular}{c}Pointwise line fitting \tnote{2} \end{tabular}} & \multicolumn{4}{c|}{\begin{tabular}{c}MRF reconstruction \end{tabular}} \\ \cline{4-11}
        
        & & & \multirow{2}{*}{\begin{tabular}{c}$t_{\mathrm{mono}}$\\ (s)\end{tabular}} & \multirow{2}{*}{\begin{tabular}{c}$t_{\mathrm{tune}}$\\ (s)\end{tabular}} & \multirow{2}{*}{\begin{tabular}{c}$\eta_{\mathrm{bk}}$\\(eV) \end{tabular}} & \multirow{2}{*}{\begin{tabular}{c}$r_{\mathrm{band}}$\\(eV) \end{tabular}} & \multirow{2}{*}{\begin{tabular}{c}$t_{\mathrm{mono}}$\\ (s)\end{tabular}} & \multirow{2}{*}{\begin{tabular}{c}$t_{\mathrm{tune}}$\\ (s)\end{tabular}} & \multirow{2}{*}{\begin{tabular}{c}$\eta_{\mathrm{bk}}$\\ (eV) \end{tabular}} & \multirow{2}{*}{\begin{tabular}{c}$r_{\mathrm{band}}$\\ (eV) \end{tabular}} \\
        
        & & & & & & & & & & \\ \hline
        
        \multirow{4}{*}{\begin{tabular}{c}WSe$_2$\\ K point \end{tabular}} & \multirow{4}{*}{900} & 1-2 & 42 & 421 & {\bf 4.6e-4} & {\bf 3.2e-4} & {\bf 6.6e-1} & {\bf 34} & 5.8e-3 & 4.1e-3 \\ \cline{3-11}
        
        & & 3-4 & 168 & 2519 & 3.6e-1 & 1.4e-1 & {\bf 1.0} & {\bf 96} & {\bf 1.1e-2} & {\bf 5.5e-3} \\ \cline{3-11}
        & & 5-8 & 412 & 11964 & 1.5e-1 & 5.3e-2 & {\bf 2.2} & {\bf 134} & {\bf 6.0e-2} & {\bf 2.1e-2} \\ \cline{3-11}
        & & 9-14 & 2792 & 78181 & 3.6e-1 & 9.6e-2 & {\bf 3.8} & {\bf 236} & {\bf 7.8e-2} & {\bf 2.1e-2} \\ \hline
        
        \multirow{4}{*}{\begin{tabular}{c} WSe$_2$ high-\\symmetry\\ line\end{tabular}} & \multirow{4}{*}{186} & 1-2 & 13 & 191 & 3.6e-1 & 2.1e-1 & {\bf 3.9e-1} & {\bf 32} & {\bf 1.1e-2} & {\bf 2.0e-2} \\ \cline{3-11}
        
        & & 3-4 & 46 & 692 & 6.2e-1 & 2.9e-1 & {\bf 2.9e-1} & {\bf 31} & {\bf 1.9e-2} & {\bf 1.8e-2} \\ \cline{3-11}
        & & 5-8 & 385 & 8858 & 5.5e-1 & 1.7e-1 & {\bf 8.6e-1} & {\bf 56} & {\bf 3.0e-2} & {\bf 1.4e-2} \\ \cline{3-11}
        & & 9-14 & 872 & 27889 & 3.3 & 8.7e-1 & {\bf 1.6} & {\bf 109} & {\bf 4.1e-2} & {\bf 1.1e-2} \\ \hline
    
    \end{tabular}
    \vspace{1ex}
    \begin{tablenotes}\footnotesize
    \item[1] Datasets are obtained from \cite{pesarxiv}.
    \item[2] Executed using the software described in \cite{Xian2021}.
    \end{tablenotes}
\end{threeparttable}
\end{table}

The computational performance of the two algorithms was evaluated using four different metrics as summarized in Supplementary Table \ref{tab:benchmark}. The timing metrics provided in the table include the average single-run computing time in each dataset as well as the total computing time of the hyperparameter tuning, which covers all grid search steps of the energy bands (indices described in $R_{\mathrm{band}}$) in every benchmarking stage. The computing time for the single runs of each dataset shows a clear advantage of the machine learning-based algorithm and the gap between the two algorithms only widens as the number of bands increases. The accuracy of the reconstruction is quantified by an RMS error averaged over all reconstructed bands and spectra, following the expression for ``band delta'' in \cite{Huhn2017}.
\begin{equation}
    \eta_{\mathrm{bk}}(E_{\mathrm{gt}}, E_{\mathrm{recon}}) = \sqrt{\frac{1}{N_b N_{\mathrm{spec}}}\sum_{i=1}^{N_b}\sum_{\textbf{k}}(E_{\mathrm{gt}, i, \textbf{k}} - E_{\mathrm{recon}, i, \textbf{k}})^2},
    \label{eq:avgbanderr}
\end{equation}
where $N_b$ is the number of bands and the subscript $i$ is the band index. The instability is quantified by the standard deviation of the residual (difference between the ground truth and reconstructed energy dispersion), $\delta E = E_{\mathrm{gt}} - E_{\mathrm{recon}}$.
\begin{equation}
    r_{\mathrm{band}}(E_{\mathrm{gt}}, E_{\mathrm{recon}}) = \sqrt{\frac{1}{N_b}\sum_{i=1}^{N_b}\sum_{\textbf{k}}(\overline{\delta E^2_{i, \textbf{k}}} - \overline{\delta E_{i, \textbf{k}}}^2)},
    \label{eq:recon_instability}
\end{equation}
where the overline indicates the mean. This metric has been used in earth sciences to quantify surface roughness \cite{Smith2014,Grohmann2011}, which may be interpreted similarly in our case. The difference is that the roughness in the reconstructed surface is largely due to the instability of the optimization, besides the quality of the data, because band dispersions are generally smooth and continuous. In the main text Fig. \ref{fig:comp_metrics}, these tabulated metrics are normalized by the number of spectra to allow comparison between datasets, as is also adopted in \cite{Xian2021}. We interpret the results in Supplementary Table \ref{tab:benchmark} in the following two aspects:
\begin{itemize}[wide, labelwidth=!, labelindent=0pt]
\item Computing time ($t_{\mathrm{mono}}$ and $t_{\mathrm{tune}}$): For the same dataset, the single-run computing time of the MRF reconstruction is about 2-3 orders of magnitude faster than distributed pointwise fitting. Even with the hyperparameter tuning included, the MRF reconstruction still runs 1-2 orders of magnitude faster, although the MRF reconstruction requires tuning one more hyperparameter than the pointwise fitting approach for each band. 
\item Reconstruction quality ($\eta_{\mathrm{band}}$ and $r_{\mathrm{band}}$): The substantially higher reconstruction error and instability for pointwise fitting are due to the lack of connectivity between neighbors come primarily from the (theoretical) initialization. Because for each band, even though a global energy shift hyperparameter is tuned, it cannot guarantee that everywhere locally the shift is optimal for band reconstruction, resulting in scrambled band indices in the local patches that the fitting fails. This scenario is a failure mode of the pointwise fitting-based reconstruction as illustrated in \cite{Xian2021} for real-world experimental data. This limitation of the pointwise fitting approach is less pronounced when the energy range overlap between bands is small, yet becomes more severe in the high-symmetry line dataset, where the strong energy range overlap and the multiple band crossing (or anti-crossing) make the reconstruction harder to resolve by tuning a single energy shift parameter. The probabilistic framework of the MRF approach largely circumvents this limitation using a physical prior that accounts for the proximity of the neighboring energy values and achieves high stability in the reconstruction.
\end{itemize}
\end{enumerate}

\section{Reconstruction for other datasets}
To test the functionality of our MRF reconstruction algorithm on other materials, we have acquired photoemission band mapping datasets from gold (Au), a metal, and bismuth tellurium selenide (Bi$_2$Te$_2$Se), a topological insulator. Due to the complexity of the electronic structure of these materials, we focus here on reconstructing a subset of the energy bands of these two materials that are pronounced within the measured energy range. Besides, we simulated the case where the electron self-energy strongly modifies the band dispersion that results in kink anomalies \cite{Giustino2008,Verdi2017,Zhang2022}.

\subsection{Near-gap electronic bands of a topological insulator (Bi$_2$Te$_2$Se)}
The dataset for Bi$_2$Te$_2$Se was measured at room temperature at the Fritz Haber Institute in Berlin using a momentum microscope (SPECS METIS 1000). The sample growth method was previously described in \cite{Mi2013}. A clean surface was prepared in vacuum by \textit{in situ} cleaving with Scotch tape. During the measurement, light excitation of 800 nm was used to examine ultrafast dynamics. The temporal features were ignored here and averaged to improve the signal-to-noise ratio of the data. The photoemission spectra of Bi$_2$Te$_2$Se near the Fermi level feature a topologically-protected surface state (SS) that intersects at the Dirac point (DP) \cite{Heremans2017} as shown in main text Fig. \ref{fig:othermat}a-b. The SS bridges the valence and conduction bands, an identifiable and prominent feature for this class of materials directly measurable via photoemission \cite{Heremans2017,Michiardi2015}.

Preprocessing of the 3D band mapping data follows the procedure for WSe$_2$ data described in the main text, except that the rotational symmetrization is only threefold, due to the symmetry of the material. We used numerical initializations from simple functions such as paraboloid and Gaussian in 2D, instead of any first-principles calculation.
The reconstructed energy dispersions were smoothed using Chambolle's total variation denoising algorithm \cite{Chambolle2004} implemented in \textsf{scikit-image} \cite{VanderWalt2014}, removing the high-frequency noise as a result of the Poisson statistics of the photoemission data. As shown in main text Fig. \ref{fig:othermat}c, the simple initializations we chose are sufficient to reconstruct the complex dispersion from the first two valence bands, the SS and parts of the first conduction band occupied by the excited electronic population. The appearance of the first conduction band for Bi$_2$Te$_2$Se is a result of photoexcitation \cite{Papalazarou2018}. The reconstructed bands show sixfold symmetry and warping in agreement with previous theoretical investigations \cite{Heremans2017,Michiardi2015}, which is more straightforwardly visualized in 2D and 3D as in main text Fig. \ref{fig:othermat}d-e. For the dispersion surfaces of the SS and the conduction band, we truncated the dispersion to a realistic energy range not far from the photon energies of the excitation light pulses.

\subsection{Bulk electronic bands of gold (Au)}
The Au dataset was measured at 100 K at the SGM-3 beamline \cite{Hoffmann2004} of the 3rd-generation synchrotron radiation facility ASTRID2 in Aarhus, Denmark. The Au samples were purchased from MaTecK GmbH with a (111) surface. The sample preparation procedure has been previously described \cite{Dendzik2016}. The photoemission data were measured along the high-symmetry direction ($\mathrm{\Gamma}$KM$\mathrm{\Gamma}$) of Au(111), which exhibits a hexagonal symmetry in the surface Brillouin zone \cite{Tusche2015} (indicated with an overbar over each symmetry label) similar to WSe$_2$. As shown in main text Fig. \ref{fig:othermat}f, the collection of energy bands present in the photoemission data for Au(111) includes the surface states (SSs), which, at sufficient momentum and energy resolution, are composed of momentum-shifted parabolas \cite{LaShell1996}. The photon energy for the photoemission measurement is $\sim$ 80 eV, which resolves the $sp$ bands and the surface states poorly but the bulk $d$ bands better. The $sp$ bands and the $d$ bands are the low-energy bulk electronic bands of Au.

Before reconstruction, the Au data has been preprocessed using contrast enhancement and intensity smoothing as described in the main text for the WSe$_2$ data before reconstruction. The reconstruction used existing DFT calculations, which feature a Au(111) slab containing five Au layers constructed according to \cite{Lin2014}, as initialization to retrieve parts of the $d$ bands that are resolvable within the current dataset. The comparison between initialization and reconstruction is shown in main text Fig. \ref{fig:othermat}g. The choice of the initialization is a consistent set of energy bands (i.e. produced by the same slab) from DFT calculations of Au(111) slabs in the energy range close to the noticeable bands in the photoemission data. Although traditionally, slab calculations along with overplotting are used to approximate the total band width, we have shown that our reconstruction approach can detect existing band-like dispersive features in these highly congested data.
\begin{figure}[hbtp!]
    \centering
    \includegraphics[width=0.9\textwidth]{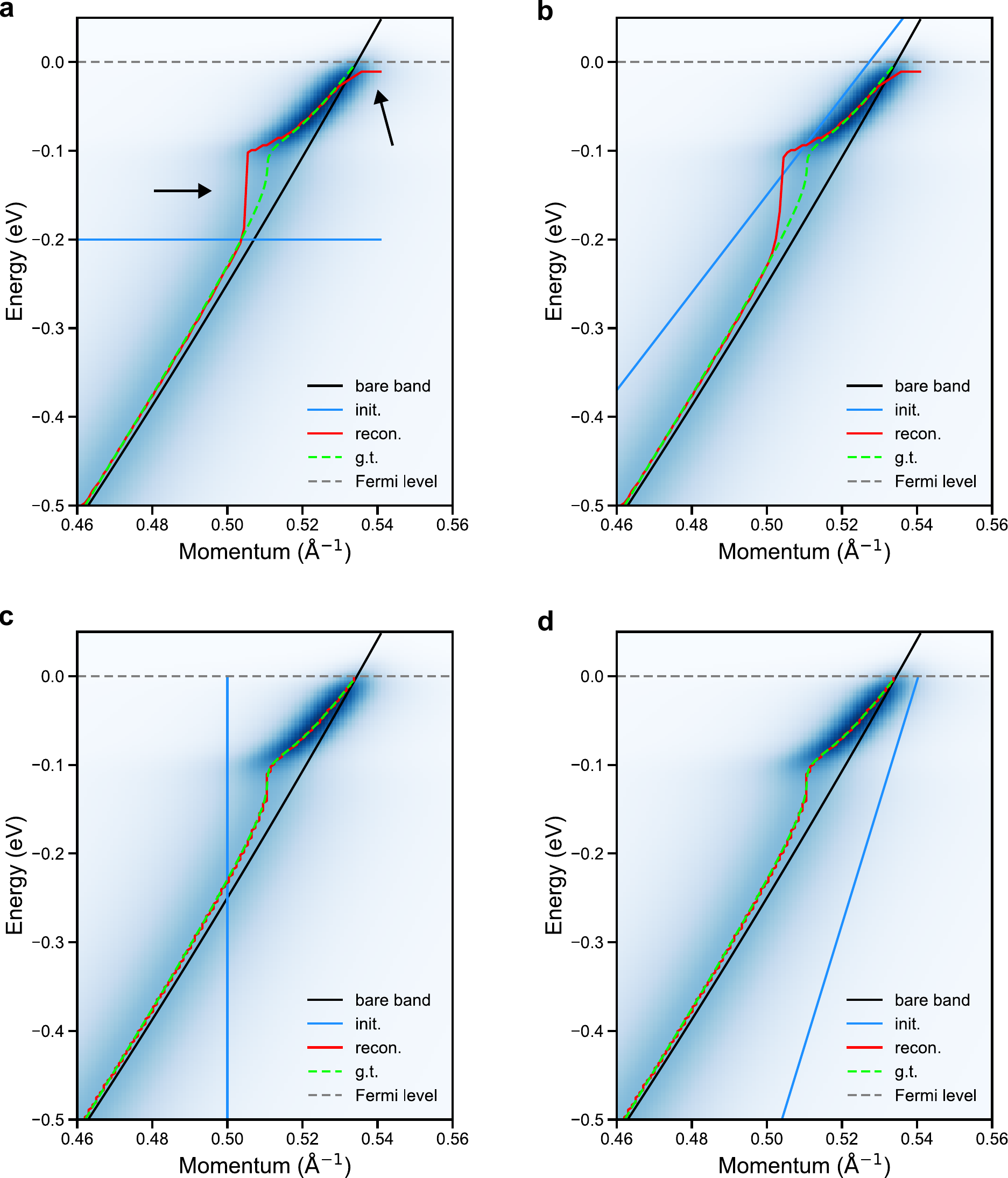} 
    \caption{\textbf{Band reconstruction involving a kink anomaly}. Band reconstruction was carried out along either the momentum (\textbf{a}, \textbf{b}) or energy (\textbf{c}, \textbf{d}) directions. Reconstruction (recon.) along the momentum direction using \textbf{a}, a flat (uninformative) initialization (init.) and \textbf{b}, an informative initialization that approximates foreknowledge of the linear bare-band dispersion yield mostly identical outcomes, which have deviations from the ground truth (g.t.) near the kink and the Fermi level, as indicated with black arrows in \textbf{a}. Reconstruction along the energy direction using \textbf{c}, a flat (uninformative) initialization and \textbf{d}, an informative initialization both yield highly accurate outcomes compared with the ground-truth quasiparticle dispersion in dashed green lines in \textbf{a}-\textbf{d}.}
    \label{fig:kinked}
\end{figure}

\subsection{Reconstructing the kink anomaly}
Kink anomalies are a kind of feature for electron-phonon interaction in photoemission signals \cite{Damascelli2003,Zhang2022} found in various materials \cite{Giustino2008,Ortenzi2009,Verdi2017}. To test the reconstruction performance, we simulated the photoemission signal for a kink anomaly using the full spectral function introduced in Eq. \eqref{eq:spsf}. The real part of the electron self-energy is calculated using the Eliashberg function \cite{Shi2004} represented as an Einstein mode (i.e. single-phonon mode with a delta-function-like frequency response) \cite{Allen1975}, which appears near the Fermi level. Further details of the computational model can be found in the Jupyter notebooks within the associated compute capsule. The presence of an Einstein mode in the spectral function results in a phonon-induced kink at around - 0.1 eV.

The outcome of the reconstruction, shown in Supplementary Fig. \ref{fig:kinked}, indicates that the MRF model can recover faithfully the quasiparticle dispersion including the shape of the kink anomaly. The reconstruction can simply be initialized with a flat line, which produces identical results from initialization with a linear dispersion that could represent prior knowledge of the algorithm user. The results show that for strongly dispersive energy bands with almost vertical dispersion along the energy direction, reconstruction along the energy direction (i.e. treating the data as a collection of momentum distribution functions) yields a better outcome. This is because the existence of the kink violates the one-to-one mapping between the band energy, $E(k)$, and the photoelectron momentum, $k$ (see Supplementary Fig. \ref{fig:kinked}a-b). In these cases, the reconstruction is still viable using the momentum distribution function as the likelihood in the MRF model, which effectively amounts to transposing the image and swapping the momentum and energy coordinates, while the same optimization algorithm described in this work for the EDC-based approach can be reused to obtain the correct quasiparticle dispersion (see Supplementary Fig. \ref{fig:kinked}c-d).

\section{Band structure calculations}
\subsection{DFT calculations}
The crystal structure of bulk WSe$_2$ with 2$H$ stacking (2$H$-WSe$_2$) belongs to the P6$_3$/mmc space group and consists of two Se-W-Se triatomic layers as shown in Supplementary Fig.~\ref{fig_ballstick}. The stacking order 
of the two hexagonal layers is -$BAB$-$ABA$- and the long $c$-axis is oriented perpendicular to the layers. Electronic structure calculations were performed within DFT using the local density approximation (LDA), the generalized-gradient approximation (GGA-PBE and GGA-PBEsol), and hybrid (HSE06) exchange-correlation functionals as implemented in FHI-aims \cite{Blum2009}. The atomic orbitals basis sets, the integration grids and the Hartree potential employed for all calculations are according to the default ``tight'' numerical settings of FHI-aims. A 16$\times$16$\times$4 uniform {\bf k}-gird was used to sample the Brillouin zone. The Broyden-Fletcher-Goldfarb-Shanno optimization algorithm was used to relax the atomic positions until the residual force component per atom was less than 10$^{-2}$ eV/\AA. Supplementary Table~\ref{Table_1} shows the optimized lattice constants, $a$ and $c$, as obtained by the evaluation of the analytical stress tensor \cite{KNUTH201533} using different exchange-correlation functionals. In all BS calculations, we included the effect of spin-orbit coupling, which is known to introduce a large splitting of the outermost valence states of bulk 2$H$-WSe$_2$~\cite{Voss_1999}.
\begin{figure}[hbtp]
    \centering
    \includegraphics[width=0.75\textwidth]{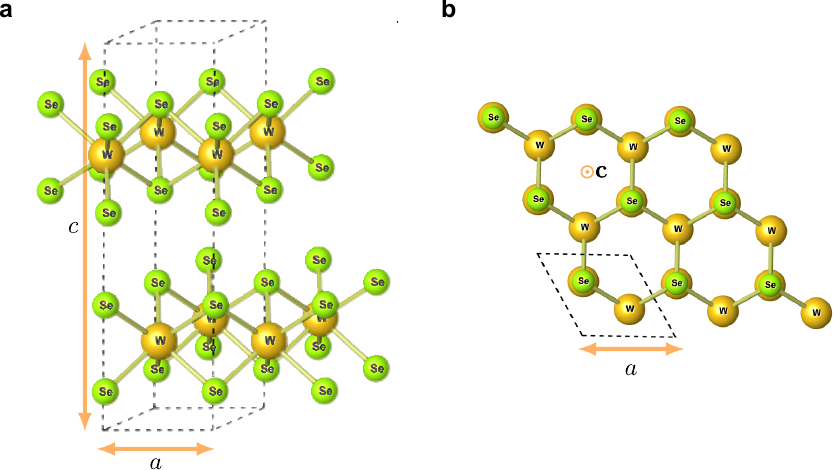}
    \hspace*{1cm}
    \caption{\textbf{Crystal structure of bulk 2$H$-WSe$_2$.} \textbf{a}, Side view and \textbf{b}, top view of the crystal structure of 2$H$-WSe$_2$. The space group of the hexagonal structure is P6$_3$/mmc with the $c$-axis oriented perpendicular to the stacking layers. In each case, the real-space unit cell is labelled by dashed black lines.
    \label{fig_ballstick}}
\end{figure}
\begin{figure}[hbtp!]
    \centering
    \includegraphics[width=0.9\textwidth]{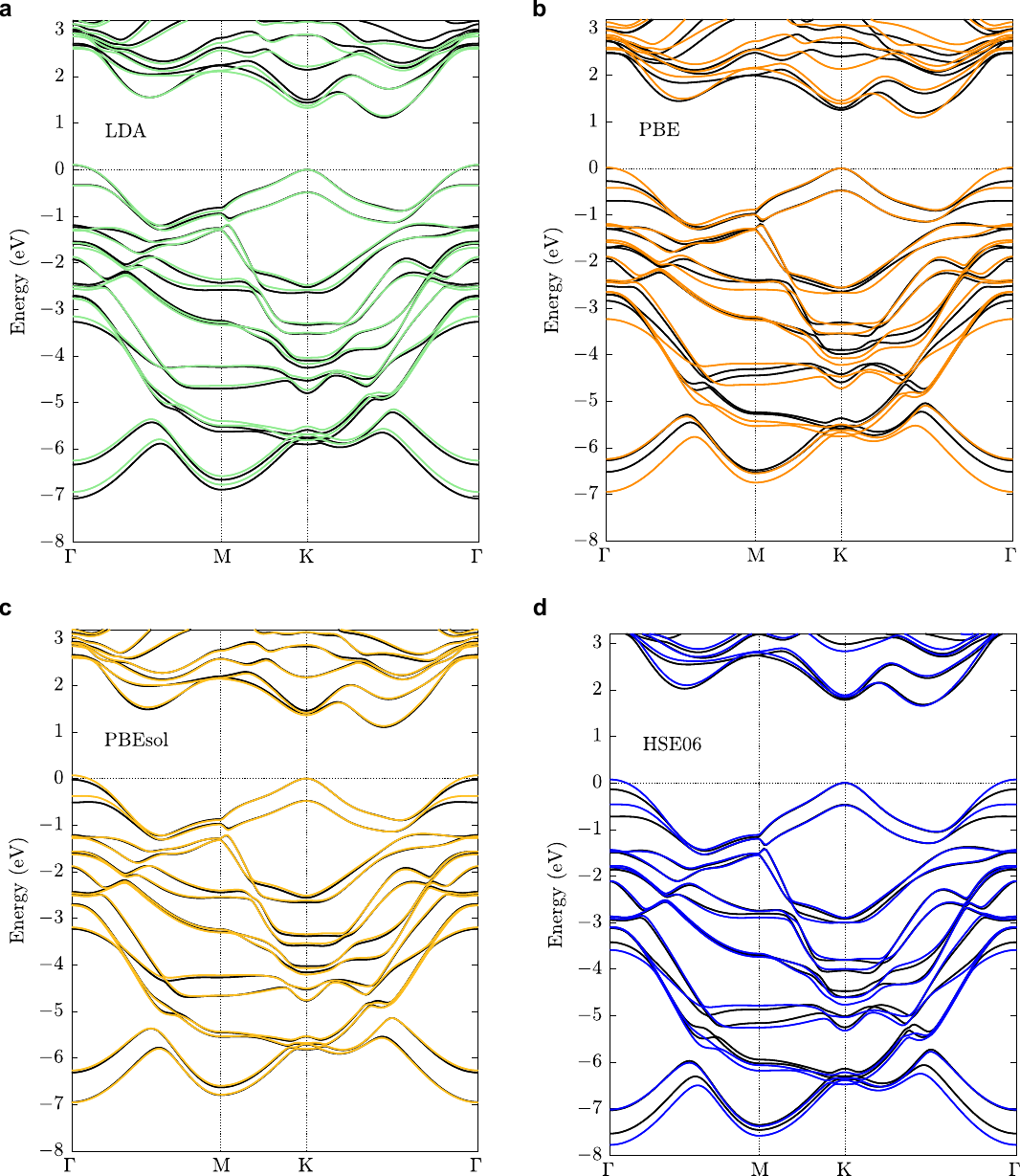}
    \caption{\textbf{Bulk electronic band structure of 2$H$-WSe$_2$}. \textbf{a}-\textbf{d}, Band structure of bulk 2$H$-WSe$_2$ along the $\Gamma$-K-M-$\Gamma$ momentum path of its Brillouin zone including the effect of spin-orbit coupling. Calculations were performed using the LDA (green, \textbf{a}), PBE (orange, \textbf{b}), PBEsol (yellow, \textbf{c}), and HSE06 (blue, \textbf{d}) exchange-correlation functionals and optimized structures (see Supplementary Table \ref{Table_1}) with the unit cell dimensions kept fixed at the experimental lattice constants. Black lines in \textbf{a}-\textbf{d} represent the corresponding calculations using fully optimized geometries. For comparison, the two band structures in each plot are rigidly shifted to align their uppermost valence state at the K high-symmetry point, where we also define as the energy zero. All band structure calculations used $k_z$ = 0.35 $\mathrm{\AA}^{-1}$.}
    \label{fig_band_strs}
\end{figure}
\begin{table}[hbtp!]
\renewcommand{\arraystretch}{1.2}
\centering
\begin{threeparttable}
    \captionof{table}{{\bf Parameters from density functional theory calculations.} Optimized lattice constants, spin-orbit splitting of the topmost valence states at the K high-symmetry point, and the band gap of bulk 2$H$-WSe$_2$ calculated within density functional theory using the LDA, PBE, PBEsol and HSE06 exchange-correlation functionals. For comparison, we also report the corresponding experimental values at room temperature.}\label{Table_1}
    \begin{tabular}{|c|c|c|c|c|c|}
        \hline
        xc-functional & LDA    &   PBE  & PBEsol  & HSE06 & Experiment \\ \hline
        $a$ ({\AA})  & 3.250  &  3.317 &  3.269  &   3.295    &   3.28 \\ \hline
        $c$ ({\AA})  & 12.827 &  14.921 & 13.211  &    13.863   &  12.98 \\ \hline
        Spin-orbit splitting at K (eV)  & 
        \begin{tabular}{c} 0.485  \tnote{1} \\ 0.490  \tnote{2} \end{tabular}  & 
        \begin{tabular}{c} 0.473  \tnote{1} \\ 0.481 \tnote{2} \end{tabular}   &
        \begin{tabular}{c} 0.476  \tnote{1} \\ 0.484 \tnote{2} \end{tabular} & 
        \begin{tabular}{c} 0.467  \tnote{1} \\ 0.480 \tnote{2} \end{tabular} & 
        0.5 \tnote{3} \\ \hline
        Band gap (eV) & 
    \begin{tabular}{c} 1.022 \tnote{1} \\ 1.052 \tnote{2} \end{tabular} & 
        \begin{tabular}{c} 1.186 \tnote{1}   \\ 1.074 \tnote{2}  \end{tabular} & 
        \begin{tabular}{c} 1.105 \tnote{1}  \\ 1.060 \tnote{2}  \end{tabular} &  
        \begin{tabular}{c} 1.679 \tnote{1}  \\ 1.582 \tnote{2}  \end{tabular} & 
         1.219 \tnote{4} \\ \hline
    \end{tabular}
    \vspace{1ex}
    \begin{tablenotes}\footnotesize
    \item[1] Fully optimized structure.
    \item[2] Optimized structure by fixing the lattice parameters to experimental values.
    \item[3] Ref. \cite{Riley2014}.
    \item[4] Ref. \cite{Kam_1984}.
    \end{tablenotes}
\end{threeparttable}
\end{table}

The calculated BSs of bulk 2$H$-WSe$_2$ using different levels of approximation for the exchange-correlation (XC) functional are shown in Supplementary Fig.~\ref{fig_band_strs}. For each XC functional, the calculations were performed on (1) fully optimized structures (black lines), and on (2) optimized structures by fixing the lattice parameters of the unit cell to the experimental values (colored lines). All calculations using different XC functionals reveal an indirect band gap with the conduction band minimum located along the $\Gamma$-K path ($\Gamma$ and K being the bulk equivalents of the $\overline{\Gamma}$ and $\overline{\mathrm{K}}$ high-symmetry points). For both sets of optimized structures, the LDA results reveal a valence band maximum at the $\Gamma$ point, compatible with experimental measurements, while the PBE, PBEsol, and HSE06 band structures obtained for fully optimized structures exhibit a valence band maximum at the K point. Nevertheless, fixing the unit cell dimensions at the experimental lattice constants reproduces the experimental behavior that the valence band maximum resides at the $\Gamma$ point. The difference between the two sets of calculations obtained using PBE, PBEsol, and HSE06 functionals is attributed to the overestimation of the lattice parameter $c$ and the residual strain along the $c$-axis \cite{Desai_2014}.  
The calculated indirect band gaps and the spin-orbit splitting of the two topmost valence states at the K point using both sets of optimized structures are shown in Supplementary Table~\ref{Table_1}.
\begin{figure}[htbp!]
    \centering
    \includegraphics[width=\textwidth]{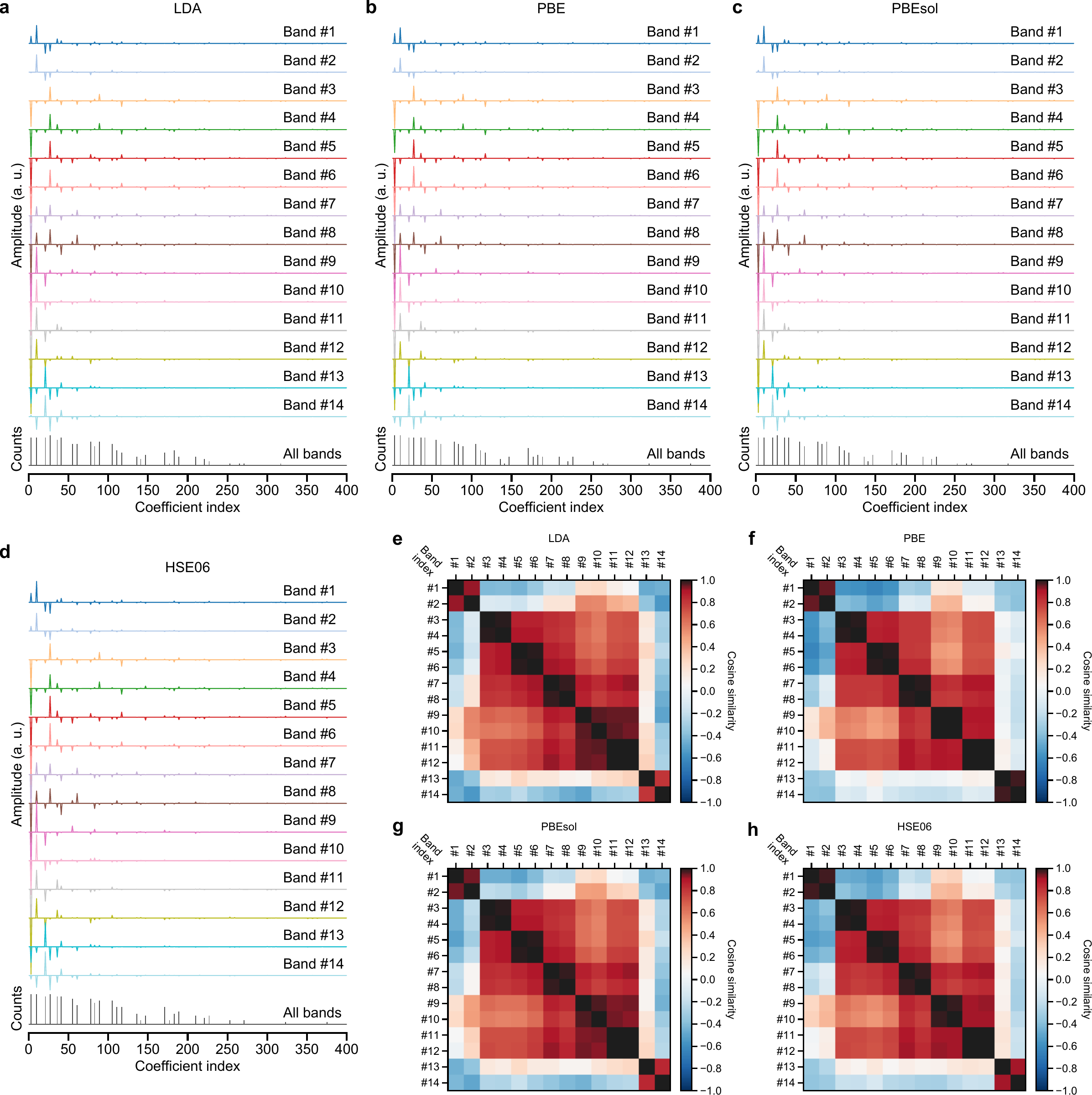}
    \caption{\textbf{Geometric featurization of the energy bands of WSe$_2$}. \textbf{a-d}, Decomposition of the 14 valence energy bands of WSe$_2$ into hexagonal Zernike polynomials for the DFT band structure calculations carried out at the levels of LDA (\textbf{a}), PBE (\textbf{b}), PBEsol (\textbf{c}), and HSE06 (\textbf{d}), respectively. Similar characteristics are seen compared with the reconstructed band structure shown in Fig. 3a in the main text, including the sparse distribution of major basis terms and the decreasing dependence on higher-order basis terms. \textbf{e-h}, Cosine similarity matrices between the 14 energy bands of WSe$_2$ for the DFT band structure calculations carried out at the levels of LDA (\textbf{e}), PBE (\textbf{f}), PBEsol (\textbf{g}), and HSE06 (\textbf{h}), respectively. The characteristics of these matrices resemble that calculated for the reconstructed band structure as shown in Fig. 3c in the main text.}
    \label{fig:bandhzps}
\end{figure}
\begin{figure}[htbp!]
    \centering
    \includegraphics[width=\textwidth]{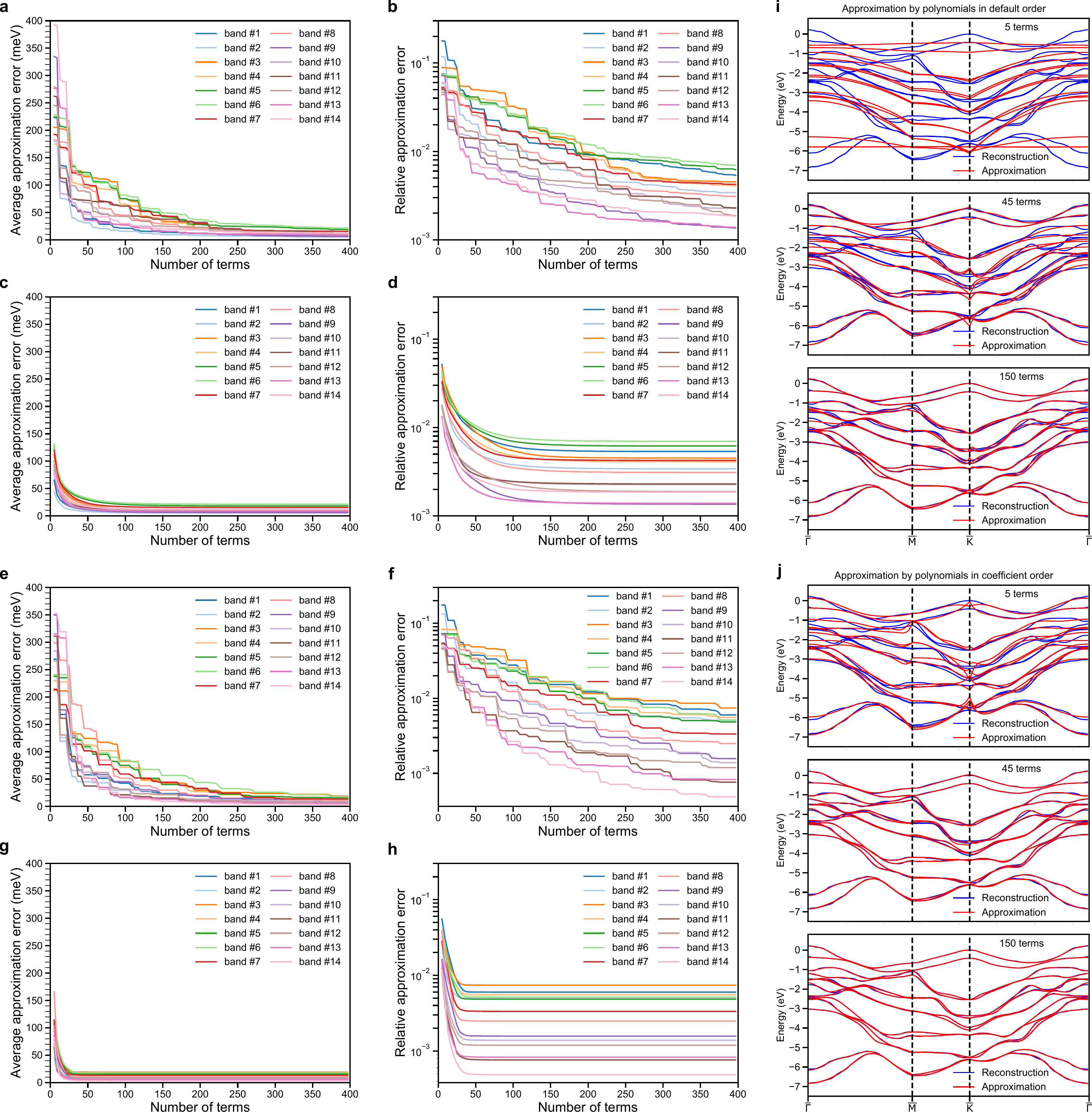} 
    \caption{\textbf{Approximation to the band structure of WSe$_2$ by a polynomial basis}. \textbf{a-j}, Demonstration of the convergence properties of the polynomial approximation using reconstructed photoemission band structure (\textbf{a-d}) and DFT band structure calculated at the LDA level (\textbf{e-h}). When summing the hexagonal Zernike polynomial in the default order, the average and relative approximation errors for the reconstructed (\textbf{a},\textbf{b}) and theoretical (\textbf{e},\textbf{f}) energy bands converge much slower than summing the polynomials in an ordering ranked by the magnitude of their coefficients (coefficient order). This observation is similar for reconstructed (\textbf{c},\textbf{d}) and theoretical (\textbf{g},\textbf{h}) energy bands. \textbf{i}-\textbf{j}, Visualization of the difference in convergence rates using the reconstructed band structure along the high-symmetry lines. The naturally-ordered polynomial basis has not yet converged with 150 terms (\textbf{i}), while the coefficient-ranked polynomials (\textbf{j}) produces an accurate approximation well within that limit.}
    \label{fig:approx}
\end{figure}

\subsection{Brillouin zone tiling}
The generation of a large and densely sampled patch of energy bands covering the first Brillouin zone and beyond is crucial for initializing the MRF model. To balance the computational cost using different XC functionals with the dense sampling similar to the experimental data grid, we used the symmetry properties of the Brillouin zone to tile the calculated momentum-space rectangular patch that covers the $\Gamma$, K and M points of the Brillouin zone. The hexagonal Brillouin zone of WSe$_2$ has a sixfold rotation symmetry axis and two independent mirror planes in the ($k_x$, $k_y$) coordinates. The initial rectangular patch is first symmetrized about the two mirror planes in the $\Gamma$-K and $\Gamma$-M directions to form a larger patch, which is then rotated by $60^{\circ}$ and $120^{\circ}$, respectively, and combined with the original mirror-symmetrized patch. The composite patch is then shifted along all six $\Gamma$-M directions by one unit cell distance and the result is cut to the required shape compatible with photoemission data.

\section{Band structure informatics}
\subsection{Global structure descriptors}
We use informatics tools for data retrieval, representation and comparison for entire bands. We extend the examples given in main text Fig. 3b to other bands of WSe$_2$ reconstructed in the present work. Supplementary Fig. \ref{fig:bandhzps} displays the band-wise comparison of dispersion surfaces within other DFT calculations. These results contain similar features as in main text Fig. 3a and c, reaffirming that the geometric featurization provides a sparse representation of the band dispersions and that the dispersion similarities are largely preserved despite the use of different exchange-correlation functionals in the DFT calculations. They may, therefore, be regarded as general features of the WSe$_2$ band structure.

In Supplementary Fig. \ref{fig:approx}, we demonstrate numerically the approximation capability of the hexagonal ZP basis set to all 14 valence bands of WSe$_2$. Despite the stark differences in energy dispersion, the approximation to reconstructed bands (Supplementary Fig. \ref{fig:approx}a-d) and theoretical band structure at the level of LDA-DFT (Supplementary Fig. \ref{fig:approx}e-h) show comparable convergence rates. Quantitatively, the approximation using hexagonal ZPs ordered by the magnitude of the corresponding coefficients (i.e. coefficient order) converges to within 10-30 meV/band within 50 polynomial basis terms, substantially faster than using the default order (see also Fig. 3b for reference).
The remaining errors are on par with the finite step size along the energy axis in the data ($\sim$ 18 meV) that results in the imperfect smoothness of the reconstructed bands. This further proves that the hexagonal ZPs can provide an accurate and sparse approximation for the band structure data. The trend of convergence between these two types of polynomial ordering is further illustrated in Supplementary Fig. \ref{fig:approx}i-j in the momentum path along high-symmetry lines of the reconstructed band structure.

\subsection{Local structure descriptors}
Local structural information includes energy gaps, effective masses, warpings, (avoided) crossings, etc. We extracted some of their associated parameters at and around three high-symmetry points ($\overline{\text{K}}$,  $\overline{\text{M}^{\prime}}$, and $\overline{\Gamma}$, see main text Fig. \ref{fig:localparams}a) and compiled the results in Supplementary Table \ref{tab:bsparams}. The dispersions and band structure parameters from the MAP reconstruction are compared with those extracted by the line-by-line fitting of the EDCs, which used the band energies from the reconstruction as initialization to improve robustness. Around $\overline{\text{K}}$, two spectral peaks corresponding to two spin-split bands were fit simultaneously, while around $\overline{\text{M}^{\prime}}$ and $\overline{\Gamma}$, four were fit simultaneously due to the spectral proximity of the first four valence bands (see Supplementary Fig. \ref{fig:theoryvsrecon}). The fitting is carried out using a linear superposition of Voigt lineshapes and the \textsf{lmfit} package \cite{Newville2019} with the reconstructed band energy as initialization (but not fixed). The fitting procedure iterates over the EDCs (e.g. a total of 50$\times$50 EDCs for the patch around $\overline{\text{M}^{\prime}}$). Unstable fits yielding erratic results (e.g. if differing greatly from neighboring values) are re-fit with either algorithmically or manually adjusted initialization. Supplementary Table \ref{tab:bsparams} shows that the local structural information from reconstruction is generally consistent with those obtained by iterative pointwise fitting while differing from DFT calculations. The deviations in the size of energy gaps at $\overline{\text{K}}$ and $\overline{\text{M}^{\prime}}$ between reconstruction and pointwise fitting lie in the same range as the momentum-averaged reconstruction errors (see Supplementary Section 2), which are due to the finite coordinate spacing in the data ($\sim$ 18 meV in energy).
\begin{table}[hbt!]
\renewcommand{\arraystretch}{1.2}
\centering
    \begin{threeparttable}
    \captionof{table}{\textbf{Band structure parameters from experiment and theory.} Effective masses of holes ($m_{\overline{\text{K}}}$), trigonal warping parameters ($C$) are extract at $\overline{\text{K}}$ point in the first two valence bands. Two directional effective masses at $\overline{\text{M}^{\prime}}$ ($m_{\overline{\text{M}^{\prime}}}$), and one at $\overline{\Gamma}$ ($m_{\overline{\Gamma}}$), are obtained for the first valence band. The energy gaps ($\Delta E$) between the first two valence bands are obtained at both $\overline{\text{K}}$ and $\overline{\text{M}^{\prime}}$ points. The number (1 or 2) in the subscript of the parameter symbols denotes the valence band index, $m_e$ is the mass of an isolated electron.}
    \label{tab:bsparams}
    \begin{tabular}{|c|c|c|c|c|c|}
    \hline
     Symmetry point &  Parameter & LDA recon. \tnote{1} & Line fitting \tnote{2} & LDA \tnote{3} & HSE06 \tnote{3} \\ \hline
     $\overline{\text{K}}$ & $m_{\overline{\text{K}},1}/m_e$ & $-$0.62 & $-$0.60 & $-$0.49 & $-$0.42 \\ \hline
     $\overline{\text{K}}$ & $m_{\overline{\text{K}},2}/m_e$ & $-$0.74 & $-$0.78 & $-$0.64 & $-$0.54 \\ \hline
     $\overline{\text{K}}$ & $C_{\overline{\text{K}},1}$ (eV$\cdot\text{\AA}^3$) & 5.3 & 5.8 & 6.2 & 4.5\\ \hline
     $\overline{\text{K}}$ & $C_{\overline{\text{K}},2}$ (eV$\cdot\text{\AA}^3$) & 4.0 & 3.9 & 3.9 & 3.2 \\ \hline
     $\overline{\text{K}}$ & $\Delta E_{\overline{\text{K}},1-2}$ (meV) & 419 & 446 & 485 & 467 \\ \hline
     $\overline{\text{M}^{\prime}}$ & $m_{\overline{\text{M}^{\prime}}-\overline{\Gamma},1}/m_e$ & 0.71 & 0.72 & 0.25 & 0.17 \\ \hline
     $\overline{\text{M}^{\prime}}$ & $m_{\overline{\text{M}^{\prime}}-\overline{\text{K}^{\prime}},1}/m_e$ & $-$1.6 & $-$1.5 & $-$1.1 & $-$0.90 \\ \hline
     $\overline{\text{M}^{\prime}}$ & $\Delta E_{\overline{\text{M}^{\prime}},1-2}$ (meV) & 352 & 338 & 127 & 48\\ \hline
     $\overline{\Gamma}$ & $m_{\overline{\Gamma},1}/m_e$ & $-$0.82 & $-$1.1 & $-$0.81 & $-$1.0 \\ \hline
    \end{tabular}
    \vspace{1ex}
    \begin{tablenotes}\footnotesize
    \item[1] Using band dispersion reconstructed globally by the proposed probabilistic machine learning algorithm with DFT calculation at the LDA level as the initialization.
    \item[2] Using band dispersion from iterative lineshape fitting of the energy distribution curves (in a region around the corresponding high-symmetry points).
    \item[3] With fully optimized structure, see Supplementary Table \ref{Table_1}.
    \end{tablenotes}
    \end{threeparttable}
\end{table}

The region extracted around $\overline{\text{K}}$ (see main text Fig. \ref{fig:localparams}d-e) contains about 10\% of the distance of $\overline{\Gamma}-\overline{\text{K}}$. Due to the strong trigonal warping (TW) effect in this class of materials, the effective masses and the TW parameters around $\overline{\text{K}}$ were fit simultaneously in 2D using the momentum-space model derived from \textbf{k}$\cdot$\textbf{p} theory \cite{Kormanyos2015}.
\begin{equation}
    E(\textbf{q}) = \frac{\hbar^2\textbf{q}^2}{2m_{\overline{\text{K}}}} + C|\textbf{q}|^3\cos(3\varphi_{\textbf{q}} + \theta) + E_0.
    \label{equ:kp}
\end{equation}
Here, \textbf{q} is the momentum vector $\textbf{k}$ recentered on a particular $\overline{\text{K}}$ (or $\overline{\text{K}^{\prime}}$) point by translation, $m_{\overline{\text{K}}}$ is the effective mass of the hole at $\overline{\text{K}}$ point, $C$ is the magnitude of the TW (named $C_{\text{3w}}$ in \cite{Kormanyos2015}), $\varphi_{\textbf{q}}$ is the polar angle in the coordinate system centered on a $\overline{\text{K}}$ (or $\overline{\text{K}^{\prime}}$) point, $\theta$ is an auxiliary fitting parameter used to accommodate the orientation of the TW with respect to the pixel coordinates defined by the rectangular region of interest, $E_0$ accounts for the energy offset. The energy gaps at $\overline{\text{K}}$ ($\Delta E_{\overline{\text{K}},1-2}$) and $\overline{\text{M}^{\prime}}$ ($\Delta E_{\overline{\text{M}^{\prime}},1-2}$) are illustrated in main text Fig. \ref{fig:localparams} (b and d), respectively. The $\overline{\text{M}^{\prime}}$ (or $\overline{\text{M}}$) point situates at a saddle point of the dispersion surface (first valence band), as shown in main text Fig. \ref{fig:localparams}b-c. Its lower symmetry (compared with $\overline{\text{K}}$, $\overline{\text{K}^{\prime}}$ and $\overline{\Gamma}$) means that the effective masses exhibits anisotropy, with opposite signs and magnitude along the $\overline{\text{M}^{\prime}}-\overline{\Gamma}$ and $\overline{\text{M}^{\prime}}-\overline{\text{K}^{\prime}}$ directions. We fit the dispersion locally using a model that also accounts for the spin-orbit interaction involving a linear momentum-dependent shift (Eq. 14 in \cite{Kormanyos2015}). The second valence band is not fitted at $\overline{\text{M}^{\prime}}$ due to the pronounced dispersion modulation by interband coupling unaccounted for in the existing saddle-shaped model. At around $\overline{\Gamma}$, a single effective mass is extracted by fitting a paraboloid to a local patch of the dispersion surface.

\printbibliography[keyword=si, heading=subbibliography, title=References]

\end{document}